\documentclass[a4paper,11pt]{article}

\usepackage{jheppub}

\usepackage[T1]{fontenc} 
\usepackage{dsfont}
\usepackage{framed}

\usepackage{caption}
\usepackage{subcaption}
\usepackage{dsfont}
\usepackage{tikz}
\usepackage[aligntableaux=center,boxsize=1.3em]{ytableau}
\usepackage{float}

\allowdisplaybreaks[1]

\def\({\left(}
\def\){\right)}
\def\[{\left[}
\def\]{\right]}
\def\<{\langle}
\def\>{\rangle}

\def\nn{\nonumber\\}
\def\pa{\partial}
\def\zb{\bar{z}}

\def\a{\alpha}
\def\b{\beta}
\def\g{\gamma}
\def\G{\Gamma}
\def\d{\delta}
\def\D{\Delta}
\def\e{\epsilon}

\def\l{\lambda}
\def\L{\Lambda}
\def\m{\mu}

\def\p{\pi}

\begin{document} 
	
\title{Anomalous dimensions from conformal field theory: Generalized \boldmath{$\phi^{2n+1}$} theories}
	
\author{Yongwei Guo}
\author{and Wenliang Li}
\emailAdd{liwliang3@mail.sysu.edu.cn}
\affiliation{School of Physics, Sun Yat-sen University, Guangzhou 510275, China}
	
\abstract{We investigate $\phi^{2n+1}$ deformations of the generalized free theory in the $\epsilon$ expansion, 
where the canonical kinetic term is generalized to a higher-derivative version.
For $n=1$, we use the conformal multiplet recombination method to determine the leading anomalous dimensions of the fundamental scalar operator $\phi$ and 
the bilinear composite operators $\mathcal J$.
Then we extend the $n=1$ analysis to the Potts model with $S_{N+1}$ symmetry 
and its higher-derivative generalization,  
in which $\phi$ is promoted to an $N$-component field.
We further examine the Chew-Frautschi plots and their $N$ dependence.
However, for each integer $n>1$, 
the leading anomalous dimensions of $\phi$ and $ \mathcal{J}$ are not fully determined and  
contain one unconstrained constant, 
which in the canonical cases can be fixed by the results from the diagrammatic method.
In all cases, we verify that the multiplet-recombination results are consistent with crossing symmetry using the analytic bootstrap methods.}
	
\maketitle 

\section{Introduction}

For a scalar field theory, the canonical kinetic term is typically proportional to $\pa_\m\phi\pa^{\m}\phi$,\footnote{Summation over repeated indices is implied.} with $\phi$ denoting the fundamental scalar.
This kinetic term can be rewritten as $\phi\,\Box\phi$ up to a total derivative, which we discard.
The expression $\phi\,\Box\phi$ can be generalized by employing a higher-order Laplacian.
In this paper, we use the word ``generalized'' to indicate a free theory with a higher-derivative kinetic term and the interacting deformations.
The action of the generalized free theory reads
\begin{align}\label{boxk-action-free}
	S \propto \int \mathrm{d}^dx \; \phi\,\Box^{k}\phi
	\,,
\end{align}
where $d$ is the spacetime dimension and $k$ is a positive integer (see \cite{Brust:2016gjy} for more details).
One can introduce interaction terms into \eqref{boxk-action-free} and investigate the infrared (IR) fixed points of the renormalization group flow (see, for example \cite{David:1992vv,Gracey:2017erc,Safari:2017irw,Safari:2017tgs}).
We assume that the IR fixed points possess conformal symmetry, 
allowing for the use of conformal field theory (CFT) methods to study the critical properties.

In a previous paper \cite{Guo:2023qtt}, we used CFT methods to study the generalized $\phi^{2n}$ theories in the $\e$ expansion.
The $\mathbb{Z}_2$-odd interactions were not addressed, as $n$ is a positive integer.
In this paper, we consider the $\mathbb{Z}_2$-odd cases.
Namely, we investigate the generalized $\phi^{2n+1}$ theories in the $\e=d_\text{u}-d$ expansion, where the upper critical dimensions are given by\footnote{It would be interesting to make sense of the corresponding nonperturbative $k>1$ theories, if such theories exist. This is an open problem, to the best our knowledge.}
\begin{align}
	d_\text{u}=
	2k \, \frac{2n+1}{2n-1}
	\,.
\end{align}
The first instance of these $\mathbb{Z}_2$-odd theories corresponds to $n=1$, i.e., the generalized $\phi^3$ theory.
We also refer to this theory as the generalized Yang-Lee edge singularity,\footnote{The $k=2$ Yang-Lee theory might be related to the zeros of the partition function of the Lifshitz theory.}
since the canonical $\phi^3$ theory describes the Yang-Lee edge singularity \cite{Fisher:1978pf}.

Before moving on to higher $n$, we also study the generalized $\phi^{3}$ theory with $S_{N+1}$ global symmetry,\footnote{One may also introduce an auxiliary field to define an O($N$) symmetric theory in $d=6-\e$ dimensions \cite{Fei:2014yja}. At large $N$, it is conjectured that the IR CFT at $d=5$ is dual to the Vasiliev higher-spin theory on $\text{AdS}_{6}$.} where we promote $\phi$ to an $N$-component field $\vec{\phi}$.
The symmetric group $S_{N+1}$ is the group of all permutations of $N+1$ objects.
It can also be viewed as the symmetry group of an $N$-simplex.
For example, $S_3$ is isomorphic to the symmetry group of an equilateral triangle, and this group consists of six elements: the identity, rotations by $2\p/3$ and $4\p/3$ about the center, and reflections about the three symmetry axes of the equilateral triangle.
The mismatch between the number of components of $\phi$ and the degree of the symmetric group originates from the lattice theory described in the next paragraph. 
In the generalized $S_{N+1}$-symmetric $\phi^{3}$ theory, the $\mathbb{Z}_2$ symmetry under the reflection $\vec{\phi} \rightarrow -\vec{\phi}$ is absent, as in the Yang-Lee case.\footnote{The symmetric group $S_{N+1}$ contains $\mathbb{Z}_2$ subgroups, but the elements of these subgroups do not act by $\vec{\phi} \rightarrow \pm \vec{\phi}$. See appendix \ref{Representation theory}.}
At $k=1$, this theory describes the $(N+1)$-state Potts model, so we also refer to the generalized $S_{N+1}$-symmetric $\phi^{3}$ theory as the generalized Potts model.\footnote{The critical point of the $k=2$ Potts model may be an $S_{N+1}$-symmetric version of an isotropic Lifshitz point.}

The $(N+1)$-state Potts model is a generalization of the Ising model \cite{Potts:1951rk}.
In the lattice spin definition,
the spin states can be represented by $N+1$ vectors in $\mathbb{R}^N$: $\vec{s}_i\in\{\vec{e}^{\,1},\vec{e}^{\,2}\ldots,\vec{e}^{\,N+1}\}$, where $i$ labels the lattice sites \cite{Zia:1975ha}.\footnote{The Potts model can also be defined using Fortuin-Kasteleyn clusters, which is equivalent to the spin definition if $N$ is a positive integer \cite{Fortuin:1971dw}.
The random cluster definition is more general and applicable to noninteger $N$.}
Here the vectors $\vec{e}^{\,\a}$ are of the same magnitude, and the angles between any pair of distinct vectors $(\vec{e}^{\,\a},\vec{e}^{\,\b})$ are equal.
For two nearest-neighbor sites, the interaction energy takes one value if the spins on both sites point in the same direction, and another value otherwise.
In other words, the interaction energy for the nearest-neighbor sites is proportional to the scalar product of the vectors.
The Hamiltonian is defined as the sum of the nearest-neighbor interaction terms:\footnote{The Ising model corresponds to $N=1$, where the two spin states are represented by two vectors pointing in opposite directions. 
}
$H_\text{Potts}=-J\sum_{\<ij\>} \vec{s}_i \cdot \vec{s}_j$, 
where $J$ is the coupling constant and $\<ij\>$ indicates the pairs of nearest-neighbor sites.
We assume that $J>0$, which corresponds to the ferromagnetic case.
At low temperatures, the spins tend to point in the same direction, and the $S_{N+1}$ symmetry is spontaneously broken, corresponding to the ordered phase.
At high temperatures, the system is in the disordered phase, which has $S_{N+1}$ symmetry.
At some temperature, the model undergoes a phase transition.
This transition is of first-order if $N$ is larger than a critical value $N_\text{c}$, and is continuous if $N\leqslant N_\text{c}$.\footnote{The critical value $N_\text{c}$ depends on the dimension $d$. It has been deduced that $N_\text{c}=3$ in 2D \cite{baxter1973potts}. For higher-dimensional cases, see the recent studies \cite{Chester:2022hzt,Wiese:2023vgq} and references therein.}
In accordance with this fact, the fixed-point coupling constants in the field-theoretic description are complex for large enough $N$, and the IR fixed points are described by complex CFTs in these cases \cite{Gorbenko:2018ncu,Gorbenko:2018dtm}.
Additionally, the Potts model is related to other models in certain limits.
For example, the $N\rightarrow\infty,0,-1$ limits correspond to the decoupled Yang-Lee theory, percolation theory, and spanning forest \cite{Jacobsen:2003qp,Caracciolo:2004hz,Deng:2006ur}.\footnote{\label{logf} The percolation theory and 2D spanning forest are associated with logarithmic CFTs \cite{Ivashkevich:1998na,Cardy:1999zp,Hogervorst:2016itc}.
For the $d>2$ spanning forest, see e.g., \cite{Safari:2020eut} and references therein.}

After discussing the $n=1$ case, we proceed to consider theories with greater integer $n$.\footnote{See \cite{Codello:2018nbe,Codello:2020mnt} for the studies of the canonical $n=2$ theory with $S_{N+1}$ symmetry. We only consider the single-field case for the $n>1$ theory in this paper.}
For $n=2$, the interaction is associated with $\phi^5$, 
and the corresponding CFT is expected to be a multicritical generalization of the Yang-Lee CFT.
More broadly, we consider the generalized $\phi^{2n+1}$ theory with $n>1$.
A notable feature is that, at leading order in the $\e$ expansion, a family of data is consistent with conformal symmetry, as far as our analysis is concerned.
This means that a large set of leading anomalous dimensions is determined up to one parameter, in contrast to the $n=1$ case where the anomalous dimensions are fixed. 
Like the $\phi^5$ theory, the $\phi^{2n+1}$ theories with higher $n$ are also associated with multicritical Yang-Lee models \cite{vonGehlen:1994rp,Lencses:2022ira,Lencses:2023evr,Lencses:2024wib}, and they provide a series of universality classes containing multicritical points with $\mathbb{Z}_2$-odd interactions.
See also \cite{Zambelli:2016cbw,Gracey:2017okb,Codello:2017qek,Codello:2017epp} for related studies of the canonical case $k=1$. 
This concludes our discussion of the theories that are investigated in this paper.

In these higher-derivative theories, we are interested in the fundamental scalar $\phi$ and the bilinear operators $\mathcal{J}_{\ell}^{(m)}$ of the schematic form $\phi\,\pa^{\ell}\Box^{m}\phi$, which are labeled by the spin $\ell$ and the number of pairs of contracted derivatives $m$.
In the free theory, the bilinear operators are (partially) conserved currents, if the spins are sufficiently large.
At $m=k-1$, the conserved currents satisfy the usual conservation conditions, i.e., they vanish when contracted with one derivative.
For $m<k-1$, the conservation law involves multiple derivatives:
\begin{align}
	\overbrace{\pa \ldots \pa}^{ c \text{ derivatives} }
	\!\!\!\! \mathcal{J}_{\ell}^{(m)}=0
	\,, \qquad
	\ell\geqslant c\equiv 2(k-m)-1\,,
\end{align}
where the indices are suppressed for simplicity.
The partially conserved currents do not vanish when we act with fewer derivatives.
These operators are related to the higher-spin symmetry of the free theory.
One can construct conserved charges from the partially conserved currents, generalizing the $k=1$ higher-spin algebra.
We refer to \cite{Brust:2016gjy} and references therein for more details.
For $k>1$, the interacting versions of the partially conserved currents play a significant role in the analytic bootstrap, 
as their twists are lower than those of the usual conserved currents. 
In the lightcone limit, 
the interacting counterparts of the partially conserved currents 
are more important than the Regge trajectories associated with the conserved currents.

To derive the anomalous dimensions of the bilinear operators, we employ the conformal multiplet recombination method initiated in \cite{Rychkov:2015naa}, which considered the canonical $\phi^4$ theory.
In \cite{Rychkov:2015naa}, it was further suggested that the results can be extended to the canonical $\phi^3$ theory at $d=6-\e$. 
\footnote{According to \cite{Rychkov:2015naa,Gliozzi:2016ysv,Gliozzi:2017gzh}, Yu Nakayama has studied the $\e$ expansion using the multiplet recombination in his unpublished notes, such as the $\e=6-d$ expansion of the $\phi^3$ theory.
} 
A key difference is the absence of $\mathbb{Z}_2$ symmetry in the $\phi^3$ theory, in contrast to the $\phi^4$ case.
Since the procedure in \cite{Rychkov:2015naa} relies on the $\mathbb{Z}_2$ symmetry, its generalization to the $\mathbb{Z}_2$-odd cases does not seem straightforward.
A similar approach based on the equation of motion was proposed to study the canonical $\phi^3$ theory \cite{Nii:2016lpa}. 
By applying the equation of motion to correlation functions, the constraints on the anomalous dimensions can be derived more readily.
In this paper, we revisit the $\phi^{2n+1}$ theory from the multiplet recombination perspective and extend the results to some spinning operators.
As a consistency check, we also verify that the results for the anomalous dimensions and operator product expansion (OPE) coefficients are consistent with crossing symmetry using the analytic bootstrap.

The multiplet recombination asserts that free multiplets recombine into long multiplets at interacting IR fixed points.
For instance, at $d=4-\e$, the Wilson-Fisher multiplet $\{\phi\}$ is a long multiplet, into which the free multiplets $\{\phi_{ \text{f} }\}$ and $\{\phi^3_{ \text{f} }\}$ recombine.
Here the subscript f implies that the operators are associated with the free theory. 
In the Gaussian limit, the descendant $\Box\phi$ is expected to become null and decouple, but a singular change in the normalization can lead to a physical operator, 
which is identified with the free-theory primary operator $\phi^3_{ \text{f} }$.
In other words, the Wilson-Fisher counterpart of $\phi^3_{ \text{f} }$ is identified as a descendant $\Box\phi$ up to a normalization factor \footnote{All physical operators in the free theory  should have counterparts in the interacting theory.}:
\begin{align}
	\lim_{\e \rightarrow 0} \( \e^{-1}\Box\phi \)
	\propto \phi^3_{ \text{f} }
	\,.
\end{align}
Despite the resemblance to the equation of motion $\Box\phi\propto\e\,\phi^3$, the multiplet recombination is conceptually different, as it can be formulated purely in the CFT language and does not require a Lagrangian description. 
The multiplet recombination above generalizes to various cases.
As in \cite{Guo:2023qtt}, we follow a suggestion of \cite{Gliozzi:2017gzh} and apply the multiplet recombination method to the $\mathbb{Z}_2$-odd deformations of the free higher-derivative CFTs. 
See \cite{Basu:2015gpa,Yamaguchi:2016pbj,Ghosh:2015opa,Raju:2015fza,Gliozzi:2016ysv,Roumpedakis:2016qcg,Soderberg:2017oaa,Behan:2017emf,Gliozzi:2017hni,Gliozzi:2017gzh,Nishioka:2022odm,Nishioka:2022qmj,Antunes:2022vtb,Guo:2023qtt} for applications of the multiplet recombination to numerous CFTs, 
and \cite{Safari:2017irw,Nii:2016lpa,Hasegawa:2016piv,Hasegawa:2018yqg,Skvortsov:2015pea,Giombi:2016hkj,Giombi:2017rhm,Giombi:2016zwa,Codello:2017qek,Codello:2018nbe,Antipin:2019vdg,Vacca:2019rsh,Giombi:2020rmc,Giombi:2020xah,Dey:2020jlc,Giombi:2021cnr,Safari:2021ocb,Zhou:2022pah,Bissi:2022bgu,Herzog:2022jlx,Giombi:2022vnz,SoderbergRousu:2023pbe} for applications with more emphasis on the equations of motion.

This paper is organized as follows.
In section \ref{Generalized Yang-Lee edge singularity}, we consider the generalized Yang-Lee edge singularity.
We first employ the multiplet recombination to compute the anomalous dimensions of $\phi$ and bilinear operators $\mathcal{J}$, and then we use the analytic bootstrap to verify that the results are consistent with crossing symmetry.
In section \ref{Generalized Potts model} we carry out a similar analysis, obtaining the anomalous dimensions in the generalized Potts model, and we check that the results are consistent with crossing symmetry as well.
Section \ref{Generalized phi-2n+1 theory} focuses on the generalized $\phi^{2n+1}$ theory with $n>1$, where we only consider the single-field case.
We determine the anomalous dimensions up to a free parameter, and then we verify the consistency between the multiplet-recombination results and crossing symmetry.
We summarize the main results in table \ref{summary}.
In appendix \ref{The action of Laplacians} we work out the action of Laplacians on the scalar three-point function.
In appendix \ref{Analytic bootstrap at subleading twist} we verify that the multiplet-recombination results in section \ref{Generalized Yang-Lee edge singularity}, \ref{Generalized Potts model}, and \ref{Generalized phi-2n+1 theory} are consistent with crossing symmetry at subleading twist.
We briefly review some representation theory of the symmetric group in appendix \ref{Representation theory}.
More details about the tensor structures in the Potts model can be found in appendix \ref{Tensor structures}.
In appendix \ref{Ratios of OPE coefficients} we compute the ratios of OPE coefficients relevant to the multiplet recombination.
In appendix \ref{An OPE coefficient in the generalized phi-2n+1 theory} we derive an OPE coefficient used in the analytic bootstrap.
In appendix \ref{logCFT} we discuss the $N\rightarrow 0$ limit of the generalized Potts model in more detail, especially the logarithmic structures. 
Appendix \ref{Generalized self-avoiding walks and loop-erased random walks} discusses some special limits in the generalized O($N$)-symmetric $\phi^{2n}$ theory.

\begin{table}[tbp]
	\centering
	\begin{tabular}{|c|c|c|}
		\hline
		Theories & $\g_\phi$ & $\g_{ \mathcal{J} }$ \\
		\hline 
		Generalized Yang-Lee edge singularity & \eqref{gamma1} & \eqref{gammaJ} \\
		Generalized Potts model & \eqref{gamma1-Potts} & \eqref{gammaJ-Potts} \\
		Generalized $\phi^{2n+1}$ theory with $n>1$ & \eqref{gamma1-phi-2n+1} & \eqref{gammaJ-phi-2n+1} \\
		\hline
	\end{tabular}
	\caption{Summary of the main results. The second column lists the anomalous dimensions of $\phi$, and the third column lists the anomalous dimensions of the bilinear operators of the schematic form $\mathcal{J}\sim\phi\,\pa^{\ell}\Box^{m}\phi$.}
	\label{summary}
\end{table}

\section{Generalized Yang-Lee edge singularity}
\label{Generalized Yang-Lee edge singularity}

Let us first consider the generalized Yang-Lee CFT.
At $d=d_{ \text{u} }-\e$, the action reads\footnote{We do not need to consider derivative interactions. The reason is that a derivative interaction should be constructed by fewer $\phi$'s, namely, two $\phi$'s, but a derivative term with two $\phi$'s is just the kinetic term, up to a total derivative.}
\begin{align}\label{Lagrangian-YL}
	S\propto\int \mathrm{d}^dx
	\( \phi\,\Box^k\phi
	+g\m^{\e/2} \phi^{3} \)
	\,,
\end{align}
where the upper critical dimension is $d_\text{u}=6k$.
The parameter $\m$ has mass dimension 1, so the coupling constant $g$ is dimensionless.
We emphasize that $g$ is allowed to be imaginary. 
Assuming that $\phi$ flips sign under space reflection $\mathcal{P}$ and the unit imaginary number $i$ changes sign under time reversal $\mathcal{T}$, the interaction term is $\mathcal{PT}$ invariant for imaginary $g$.
See \cite{Bender:2012ea,Bender:2013qp} for more details on the canonical case with $k=1$.

In this section, the anomalous dimensions of $\phi$ and bilinear operators $\mathcal{J}$ are derived from the conformal multiplet recombination.
Furthermore, we verify the consistency of the results with crossing symmetry using the analytic bootstrap.

\subsection{Multiplet recombination}
\label{Multiplet recombination}

In the spirit of CFT approaches, we do not refer to the action \eqref{Lagrangian-YL}.
Instead, the Yang-Lee CFT is characterized by
\begin{align}\label{boxk-phi}
	\lim_{\e \rightarrow 0} \(
	\a^{-1} \Box^{k}\phi \)
	=\phi^{2}_{ \text{f} }
	\,,
\end{align}
where $\a=\a(\e)$ vanishes in the Gaussian limit, i.e., $\lim_{\e \rightarrow 0}\a=0$.
The above equation means that we identify the descendant $\a^{-1}\Box^{k}\phi$ as the deformed version of $\phi^{2}_\text{f}$, and the normalization $\a^{-1}$ is introduced to ensure a finite Gaussian limit.
In other words, the free multiplets $\{ \phi_{ \text{f} } \}$ and $\{ \phi^{2}_{ \text{f} } \}$ recombine into a long multiplet $\{ \phi \}$.

To study the Yang-Lee CFT, the strategy is to examine the implications of \eqref{boxk-phi} for correlation functions.
More specifically, assuming that the $\e\rightarrow 0$ limit is smooth, the condition \eqref{boxk-phi} implies that the Gaussian limit of the correlators in the interacting theory should match the free correlators.
Such relations are referred to as matching conditions.
Since the deformed CFT data is encoded in the correlation functions, the matching conditions may provide some constraints on the data.

Our focus will be on the two-point and three-point functions.
We study the matching conditions for the scalar correlators first.
As we will see below, these conditions determine the anomalous dimension $\gamma_\phi$ and the normalization $\a$, as well as the OPE coefficient associated with $\phi\in\phi\times\phi$, at leading order in the $\e$ expansion.
Then, we consider the matching condition for the spinning correlator involving a primary bilinear operator.\footnote{The spin of the bilinear operator can be zero.}
Using the results for $\g_\phi$ and $\a$, we obtain the leading anomalous dimension of the bilinear operator.

The matching condition associated with the two-point function reads
\begin{align}\label{boxk-2pt}
	\lim_{\e \rightarrow 0} \( \a^{-2} 
	\< \Box^{k}\phi(x_{1})\Box^{k}\phi(x_{2}) \> \)
	=\< \phi^{2}_{ \text{f} }(x_{1})
	\phi^{2}_{ \text{f} }(x_{2}) \>
	\,.
\end{align}
Namely, the two-point function of the descendant $\a^{-1}\Box^{k}\phi$ should reduce to the free correlator of $\phi^{2}_{ \text{f} }$ in the Gaussian limit.
This leads to the constraint
\begin{align}\label{boxk-2pt-result}
	\lim_{\e \rightarrow 0} \[
	\a^{-2} \, 4^{2k} \(\D_{ \smash{\phi} }\)_{2k}
	\(\D_{ \smash{\phi} }-\frac{d-2}{2}\)_{2k}
	\frac{1}{ x_{12}^{ 2\D_{ \smash{\phi} }+4k } } \, \]
	=\frac{2}{ x_{12}^{ 2\D_{ \smash{\phi^2_{ \text{f} } } } } }
	\,,
\end{align}
where $(x)_{y}\equiv\frac{\G(x+y)}{\G(x)}$ is the Pochhammer symbol and $x_{ij}\equiv|x_{i}-x_{j}|$ is the distance between two operators.
We have chosen the normalization
\begin{align}\label{normalization}
	\< \phi(x_1)\phi(x_2) \> = \frac{1}{x_{12}^{ \smash{2\D_{\phi} } }}
	\,.
\end{align}
On the left-hand side of \eqref{boxk-2pt}, we act on $\< \phi(x_{1})\phi(x_{2}) \>$ with the Laplacians, obtaining the two-point function of the descendant $\Box^{k}\phi$.
The result follows from the identity
\begin{align}\label{box-identity}
	\Box_{x_1} \frac{1}{ x_{12}^{2\D} }
	=4\D(\D-d/2+1) \frac{1}{ x_{12}^{2\D+2} }
	\,.
\end{align}
On the right-hand side of \eqref{boxk-2pt}, the free correlator is given by Wick contractions. \footnote{We assume that the Gaussian limit is smooth, so the normalization in the free theory is also fixed by \eqref{normalization}, i.e., $\< \phi_\text{f}(x_1)\phi_\text{f}(x_2) \> = 1/x_{12}^{ \smash{2\D_{\phi_\text{f} } } }$.}
The functional forms in $x_{12}$ indeed match in \eqref{boxk-2pt-result}, since
\begin{align}
	\lim_{\e \rightarrow 0} \(
	2\D_{ \smash{\phi} }+4k \)
	=2\D_{ \smash{\phi_{ \text{f} } } }+4k
	=8k
	=2\D_{ \smash{\phi^2_{ \text{f} } } }
	\,.
\end{align}
The condition \eqref{boxk-2pt-result} thus imposes a constraint on $\D_{ \smash{\phi} }$ and $\a$.
We define the anomalous dimension of $\phi$ as
\begin{align}\label{gamma1-definition}
	\g_{\phi}=\D_{ \smash{\phi} }-\frac{d-2k}{2}
	\,,
\end{align}
and substitute $d=6k-\e$ into \eqref{boxk-2pt-result}.
Omitting the functional forms in $x_{12}$, we are left with
\begin{align}\label{alpha-gamma1}
	(-1)^{k-1}4^{2k}(k-1)!\,k!\(2k\)_{2k}
	\lim_{\e \rightarrow 0} \( \a^{-2} \g_{\phi} \)=2
	\,,
\end{align}
where $\g_{\phi}$ arises from the second Pochhammer symbol in \eqref{boxk-2pt-result}.
For the moment, we note that $\g_{\phi}$ is of order $\a^{2}$.
We will revisit the constraint \eqref{alpha-gamma1} later.

Now we consider three-point functions.
The first matching condition is
\begin{align}\label{boxk-3pt}
	\lim_{\e \rightarrow 0} \( \a^{-1}
	\< \Box^{k}\phi(x_{1})\phi(x_{2})\phi(x_{3}) \> \)
	=\< \phi^{2}_{ \text{f} }(x_{1})
	\phi_{ \text{f} }(x_{2})
	\phi_{ \text{f} }(x_{3}) \>
	\,.
\end{align}
As before, the right-hand side is computed using Wick contractions. 
On the left-hand side, we act on $\< \phi(x_{1})\phi(x_{2})\phi(x_{3}) \>$ with the Laplacians.
The result is worked out in appendix \ref{The action of Laplacians}.
We obtain
\begin{align}\label{boxk-3pt-result}
	\lim_{\e \rightarrow 0} \[ \a^{-1} \sum^{k}_{ \substack{i,j=0 \\
	i+j \geqslant k } } \( B_{i,j} \, C_{i,j}^{ \D_{ \smash{\phi} },\D_{ \smash{\phi} } }
	\frac{ x_{23}^{2(i+j-k)} }{ 
	x_{12}^{2i \vphantom{j} } x_{13}^{ 2j } } \)
	\frac{ \l_{\phi\phi\phi} }{ x_{12}^{ \D_{ \smash{\phi} } } 
	x_{13}^{ \D_{ \smash{\phi} } } x_{23}^{ \D_{ \smash{\phi} } } } \]
	=\frac{2}{ x_{12}^{ \D_{ \smash{\phi^2_{ \text{f} } } } } 
	x_{13}^{ \D_{ \smash{\phi^2_{ \text{f} } } } } }
	\,,
\end{align}
where $\l_{\phi\phi\phi}$ is the OPE coefficient of $\phi \in \phi \times \phi$.
The coefficients $B_{i,j}$ and $C_{i,j}^{a,b}$ are given in \eqref{Bij} and \eqref{Cij}.
Due to the last Pochhammer symbol in \eqref{Cij}, $C_{i,j}^{ \D_{ \smash{\phi} },\D_{ \smash{\phi} } }$ is proportional to the anomalous dimension $\g_{\phi}$, except for the $i=j=k$ case where $C_{i,j}^{ \D_{ \smash{\phi} },\D_{ \smash{\phi} } }$ is finite in the Gaussian limit.
In other words, the dominant term in the $\e\rightarrow 0$ limit is given by $i=j=k$.
Then, one can check that the functional forms match in \eqref{boxk-3pt-result}.
We obtain a constraint on $\l_{\phi\phi\phi}$ and $\a$,
\begin{align}\label{alpha-lambda}
	(-1)^{k}2^{2k}\(k\)_{k}^{2}
	\lim_{\e \rightarrow 0} \( \a^{-1}\l_{\phi\phi\phi} \)=2
	\,,
\end{align}
indicating that the OPE coefficient $\l_{\phi\phi\phi}$ is of order $\a$.
So far, we have obtained two constraints, i.e., \eqref{alpha-gamma1} and \eqref{alpha-lambda}, 
but they involve three unknown parameters: $\a$, $\g_\phi$, and $\l_{\phi\phi\phi}$. 
We would like to deduce another constraint that does not involve additional unknown parameters, 
so we consider the three-point function $\<\phi(x_{1})\phi(x_{2})\phi(x_{3}) \>$ again. 
In \eqref{boxk-3pt}, we have considered the matching conditions associated with one type of $\Box^k$. 
It is natural to introduce more $\Box^k$'s. 
However, as a result of the $\mathbb{Z}_2$ symmetry of the free theory, 
the free correlator $\< \phi^{2}_{ \text{f} }(x_{1})\phi^{2}_{ \text{f} }(x_{2})\phi_{ \text{f} }(x_{3}) \>$ vanishes 
and thus 
the matching condition associated with $\< \Box^{k}\phi(x_{1})\Box^{k}\phi(x_{2})\phi(x_{3}) \>$ is not 
particularly useful. 
If we consider three types of $\Box^k$'s,  
the corresponding free-theory correlator respects the $\mathbb{Z}_2$ symmetry, 
so this is a useful matching condition,
\begin{align}
	\lim_{\e \rightarrow 0} \( \a^{-3}
	\< \Box^{k}\phi(x_{1})\Box^{k}\phi(x_{2})\Box^{k}\phi(x_{3}) \> \)
	=\< \phi^{2}_{ \text{f} }(x_{1})
	\phi^{2}_{ \text{f} }(x_{2})
	\phi^{2}_{ \text{f} }(x_{3}) \>
	\,.
\end{align}
The action of the Laplacians is derived in appendix \ref{The action of Laplacians}. 
The explicit result associated with three types of $\Box^k$'s can be found in \eqref{3-box}.  
As before, the functional forms match, leaving us with a constraint on $\g_{\phi}$, $\l_{\phi\phi\phi}$, and $\a$:
\begin{align}\label{3box n=1}
	(-1)^{k-1}2^{12k-5}(3k-1)!\(1/2\)_{k}^3
	\lim_{\e \rightarrow 0} \[
	\a^{-3} (\e-6\g_{\phi}) \l_{\phi\phi\phi} \]=8
	\,.
\end{align}
If we divide the left-hand side by those of \eqref{alpha-gamma1} and \eqref{alpha-lambda}, then the right-hand side becomes 2. We have
\begin{align}\label{epsilon-gamma}
	\frac{(-1)^{k}2^{4k-4}\(1/2\)_{k}^{2} }{ k!\(3k\)_{k} }
	\[ \, \lim_{\e \rightarrow 0} \(
	\e\g_{\phi}^{-1}\)-6 \]=2
	\,.
\end{align}
A special case occurs if the coefficient outside the square brackets is $-1/3$.
However, this is impossible for positive integer $k$.
The reason is as follows.
The absolute value of the coefficient increases with $k$.
%, since the ratio of the coefficient for $k=k_0+1$ to that for $k=k_0$ is $\frac{54k_0^3+81k_0^2+39k_0+6}{32k_0^3+64k_0^2+38k_0+6}>1$.
This absolute value is smaller than $1/3$ for $k=4$, and slightly larger than $1/3$ for $k=5$.
This means that the coefficient cannot be $-1/3$.
We conclude that $\lim_{\e\rightarrow0}(\e\,\g_\phi^{-1})$ is finite for positive integer $k$ 
and $\g_\phi$ is of order $\e$.
It follows that $\a$ and $\l$ are of order $\e^{1/2}$. 
Moreover, we derive the explicit expression for the leading anomalous dimension
\begin{align}\label{gamma1}
	\g_{\phi}=\frac{\e}{ 2\(3+\frac{ (-1)^{k}k!\(3k\)_{k} }
	{2^{4k-4}\(1/2\)_{k}^{2} }\) }
	+\ldots
	\,,
\end{align}
which generalizes the $k=1$ result in \cite{Fisher:1978pf}.
Here and below, we use the ellipsis to denote the higher-order terms in $\e$.
\footnote{In the canonical case $k=1$, the higher-order terms in $\g_\phi$ are known to be associated with integer powers of $\e$ \cite{deAlcantaraBonfim:1981sy}. 
It is natural to expect that only integer powers of $\e$ appear in %the higher order terms for 
the generic $k>1$ counterparts as well. 
To verify this, it would be interesting to carry out the explicit computation for the $k>1$ theories in the more standard diagrammatic approach.}
For $k=1$, the anomalous dimension of $\phi$ is $\g_{\phi}=-\e/18+O(\e^2)$, which is negative and implies that the canonical Yang-Lee theory is nonunitary.\footnote{For $k>1$, the generalized free theory itself is nonunitary 
due to the violation of the unitarity bound.} 
The anomalous dimension $\g_{\phi}$ is particularly large for $k=5$, as the denominator in \eqref{gamma1} is small in this case. 
In addition, we also obtain
\begin{align}\label{alpha2}
	\a^{2}&=-\frac{ (2k)! \, (3k-1)! \, \e }
	{ 2^{5-4k}k!+6(-1)^{k}\frac{ \(1/2\)_{k}^{2} }{ \(3k\)_{k} } }
	+\ldots
	\,,
	\\
	\label{lambda}
	\l_{\phi\phi\phi}&=\frac{(-1)^{k}\a}{ 2^{6k-5}(3/2)_{k-1}^{2} }+\ldots
\end{align}
At leading order, the OPE coefficient $\l_{\phi\phi\phi}$ and $\a$ are imaginary if $k$ is even or $k\in\{1,3\}$.

Now we consider the spinning correlator involving a generic primary bilinear operator $\mathcal{J}_{\ell}^{(m)} \sim \phi\pa^{\ell}\Box^{m}\phi$.\footnote{The operators $\mathcal{J}_{\ell}^{(m)}$ should be nondegenerate. See, for example, footnote 12 of \cite{Guo:2023qtt}.}
The matching condition is
\begin{align}\label{boxk-current}
	\lim_{\e \rightarrow 0} \( \a^{-2}
	\< \Box^{k}\phi(x_{1})\Box^{k}\phi(x_{2})\mathcal{J}_{\ell}^{(m)}(x_{3},z) \> \)
	=\< \phi^{2}_{ \text{f} }(x_{1})
	\phi^{2}_{ \text{f} }(x_{2})
	\mathcal{J}^{(m)}_{ \ell,\text{f} }(x_{3},z) \>
	\,.
\end{align}
We have used the index-free notation where the spin indices are contracted with those of the polarization vector $z\in\mathbb{C}^{d}$.
We simplify the calculation further by focusing on the leading term in the $x_{3} \rightarrow \infty$ expansion \cite{Gliozzi:2017gzh}.
This makes it easier to extract the constraint given by \eqref{boxk-current}, since the action of the Laplacians on the leading term is simpler.
See section 3 of \cite{Guo:2023qtt} for more details on the treatment of the spinning correlator.
We have
\begin{align}\label{boxk-current-result}
	&\lim_{\e\rightarrow 0}\Bigg[ \a^{-2}4^{2k}
	\(\D_{ \smash{\phi} }-\frac{ \D_{ \mathcal{J} } }{2}+\frac\ell 2\)_{2k}
	\(\D_{ \smash{\phi} }-\frac{ \D_{ \mathcal{J} } }{2} -\frac \ell 2-\frac{d-2}{2} \)_{2k} \nn
	& \times \frac{ \l_{ \phi\phi\mathcal{J} } \(z \cdot x_{12}\)^{\ell}  }
	{ x_{12}^{ 2\D_{ \smash{\phi} }+4k-\D_{ \mathcal{J} }+\ell }|x_{3}|^{ 2\D_{\mathcal{J} } } }
	+O ( \, |x_{3}|^{-2\D_{\mathcal{J} }-1 } ) \Bigg]
	=\frac{ \l_{ \phi^2 \phi^2 \mathcal{J},\text{f} } \(z \cdot x_{12}\)^{\ell}  }
	{ x_{12}^{ 2\D_{ \smash{ \phi^2_\text{f} } }-\D_{ \smash{ \mathcal{J}_\text{f} } }+\ell }|x_{3}|^{ 2\D_{ \smash{ \mathcal{J}_\text{f} } } } }
	+O ( \, |x_{3}|^{-2\D_{ \smash{ \mathcal{J}_\text{f} } }-1 } )
	\,,
\end{align}
where $\l_{ \phi\phi\mathcal{J} }$ and $\l_{ \phi^2 \phi^2 \mathcal{J}, \text{f} }$ are the OPE coefficients of $\mathcal{J}\in \phi \times \phi$ and $\mathcal{J}_{ \text{f} } \in \phi^{2}_{ \text{f} } \times \phi^{2}_{ \text{f} }$.
The anomalous dimensions of bilinear operators are defined by
\begin{align}\label{gammaJ-definition}
	\g_{ \mathcal{J} }=\D_{ \mathcal{J} }-(d-2k+2m+\ell)
	\,,
\end{align}
where the terms in parentheses correspond to the classical dimension of $\mathcal{J}_{\ell}^{(m)}$.
The functional forms in \eqref{boxk-current-result} match.
We then write
\begin{align}\label{gammaJ-constraint}
	(-1)^m 4^{2k}(2k-m-1)!m!(k+\ell+m)_{2k}
	\lim_{\e \rightarrow 0} 
	\[ \a^{-2}\( \g_{\phi}-\frac{ \g_{ \mathcal{J} } }{2} \) \]
	=\frac{ \l_{ \phi^2 \phi^2 \mathcal{J},\text{f} } }{ \l_{ \phi\phi\mathcal{J},\text{f} } }
	\,,
\end{align}
where $\lim_{\e \rightarrow 0}\l_{ \phi\phi\mathcal{J} }=\l_{ \phi\phi\mathcal{J},\text{f} }$ is used.
Using the ratio of OPE coefficients $\frac{ \l_{ \phi^2 \phi^2 \mathcal{J},\text{f} } }{ \l_{ \phi\phi\mathcal{J},\text{f} } }=4$ derived in appendix \ref{Ratios of OPE coefficients}, we obtain
\begin{align}\label{gammaJ}
	\g_{ \mathcal{J}^{(m)}_\ell }= 2\g_{\phi}
	-\frac{(-1)^m \a^{2} }{ 2^{4k-3}m!(2k-m-1)!(k+\ell+m)_{2k} }
	+\ldots
	\,,
\end{align}
where $\g_{\phi}$ and $\a^{2}$ are given in \eqref{gamma1} and \eqref{alpha2}.
This is the $k\geqslant 1$ generalization of the $k=1$ result in \cite{Giombi:2016hkj}.
The stress tensor is the spin-2 current on the highest trajectory $m=k-1$, and its anomalous dimension vanishes to order $\e$, as expected.

The spin-0 operator on the lowest trajectory $\mathcal{J}_{\ell=0}^{(m=0)}=\phi^{2}$ is not a primary, so the matching condition \eqref{boxk-current} does not apply in this case.
Nevertheless, we can set $\ell=m=0$ in \eqref{gammaJ}.
It turns out that we obtain the shadow of $\phi$,
\begin{align}\label{shadow-YL}
	\D_{ \smash{\phi} }+\D_{ \mathcal{J} }\big|_{\ell=m=0}=d+0\e+\ldots
	\,,
\end{align}
which holds for $k \geqslant 1$.
The $k=1$ case was already noticed in \cite{Goncalves:2018nlv}.

\subsection{Analytic bootstrap}
\label{Analytic bootstrap}

In this subsection, we verify the consistency between the results above and crossing symmetry.
In the lightcone limit of the crossing equation for the four-point function $\< \phi \phi \phi \phi \>$, the anomalous dimensions of the bilinear operators are read off from the terms with an enhanced singularity in the crossed channel.\footnote{The meaning of the enhanced singularity is as follows. An expression with an enhanced singularity is power-law divergent after repeated action of the conformal Casimir operator (see, e.g., \cite{Alday:2015eya}).}
The results are consistent with those derived from the multiplet recombination.

The four-point function of identical scalars takes the form
\begin{align}
	\< \phi(x_{1})\phi(x_{2})\phi(x_{3})\phi(x_{4}) \>
	=\frac{ g(z,\zb) }{ x_{12}^{ 2\D_{ \smash{\phi} } }x_{34}^{ 2\D_{ \smash{\phi} } } }
	\,,
\end{align}
where $z$ and $\zb$ are defined by
\begin{align}\label{z-zb}
	z \zb=\frac{ x^2_{12}x^2_{34} }{ x^2_{13}x^2_{24} }
	\,, \qquad
	(1-z)(1-\zb)=\frac{ x^2_{14}x^2_{23} }{ x^2_{13}x^2_{24} }
	\,.
\end{align}
The crossing equation is
\begin{align}\label{crossing}
	g(z,\zb)=\frac{ z^{ \D_{ \smash{\phi} } }\zb^{ \D_{ \smash{\phi} } } }
	{ (1-z)^{ \D_{ \smash{\phi} } }(1-\zb)^{ \D_{ \smash{\phi} } } } g(1-\zb,1-z)
	\,.
\end{align}
The conformal block decomposition of the left-hand side reads
\begin{align}\label{block-decomposition}
	g(z,\zb)=\sum_{ \mathcal{O} } P_{ \mathcal{O} } G_{\mathcal{O}}(z,\zb)
	\,,
\end{align}
where $P_{ \mathcal{O} }$ is the squared OPE coefficient of $\mathcal{O}$ and $G_{ \mathcal{O} }$ is the conformal block associated with $\mathcal{O}$.
We consider the lightcone limit $z\rightarrow 0$.
The leading term of the conformal block in the lightcone expansion is
\begin{align}\label{block-generic}
	G_{ \mathcal{O} }(z,\zb)=z^{ (\D_{ \mathcal{O} }-\ell)/2 } k_{h}(\zb) + O (z^{ (\D_{ \mathcal{O} }-\ell)/2+1 } )
	\,.
\end{align}
Here $h=\frac{\D_{ \mathcal{O} }+\ell}{2}$ is the conformal spin and $k_{h}(\zb)$ denotes the SL(2,$\mathbb{R}$) block
\begin{align}
	k_{h}(\zb)=\zb^{h} {}_2F_1(h,h;2h;\zb)
	\,,
\end{align}
where ${}_2F_1(a,b;c;\zb)$ is the Gaussian hypergeometric function.
In this convention, the OPE coefficients have a different normalization from those in the multiplet recombination approach.
They are related by two-point function coefficients $\l_{ \mathcal{O}\mathcal{O}\mathds{1} }$:
\begin{align}\label{OPE-normalization}
	P_{ \mathcal{O} }= \frac{ \l_{ \phi\phi\mathcal{O} }^2 }{ \l_{ \mathcal{O}\mathcal{O}\mathds{1} } }
	\,.
\end{align}
The right-hand side of \eqref{crossing} has the conformal block decomposition
\begin{align}\label{expand-g}
	g(1-\zb,1-z)=1+P_{\phi} G_{\phi}(1-\zb,1-z)+\ldots
	\,,
\end{align}
where 1 is the contribution of the identity operator and the second term is the contribution of $\phi$.
The ellipsis represents terms of higher order in $\e$, and they will be omitted below.
At the leading order in the $z \rightarrow 0$ expansion, the right-hand side of \eqref{crossing} implies an enhanced singularity in the limit $\zb\rightarrow 1$.
This behavior matches the infinite sums of the SL(2,$\mathbb{R}$) blocks on the left-hand side of \eqref{crossing} with $\mathcal{O}=\mathcal{J}^{(m=0)}_{\ell}$, thanks to the asymptotic relation \cite{Simmons-Duffin:2016wlq}
\begin{align}\label{match-singularity}
	\sum_{\ell=0,2,4,\ldots} 2S_{p}(h) \, k_{h}(\zb) \sim \( \frac{1-\zb}{\zb} \)^{p}
	\,, \qquad
	(\zb \rightarrow 1)
	\,,
\end{align}
where the coefficient is
\begin{align}\label{S-formula}
	S_{p}(h)=\frac{ \G(h-p-1) }{ \G(-p)^{2}\G(h+p+1) } \frac{ \G(h)^{2} }{ \G(2h-1) }
	\,.
\end{align}
The analysis can be extended to higher-twist bilinear operators, i.e., to the case of $m>0$.
In appendix \ref{Analytic bootstrap at subleading twist}, we also verify the multiplet-recombination results at $m=1$ using the analytic bootstrap.

Let us expand the left-hand side of \eqref{crossing} in the anomalous dimension
\begin{align}\label{gamma-crossing}
	\tilde{\g}_{ \mathcal{J} }=\D_{ \mathcal{J} }-\(2\D_{ \smash{\phi} }+2m+\ell\)
	\,,
\end{align}
where $\D_{ \smash{\phi} }$ is the full scaling dimension of $\phi$.
The conformal block associated with $\mathcal{J}_{\ell}^{(m=0)}$ becomes
\begin{align}\label{Jblock-expand}
	G_{ \mathcal{J} }(z,\zb)=z^{ \D_{ \smash{\phi} } } \( k_{h}(\zb)+\frac{1}{2}\tilde{\g}_{ \mathcal{J} } k_{h}(\zb)\log z+ O( \tilde{\g}_{ \mathcal{J} }^{2} ) \) + O(z^{ (\D_{ \mathcal{J} }-\ell)/2+1 } )
	\,,
\end{align}
where we have suppressed the indices associated with $\mathcal{J}_{\ell}^{(0)}$ for simplicity.
Consider the right-hand side of the crossing equation \eqref{crossing} at leading order in the lightcone expansion.
The identity operator contributes to the enhanced singularity at order $\e^{0}$.
The corresponding term is $(\frac{1-\zb}{\zb})^{-\D_{ \smash{ \phi_\text{f} } } }$, with $\D_{ \smash{ \phi_\text{f} } }=2k$.
The free squared OPE coefficients of $\mathcal{J}^{(0)}_{\ell}$ can be derived using \eqref{match-singularity}:
\begin{align}\label{free-OPE}
	P_{ \smash{ \mathcal{J}^{(0)}_{\ell} },\text{f} } = 2S_{-2k}(h)
	=2\frac{ (2k)_{\ell}^{2} }{ \ell! \, (4k+\ell-1)_{\ell} }
	\,.
\end{align}
The leading correction on the right-hand side of \eqref{crossing} is of order $\e^{1}$, as $P_\phi \sim \e^1$ according to \eqref{lambda} and \eqref{OPE-normalization}.
So we conclude that $\tilde{\g}_{ \mathcal{J} } \sim \e^1$ in \eqref{Jblock-expand}, which is necessary for the consistency of \eqref{crossing}.
At order $\e^1$, the enhanced singularity is associated with the conformal block $G_{\phi}(1-\zb,1-z)$ in the crossed channel.
To extract the anomalous dimensions $\tilde{\g}_{ \mathcal{J} }$ in \eqref{Jblock-expand}, we consider the $\log z$ part of a scalar block \cite{Li:2020ijq}
\begin{align}\label{scalar-block}
	G_{ \mathcal{O} }(1-\zb,1-z)\Big|_{ \log z \text{ part} } \!\! =-\frac{ \G(\D_{ \smash{ \mathcal{O} } }) }{ \G(\D_{ \smash{ \mathcal{O} } }/2)^{2} }(1-\zb)^{ \frac{\D_{ \smash{ \mathcal{O} } } }{2} } {}_2F_1 \! \( \frac{ \D_{ \smash{ \mathcal{O} } } }{2}, \frac{ \D_{ \smash{ \mathcal{O} } } }{2};\D_{ \smash{ \mathcal{O} } }-\frac{d-2}{2};1-\zb \) +O(z)
	\,,
\end{align}
where $\mathcal{O}$ is a scalar operator.
We have
\begin{align}\label{phi-contribution}
	&\frac{ \zb^{ \D_{ \smash{\phi} } } }{ (1-\zb)^{ \D_{ \smash{\phi} } } }
	P_{\phi}
	G_{\phi}(1-\zb,1-z)\Big|_{ \log z \text{ part} }
	\nn=\;&P_{\phi} \frac{\G(2k)}{ \G(k)^{2} } \sum_{i=0}^{k-1}(-1)^{i+1}
	\frac{ (1-2k)_{i}(k)_{i} }{ i! \, (1-k)_{i} }
	\( \frac{1-\zb}{\zb} \)^{i-k} +\text{Casimir-regular terms}
	 +\ldots
	\,,
\end{align}
where the Casimir-regular terms do not contribute to the enhanced singularity.
Using the OPE coefficient \eqref{lambda} derived from the multiplet recombination, we obtain the anomalous dimension
\begin{align}\label{crossing gamma YL}
	\tilde{\g}_{ \mathcal{J}_\ell^{(0)} }=-\frac{\a^2}{2^{4 k-3}(2 k-1)! (k+\ell)_{2 k}}+\ldots
	\,,
\end{align}
which agrees with \eqref{gammaJ} at $m=0$, confirming the consistency of the multiplet-recombination results \eqref{lambda} and \eqref{gammaJ} with crossing symmetry.
Note that this agreement holds for arbitrary $\a$.\footnote{In the generalized $\phi^{2n}$ theories, we used the analytic bootstrap to verify the multiplet-recombination results as well \cite{Guo:2023qtt}.
The results from the two approaches also agree for arbitrary $\a$.
}

Due to the argument $\D_{ \smash{\mathcal{O} } }-\frac{d-2}{2}$, the hypergeometric function in \eqref{scalar-block} with $\mathcal{O}=\phi$ diverges as $1/\g_{\phi}$ in the Gaussian limit.
Nonetheless, the conformal block decomposition \eqref{expand-g} makes sense in the $\e$ expansion, because $P_{\phi}/\g_{\phi}$ is finite as $\e \rightarrow 0$.
Using \eqref{gamma1}, we find that the finite term is exactly equal to the contribution of $\phi^2_{\text{f} }$ in the free theory:
\begin{align}
	\lim_{\e\rightarrow 0} \[
	P_{\phi} G_{\phi}(1-\zb,1-z) \]
	= P_{ \phi^2_{ \text{f} } } \, 
	G_{\phi^2_{\text{f} } }(1-\zb,1-z)
	\,.
\end{align}
This is a manifestation of the multiplet recombination in the conformal block expansion.

\section{Generalized Potts model}
\label{Generalized Potts model}

We can extend the $\phi^3$ discussion to the case of an $N$-component field $\phi_{a}$, where $a=1,2,\ldots,N$.
The action for the generalized ($N+1$)-state Potts model is 
\begin{align}
	S \propto \int \mathrm{d}^dx \(
	\sum_{a=1}^{N}\phi_{a}\Box^{k}\phi_{a} + g\m^{\e/2} \sum_{a,b,c=1}^{N} T_{abc} \, \phi_{a}\phi_{b}\phi_{c} \)
	\,,
\end{align}
where $T_{abc}$ is an $S_{N+1}$-invariant tensor\footnote{The rank-3 invariant tensor of $S_{N+1}$ is unique up to a scalar multiple \cite{wallace1978spin}.}
\begin{align}\label{T-definition}
	T_{abc}=\sum_{\a=1}^{N+1} e_{a}^{\a} e_{b}^{\a} e_{c}^{\a}
	\,.
\end{align}
As mentioned in the Introduction, the vectors $e_{a}^{1},e_{a}^{2}\ldots,e_{a}^{N+1}$ have the same magnitude and any two distinct vectors $e_{a}^{\a}$ and $e_{a}^{\b}$ form the same angle.
In fact, these vectors are associated with the vertices of an $N$-simplex, whose symmetry group is $S_{N+1}$. 
The field $\phi_a$ lives in an $N$-dimensional irreducible representation of $S_{N+1}$ called the standard representation.
More details can be found in appendix \ref{Representation theory}.

In this section, we obtain the anomalous dimension of $\phi_{a}$ and bilinear operators $\mathcal{J}_{\ell}^{(m)}$ using the multiplet recombination.
For positive integers $N>2$, there are four types of bilinear operators, due to the tensor product decomposition \cite{murnaghan1938analysis}
\begin{align}\label{tensor-decomposition}
	(N,1)\otimes(N,1)=(N+1)+(N,1)+(N-1,2)+(N-1,1^{2})
	\,.
\end{align}
Here the partitions of $N+1$ boxes represent the Young diagrams:
\begin{align}
	(N+1)&=
	\overbrace{
		\begin{ytableau}
			{} & {} & \none[\ldots] & {}
	\end{ytableau} }^{N+1 \text{ columns} }
	\,,
	\qquad\qquad\quad\;\;
	(N,1)=
	\overbrace{
		\begin{ytableau}
			{} & {} & \none[\ldots] & {} \\
			{}
	\end{ytableau} }^{N \text{ columns} }
	\,,
	\nn
	{}
	\nn
	(N-1,2)&=
	\overbrace{
		\begin{ytableau}
			{} & {} & {} & \none[\ldots] & {} \\
			{} & {}
	\end{ytableau} }^{N-1 \text{ columns} }
	\,,
	\qquad
	(N-1,1^{2})=
	\overbrace{
		\begin{ytableau}
			{} & {} & \none[\ldots] & {} \\
			{} \\
			{}
	\end{ytableau} }^{N-1 \text{ columns} }
	\,.
\end{align}
We refer to these four representations as the trivial, standard, symmetric traceless and antisymmetric representations (see appendix \ref{Representation theory} for an explanation of these names).
The general results for the anomalous dimensions depend on $N$, and some particularly interesting cases are the percolation theory ($N\rightarrow 0$) and the spanning forest ($N\rightarrow -1$).
As in section \ref{Analytic bootstrap}, we verify the recombination results using the analytic bootstrap in section \ref{Analytic bootstrap Potts}.

\subsection{Multiplet recombination}
\label{Multiplet recombination Potts}

The multifield version of \eqref{boxk-phi} takes the form
\begin{align}\label{boxk-phi-Potts}
	\lim_{\e \rightarrow 0} \(
	\a^{-1} \Box^{k}\phi_{a} \)
	=\sum_{b,c=1}^{N} T_{abc} \, \phi_{b,\text{f}}\,\phi_{c,\text{f}}
	\,,
\end{align}
where $T_{abc}$ is given in \eqref{T-definition}.
Below, we derive the anomalous dimensions of $\phi_a$ and the four types of bilinear operators using \eqref{boxk-phi-Potts}.
The procedure is similar to that of section \ref{Multiplet recombination}.
The difference is that here the correlators have tensor structures related to the $S_{N+1}$ symmetry, and the constraint equations are modified by $N$-dependent factors.

The matching condition analogous to \eqref{boxk-2pt} reads
\begin{align}\label{2pt-Potts}
	\lim_{\e \rightarrow 0} \( \a^{-2} 
	\< \Box^{k}\phi_{a_{1} }(x_{1})\Box^{k}\phi_{a_{2} }(x_{2}) \> \)
	=\< K_{a_1, \text{f} } (x_{1})
	K_{a_2, \text{f} }(x_{2}) \>
	\,,
\end{align}
where we have defined
\begin{align}
	K_{a,\text{f}}\equiv\sum_{b,c=1}^{N} T_{abc} \, \phi_{b,\text{f}}\,\phi_{c,\text{f}}
	\,.
\end{align}
The left-hand side of \eqref{2pt-Potts} is nearly identical to that of \eqref{boxk-2pt-result}.
The difference is that here we have a tensor structure $\d_{ a_1 a_2 }$, which is the only rank-2 $S_{N+1}$-invariant tensor (up to a scalar multiple).\footnote{The uniqueness can be seen from the tensor product decomposition \eqref{tensor-decomposition}, where the trivial representation occurs only once on the right-hand side \cite{wallace1978spin}.}
The correlator on the right-hand side is again given by Wick contractions.
We obtain
\begin{align}
	\lim_{\e \rightarrow 0} \[
	\a^{-2} \, 4^{2k} \(\D_{ \smash{\phi} }\)_{2k}
	\(\D_{ \smash{\phi} }-\frac{d-2}{2}\)_{2k}
	\frac{ \d_{ a_1 a_2 } }{ x_{12}^{ 2\D_{ \smash{\phi} }+4k } } \, \]
	=\frac{ 2(N-1)(N+1)^{2} \, \d_{ a_1 a_2 } }{ x_{12}^{ 2\D_{ \smash{K_{ \text{f} } } } } }
	\,,
\end{align}
where $\D_{ K_\text{f} }$ denotes the scaling dimension of $K_{ a,\text{f} }$.
We have used \eqref{e-relation} and \eqref{T-definition}, which lead to $\d_{ a_1 a_2 }$ and the $N$-dependent coefficient on the right-hand side.
The functional forms and the tensor structures match, so we are left with the constraint
\begin{align}\label{alpha-gamma1-Potts}
	(-1)^{k-1}4^{2k}(k-1)!\,k!\(2k\)_{2k}
	\lim_{\e \rightarrow 0} \( \a^{-2} \g_{\phi} \)=2(N-1)(N+1)^{2}
	\,,
\end{align}
where $\g_{\phi}$ is the anomalous dimension of $\phi_{a}$ defined by \eqref{gamma1-definition}.

For three-point functions, we consider the matching condition
\begin{align}
	\lim_{\e \rightarrow 0} \( \a^{-1}
	\< \Box^{k}\phi_{ a_{1} }(x_{1})\phi_{ a_{2} }(x_{2})\phi_{ a_{3} }(x_{3}) \> \)
	= \< K_{a_1,\text{f}}(x_{1}) \,
	\phi_{ a_{2},\text{f} }(x_{2})\phi_{ a_{3},\text{f} }(x_{3}) \>
	\,.
\end{align}
On the left-hand side, the tensor structure is given by $T_{a_1 a_2 a_3}$, which is the only rank-3 $S_{N+1}$-invariant tensor (up to a scalar multiple).
On the right-hand side, the free correlator also has the tensor structure $T_{a_1 a_2 a_3}$.
We obtain
\begin{align}
	\lim_{\e \rightarrow 0} \[ \a^{-1} \sum^{k}_{ \substack{i,j=0 \\
			i+j \geqslant k } } \( B_{i,j} \, C_{i,j}^{ \D_{ \smash{\phi} },\D_{ \smash{\phi} } }
	\frac{ x_{23}^{2(i+j-k)} }{ 
		x_{12}^{2i \vphantom{j} } x_{13}^{ 2j } } \)
	\frac{ \l_{\phi\phi\phi} T_{a_1 a_2 a_3} }{ x_{12}^{ \D_{ \smash{\phi} } } 
		x_{13}^{ \D_{ \smash{\phi} } } x_{23}^{ \D_{ \smash{\phi} } } } \]
	=\frac{ 2\,T_{a_1 a_2 a_3} }{ x_{12}^{ \D_{ \smash{ K_{ \text{f} } } } } 
		x_{13}^{ \D_{ \smash{ K_{ \text{f} } } } } }
	\,,
\end{align}
where $B_{i,j}$ and $C_{i,j}$ are the same as in \eqref{Bij} and \eqref{Cij}.
The constraint on $\l_{\phi\phi\phi}$ and $\a$ takes the same form as \eqref{alpha-lambda}:
\begin{align}\label{alpha-lambda-Potts}
	(-1)^{k}2^{2k}(k)_{k}^{2}
	\lim_{\e \rightarrow 0} \( \a^{-1}\l_{\phi\phi\phi} \)=2
	\,.
\end{align}
Next, we consider the matching condition
\begin{align}
	\lim_{\e \rightarrow 0} \( \a^{-3}
	\< \Box^{k}\phi_{a_1}(x_{1})\Box^{k}\phi_{a_2}(x_{2})\Box^{k}\phi_{a_3}(x_{3}) \> \)
	=\<K_{a_1,\text{f}}(x_{1})
	K_{a_2,\text{f}}(x_{2})
	K_{a_3,\text{f}}(x_{3}) \>
	\,,
\end{align}
Again, the action of the Laplacians is given in \eqref{3-box}.
In the $\e$ expansion, this leads to the constraint
\begin{align}\label{alpha-gamma1-lambda-Potts}
	(-1)^{k-1}2^{12k-5}(3k-1)!\(1/2\)_{k}^3
	\lim_{\e \rightarrow 0} \[
	\a^{-3} (\e-6\g_{\phi}) \l_{\phi\phi\phi} \]=8(N-2)(N+1)^{2}
	\,.
\end{align}
Combining \eqref{alpha-gamma1-Potts}, \eqref{alpha-lambda-Potts} and \eqref{alpha-gamma1-lambda-Potts}, we obtain the anomalous dimension\vspace{0.85em}
\begin{align}\label{gamma1-Potts}
	\g_{\phi}=\frac{\e}{ 2\(3+\frac{N-2}{N-1}\frac{ (-1)^{k}k!(3k)_k }{ 2^{4k-4}(1/2)_k^2}\) }
	+\ldots
\end{align}
At $k=1$, the result agrees with that from the renormalization group analysis \cite{Priest:1976zz,Amit:1976pz}.
We also obtain $\a$ and the OPE coefficient of $\phi$:
\begin{align}
	\label{alpha-Potts}
	\a^2&=-\frac{ (2k)! \, (3k-1)! \, \e }
	{ (N+1)^{2}\( (N-2)2^{5-4k}k!+6(N-1)(-1)^k\frac{(1/2)_k^2}{(3k)_k} \) }
	+\ldots \,, \\
	\label{lambda-Potts}
	\l_{\phi\phi\phi}&=\frac{(-1)^{k}\a}{ 2^{6k-5}(3/2)_{k-1}^{2} }+\ldots
	\,.
\end{align}
The expression \eqref{gamma1-Potts} is very similar to \eqref{gamma1}, but here an $N$-dependent factor appears in the second term in the denominator.\footnote{Interestingly, the anomalous dimension of $\phi$ is $k$-independent to order $\e$ at $N=2$: $\g_{\phi}=\e/6+\ldots$
For $N=1$, we have $\g_\phi=0 \e+\ldots$ due to $T_{111}=0$. 
See \cite{Fang:2022ufx,Fang:2022tav} and references therein 
for the Fortuin-Kasteleyn representation of the Ising model above the standard upper critical dimension, i.e., $d>4$. }
As mentioned in the Introduction, the $N \rightarrow \infty$ limit yields the decoupled Yang-Lee theory.
One can check that
\begin{align}\label{infinte-N-gamma1}
	\lim_{N \rightarrow \infty}
	\g_{\phi}^{
	\text{Potts}
	}=\g_{\phi}^{
	\text{YL}
	}
	\,,
\end{align}
where $\g_{\phi}^{\text{Potts} }$ and $\g_{\phi}^{\text{YL} }$ are given in \eqref{gamma1-Potts} and \eqref{gamma1}.
Moreover, we can take the $N\rightarrow 0,-1$ limits of \eqref{gamma1-Potts} to study the  generalized percolation theory and spanning forest.
The anomalous dimensions are
\begin{align}
	\g_{\phi}\big|_{N \rightarrow 0}
	=\frac{\e}{
	2\( 3+\frac{(-1)^k k! (3k)_k}{2^{4k-5} (1/2)_k^2} \)
	}+\ldots
	\,, \qquad
	\g_{\phi}\big|_{N \rightarrow -1}
	=\frac{\e}{
	2\( 3+\frac{3(-1)^k k! (3k)_k}{2^{4k-3} (1/2)_k^2} \)
	}+\ldots
	\,.
\end{align}
At $k=1$, they lead to the critical exponents of percolation theory $\eta_\text{percolation}=-\e/21+O(\e^2)$, and spanning forest $\eta_\text{spanning forest}=-\e/15+O(\e^2)$.

Note that \eqref{gamma1-Potts} is divergent for $N=N_\text{c}$, 
where the critical value is
\begin{align}
N_\text{c}=1+\frac{1}{1+\frac{3 (-1)^k 2^{4k-4} (1/2)_k^2}{k! (3k)_k}}
\,.
\end{align}
For $k=1$, we have $N_\text{c}=7/3$. 
In fact, the constraints \eqref{alpha-gamma1-Potts}, \eqref{alpha-lambda-Potts}, and \eqref{alpha-gamma1-lambda-Potts} have no interacting solution at $N=N_\text{c}$ if $\gamma_\phi$ is of order $\e^1$ or higher. 
The Potts analog of \eqref{epsilon-gamma} is 
\begin{align}
\lim_{\e \rightarrow 0}\left( \e\g_{\phi}^{-1}\right)=
\left(6+\frac{(-1)^k2^{3+2k}k!(k+1/2)_k}{(k)_{2k}}
\right) \frac {N-N_\text{c}}{N-1}\,,
\end{align}
so the way out is to assume that $\gamma_\phi$ is of lower order than $\e^1$.  
\footnote{Below \eqref{epsilon-gamma}, we argue that $\g_\phi$ should be of first order in $\e$. 
The Yang-Lee result does not apply to the discussion here due to the important finite-$N$ effects, 
which are not captured by the large $N$ limit associated with the Yang-Lee case.
}
Then we can omit the $\e$ term in \eqref{alpha-gamma1-lambda-Potts}.  
In this way, \eqref{alpha-gamma1-Potts}, \eqref{alpha-lambda-Potts}, and \eqref{alpha-gamma1-lambda-Potts} 
are consistent with an interacting solution $\a\neq 0$.   
However, as one of the three constraints becomes redundant, 
there remains a free parameter in $(\g_\phi, \a^2, \l_{\phi\phi\phi})$. 
According to \eqref{alpha-gamma1-Potts}, $\a^2$ is also of lower order than $\epsilon^{1}$ at $N=N_\text{c}$. 
If $\a$ takes the form $(-\e)^{\theta}$ with $\theta<\frac 1 2 $, 
then this coefficient can have both real and imaginary parts,  
which is different from a purely imaginary number $(-\e)^{1/2}$ in the canonical Yang-Lee case. 
\footnote{We assume that the implicit coefficient of $(-\e)^{\theta}$ is a real number. }
A complex $\a$ is associated with a complex CFT, rather than a real or an imaginary CFT. 
\footnote{As a basic example of imaginary CFTs, 
the canonical Yang-Lee CFT can be formulated as a real CFT using the rotation $\phi\rightarrow i\phi$. }
For the canonical case $k=1$, we can make use of the explicit results from the diagrammatic calculation. 
The two-loop $\beta$ function from \cite{deAlcantaraBonfim:1981sy} 
implies that the leading fixed-point coupling constant $g$ takes the form $(-\e)^{1/4}$, so $\alpha\sim g$ is indeed complex. 
In addition, the complex CFTs can arise for other values of $N$ at lower $d$. 
We refer to \cite{Wiese:2023vgq} and references therein for more details on the canonical case $k=1$.

Now, we consider the bilinear operators.
The matching condition reads
\begin{align}\label{boxk-current-Potts}
	\lim_{\e \rightarrow 0} \( \a^{-2}
	\< \Box^{k}\phi_{ a_{1} }(x_{1})\Box^{k}\phi_{ a_{2} }(x_{2})\mathcal{J}_{\ell}^{(m)}(x_{3},z) \> \) 
	=\<   K_{a_1,\text{f}} (x_{1})
	K_{a_2,\text{f}} (x_{2})
	\mathcal{J}^{(m)}_{ \ell,\text{f} }(x_{3},z) \>
	\,,
\end{align}
which holds for all four types of bilinear operators.
We obtain an equation similar to \eqref{boxk-current-result}:
\begin{align}
	&\lim_{\e\rightarrow 0}\Bigg[ \a^{-2}4^{2k}
	\(\D_{ \smash{\phi} }-\frac{ \D_{ \mathcal{J} } }{2}+\frac\ell 2\)_{2k}
	\(\D_{ \smash{\phi} }-\frac{ \D_{ \mathcal{J} } }{2} -\frac \ell 2-\frac{d-2}{2} \)_{2k} \nn
	& \times \frac{ \l_{ \phi\phi\mathcal{J} } \(z \cdot x_{12}\)^{\ell} \mathbf{t} }
	{ x_{12}^{ 2\D_{ \smash{\phi} }+4k-\D_{ \mathcal{J} }+\ell }|x_{3}|^{ 2\D_{\mathcal{J} } } }
	+O ( \, |x_{3}|^{-2\D_{\mathcal{J} }-1 } ) \Bigg]
	=\frac{ \l_{ KK\mathcal{J},\text{f} } \(z \cdot x_{12}\)^{\ell} \mathbf{t} }
	{ x_{12}^{ 2\D_{ \smash{ K_{ \text{f} } } }-\D_{ \mathcal{J}_\text{f} }+\ell }|x_{3}|^{ 2\D_{\mathcal{J}_\text{f} } } }
	+O ( \, |x_{3}|^{-2\D_{\mathcal{J}\smash{, \text{f} } }-1 } )
	\,,
\end{align}
where $\l_{ KK\mathcal{J},\text{f} }$ denotes the OPE coefficient of $\mathcal{J}_{ab,\text{f} } \in K_{a,\text{f} } \times K_{b,\text{f} }$.
The tensor structures $\mathbf{t}$ on both sides match, because the operators on both sides of \eqref{boxk-current-Potts} live in the same $S_{N+1}$ representations.
The constraint on the anomalous dimensions can be written collectively as
\begin{align}\label{boxk-current-result-Potts}
	(-1)^{m}4^{2k}m!(2k-m-1)! (k+\ell+m)_{2k}
	\lim_{\e \rightarrow 0} 
	\[ \a^{-2}\( \g_{\phi}-\frac{ \g_{ \mathcal{J} } }{2} \) \]
	=\frac{ \l_{ KK\mathcal{J},\text{f} } }{ \l_{ \phi\phi\mathcal{J},\text{f} } }
	\,,
\end{align}
which is of the same form as \eqref{gammaJ-constraint}.
The difference is that here the ratio of free OPE coefficients depends on $N$.
The result is
\begin{align}\label{gammaJ-Potts}
	\g_{ \mathcal{J}^{(m)}_\ell } = 2\g_{\phi}
	-\frac{ (-1)^m R \, \a^{2} }{ 2^{4k-1}m!(2k-m-1)!(k+\ell+m)_{2k} }
	+\ldots
	\,,
\end{align}
where the ratio $R\equiv\frac{ \l_{ KK\mathcal{J},\text{f} } }{ \l_{ \phi\phi\mathcal{J},\text{f} } }$ is worked out in appendix \ref{Ratios of OPE coefficients}:
\begin{align}\label{ratio}
	R=
	\begin{cases}
		\; 4(N-1)(N+1)^{2} \qquad & \text{trivial} \\
		\; 4(N-2)(N+1)^{2}  & \text{standard} \\
		\; -4(N+1)^{2} & \text{symmetric traceless} \\
		\; -4(N+1)^{2} & \text{antisymmetric}
		\,,
	\end{cases}
\end{align}
For the trivial and standard representations, the $N\rightarrow \infty$ limit of \eqref{gammaJ-Potts} leads to the anomalous dimension \eqref{gammaJ} in the Yang-Lee theory.
For the spin-0 trivial and symmetric traceless operators in the canonical case, the results agree with \cite{theumann1984crossover}. 
On the highest trajectory $m=k-1$, the spin-2 current in the trivial representation $(N+1)$ is the stress tensor.
As expected, the anomalous dimension of the stress tensor is $0\e+\ldots$ according to \eqref{gammaJ-Potts}.
At $k=1$, the set of leading anomalous dimensions $\{\g_\phi,\g_{\mathcal{J}}\}$ contains negative elements when $N>7/3$, 
which indicates that the canonical Potts model is nonunitary near the upper critical dimension. 
Moreover, there exists a shadow relation in the standard representation,
\begin{align}\label{shadow-Potts}
	\D_{ \smash{\phi} }+\D_{ \mathcal{J}_{ \text{standard} } }\big|_{ \ell=m=0 }=d+0\e+\ldots
	\,,
\end{align}
which applies to any positive integer $k$.
In the $N\rightarrow\infty$ limit, this equation is related to \eqref{shadow-YL} in the generalized Yang-Lee theory. 
If we take the limit $N\rightarrow 0$, 
there are degeneracies in the scaling dimensions, 
leading to logarithmic structures (see appendix \ref{logCFT}).\footnote{We also discuss some special limits of the generalized O($N$)-symmetric $\phi^{2n}$ theory in appendix \ref{Generalized self-avoiding walks and loop-erased random walks}.} 
In the limit $N \rightarrow -1$, the ratio of OPE coefficients \eqref{ratio} vanishes, but the formal product $R\,\a^2$ in \eqref{gammaJ-Potts} remains finite because $\a^2$ diverges as $1/(N+1)^2$. 

Below, we examine the Chew-Frautschi plots to gain further insights into the Regge trajectories of the (generalized) Potts model.
According to the procedure in \cite{Caron-Huot:2022eqs}, 
the analytic continuation of the trajectories is given by 
\begin{align}\label{trajectory-Potts}
	&( \D-d/2 )^2 \nn
	=\;&\(
	d-2k+2m+\ell+\g_{
	\mathcal{J}
	}-d/2
	\)^2 \nn
	=\;&\(
	k+2m+\ell+2\g_1
	-\frac{ (-1)^mR \, \a^2 }{
	2^{4k-1}m!(2k-m-1)!(k+m+\ell)_{2k}
	}-\frac{\e}{2}
	+\ldots
	\)^2
	\,.
\end{align}
Using this relation between $\D$ and $\ell$, we plot the Regge trajectories of the bilinear operators $\mathcal{J}^{(m)}_{\ell}$.

\begin{figure}[h]
	\centering
	\includegraphics[width=.9\textwidth,origin=c,angle=0]{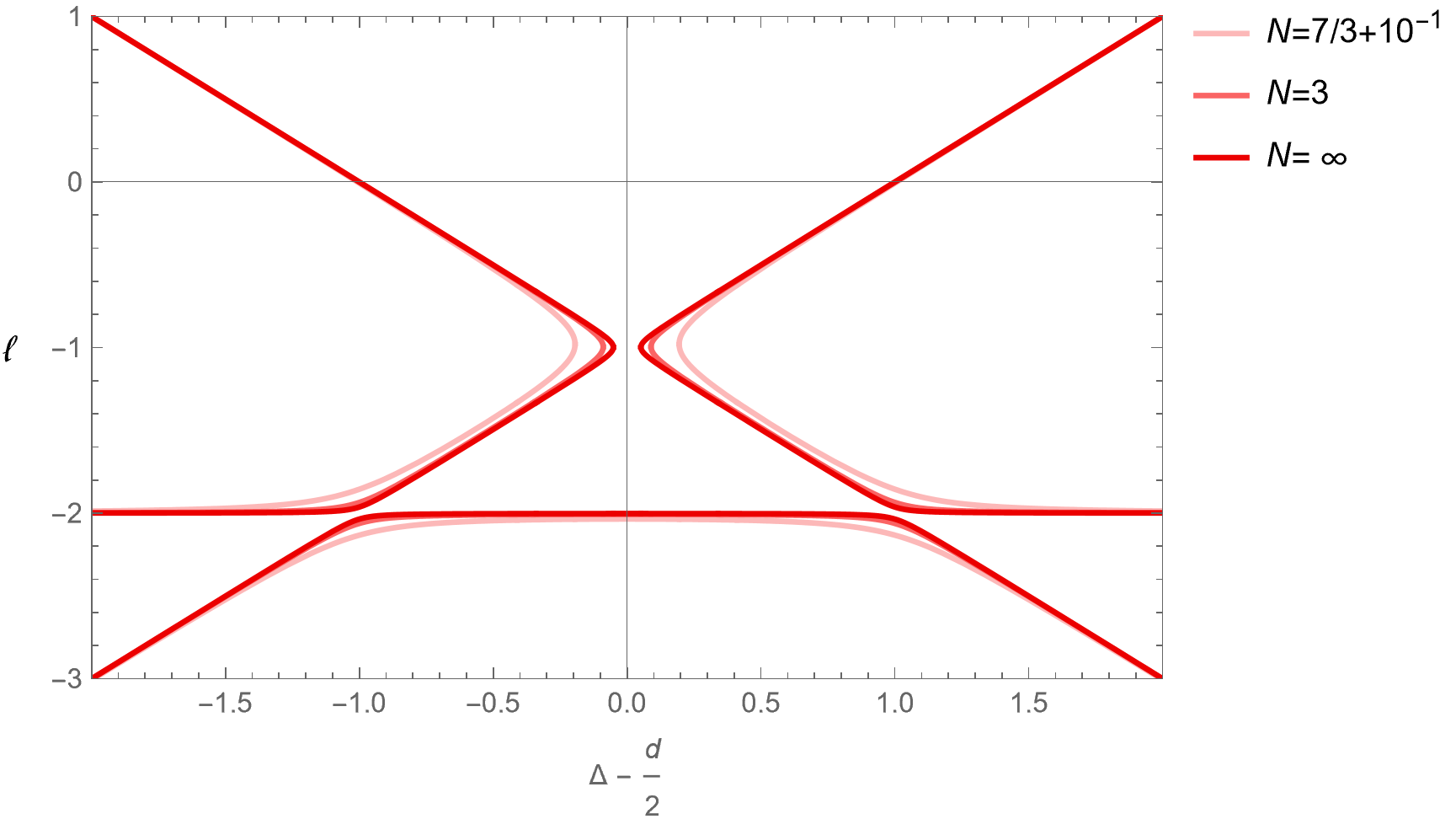}
	\caption{Regge trajectories of $\mathcal{J}^{(m=0)}_{\ell,\text{trivial} }$. The $N=\infty$ case corresponds to the decoupled Yang-Lee theory. The plot is made at $k=1$ and $\e=10^{-3}$.}
	\label{CF-largeN}
\end{figure}

\begin{figure}[h]
	\centering
	\includegraphics[width=.9\textwidth,origin=c,angle=0]{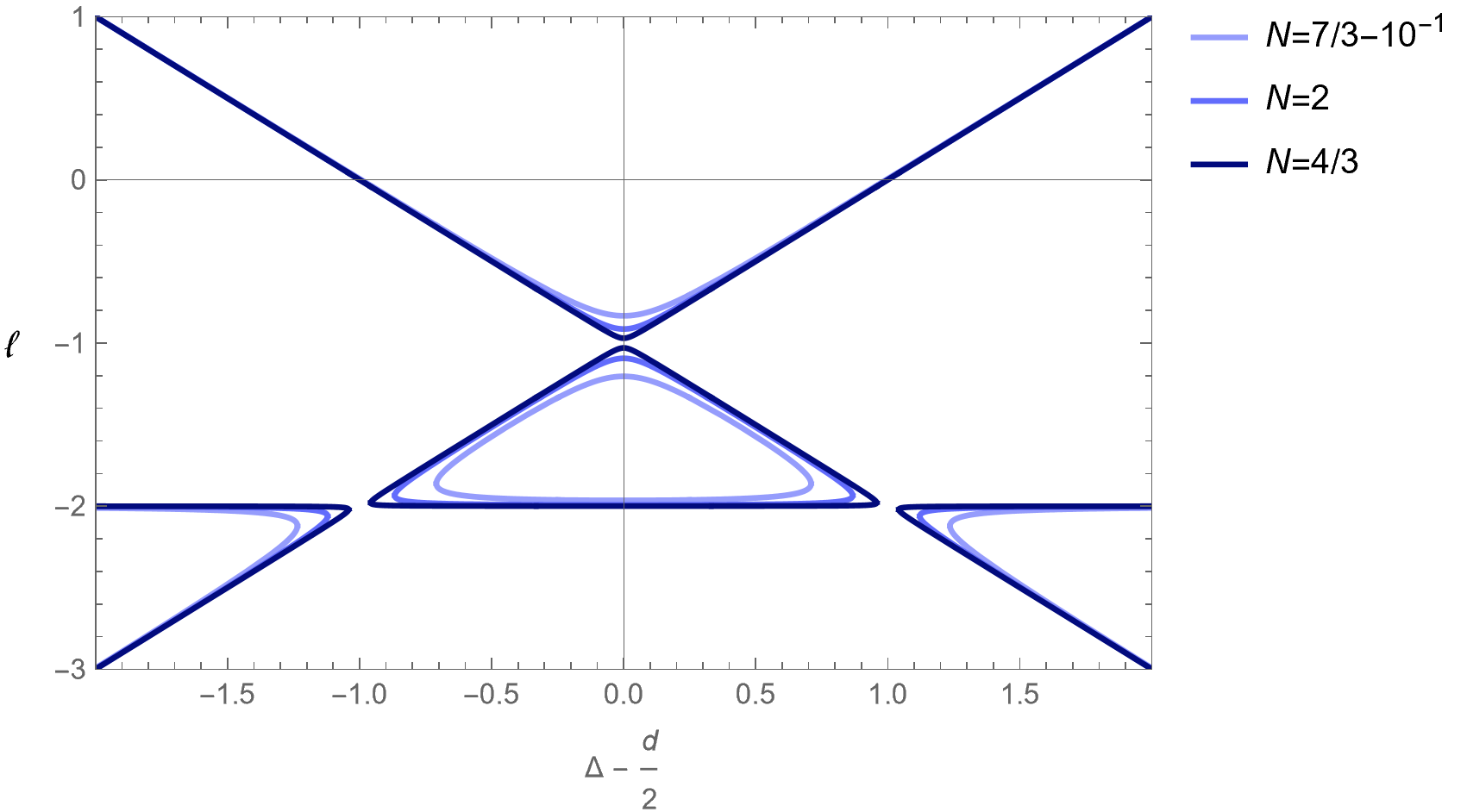}
	\caption{Regge trajectories associated with $\mathcal{J}^{(m=0)}_{\ell, \,\text{trivial} }$. The trajectory in the interacting theory approaches the free-theory trajectory as $N \rightarrow 1$. We used $k=1$ and $\e=10^{-3}$.}
	\label{CF-smallN}
\end{figure}

\begin{figure}[h]
	\centering
	\includegraphics[width=.7\textwidth,origin=c,angle=0]{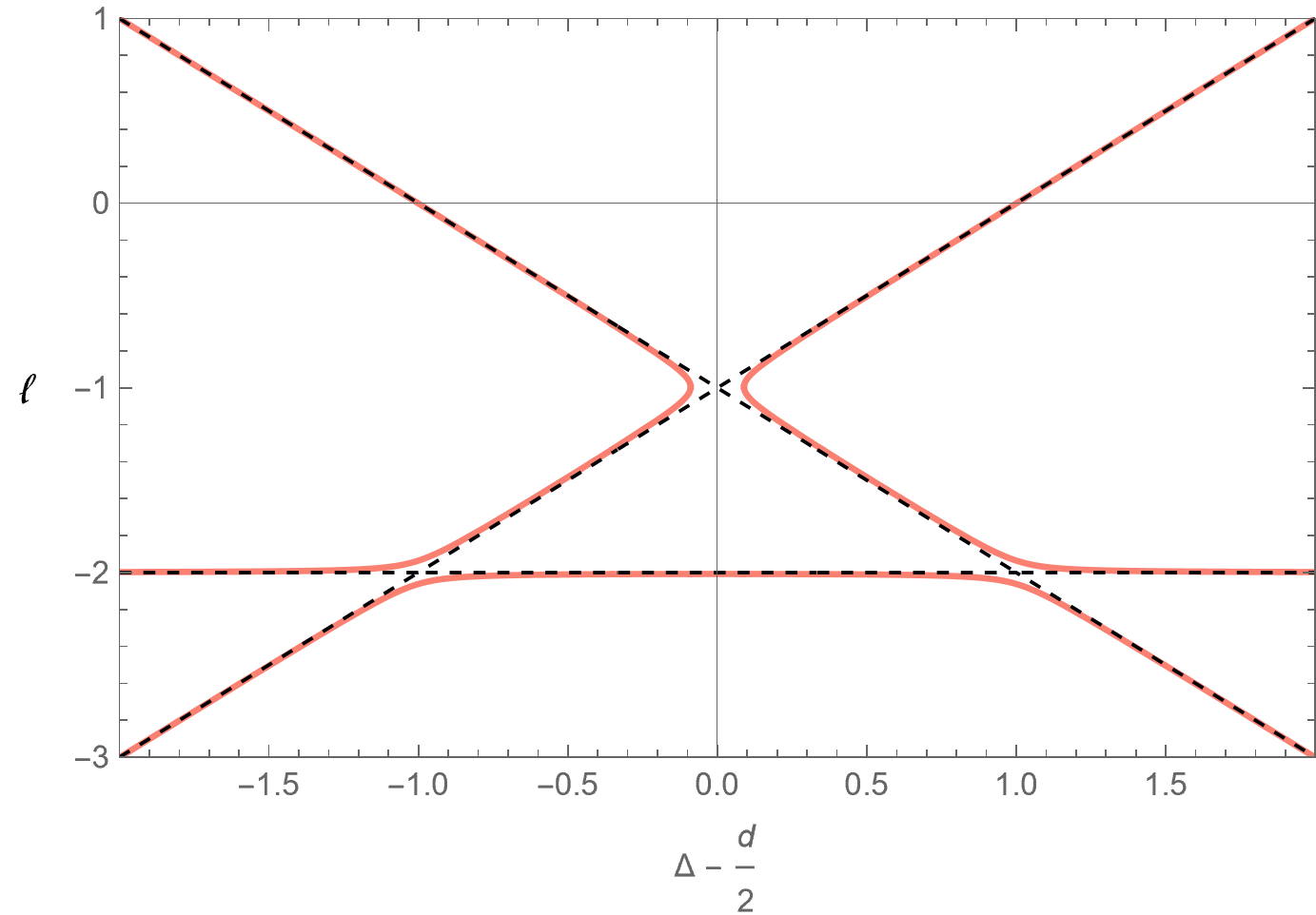}
	\caption{Chew-Frautschi plot for $\mathcal{J}^{(m=0)}_{\ell, \,\text{trivial} }$ at $N=3$. The dashed lines indicate the $\e\rightarrow0$ limit of the Regge trajectory. We set $k=1$ and $\epsilon=10^{-3}$.}
	\label{CF-N3}
\end{figure}

We first focus on the canonical case $k=1$.
A Chew-Frautschi plot associated with bilinear operators in the trivial representation is given in figure \ref{CF-N3}.
The plot varies with $N$, as shown in figure \ref{CF-largeN}.
The trajectories intersect with the horizontal axis.
Substituting $\ell=0$ into \eqref{trajectory-Potts}, we find two solutions
\begin{align}\label{CF-Delta}
	\D^*_\pm-\frac d 2=
	\pm\(
	1+\frac{3R'-11N+23}{6(3N-7)} \, \e
	+\ldots
	\)
	\,,
\end{align}
where we have defined $R'\equiv R/(N+1)^2$.
The solution $\D^*_-$ in the standard representation is precisely the scaling dimension of $\phi_a$, and $\D^*_+$ is given by $\D_{ \mathcal{ J_{ \text{standard} } } }|_{ \ell=m=0 }$.
This is consistent with the shadow relation \eqref{shadow-Potts}, since $\D^*_\pm$ are related by the shadow transformation $\D \rightarrow d-\D$.

\begin{figure}[h]
	\centering
	\includegraphics[width=.7\textwidth,origin=c,angle=0]{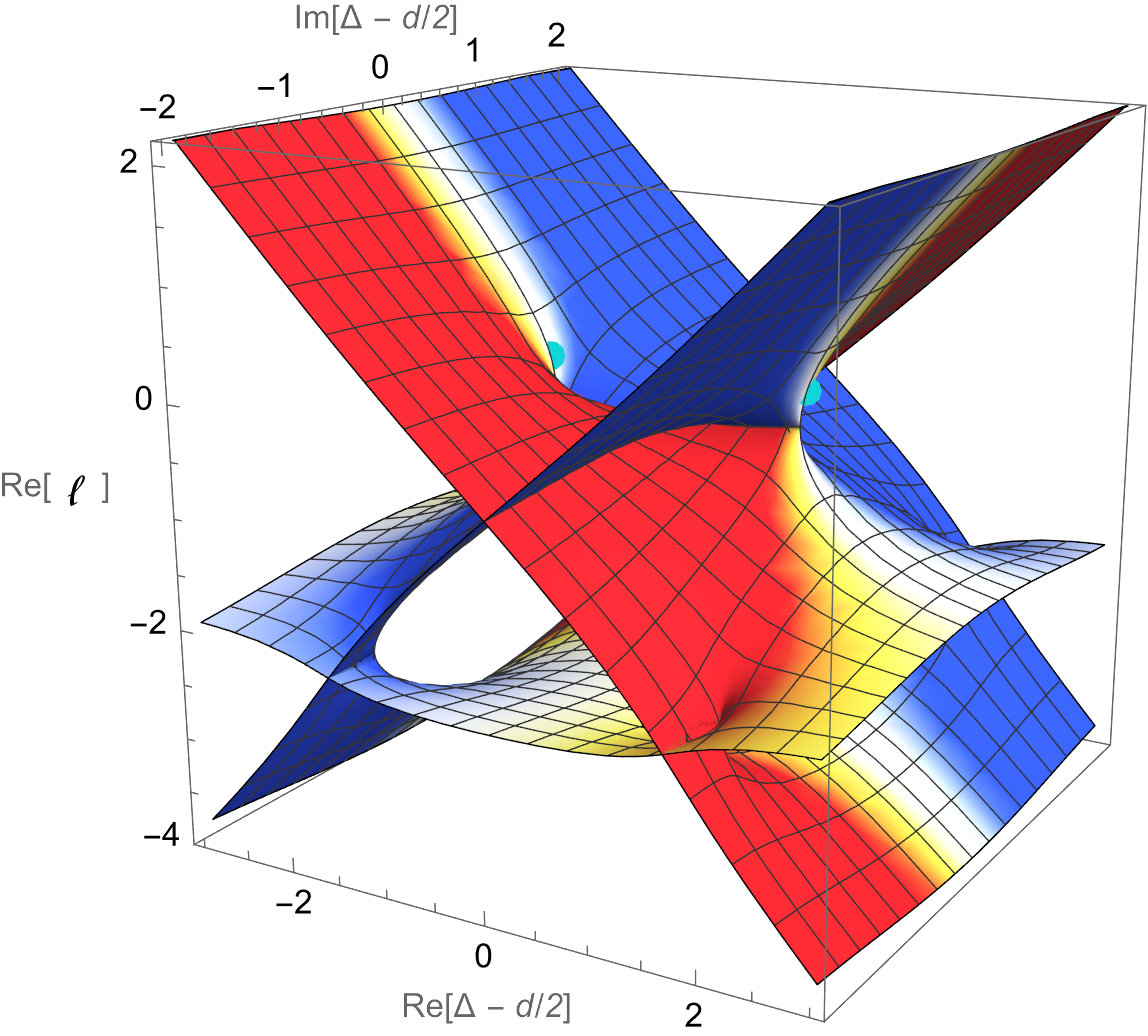}
	\caption{An $\mathbb{R}^3$ projection of the $\mathbb{C}^2$ Chew-Frautschi plot for $\mathcal{J}^{(m=0)}_{\ell, \,\text{trivial} }$ at $N=3$. The surfaces are the Riemann surfaces of the functions $\ell(\D)$. The imaginary part of $\ell$ is shown by color. Red regions correspond to $\text{Im}[\ell]>0$, and blue regions are associated with $\text{Im}[\ell]<0$. The two turquoise points represent two solutions to $\ell=0$, which are $\D_{\pm}$ in \eqref{CF-Delta}. The plot is symmetric with respect to the plane $\text{Re}[\D-d/2]=0$. We choose the parameters $k=1$ and $\e=0.3$.}
	\label{CF-3d}
\end{figure}

\begin{figure}[h]
	\centering
	\includegraphics[width=.7\textwidth,origin=c,angle=0]{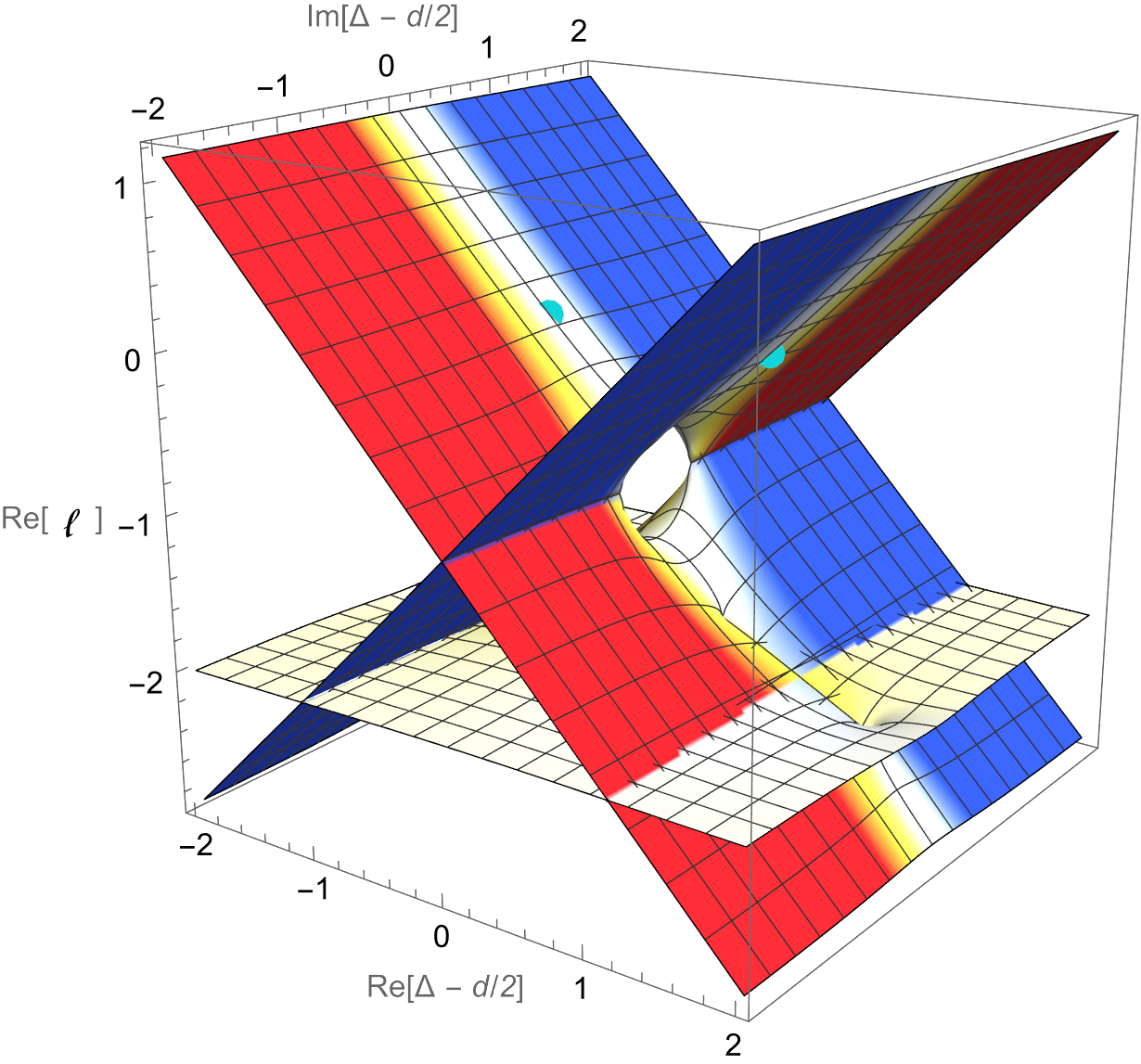}
	\caption{A 3D Chew-Frautschi plot for $\mathcal{J}^{(m=0)}_{\ell, \,\text{trivial} }$ at $N=6/5$. Color coding indicates the imaginary part of $\ell$: red for $\text{Im}[\ell] > 0$ and blue for $\text{Im}[\ell] < 0$. The turquoise points are the solutions to $\ell = 0$, given by $\Delta_{\pm}$ in \eqref{CF-Delta}. The plot is symmetric relative to the plane $\text{Re}[\D-d/2]=0$. The parameters are $k=1$ and $\e=0.3$.}
	\label{CF-3d-smallN}
\end{figure}

\begin{figure}[h]
	\centering
	\includegraphics[width=.61\textwidth,origin=c,angle=0]{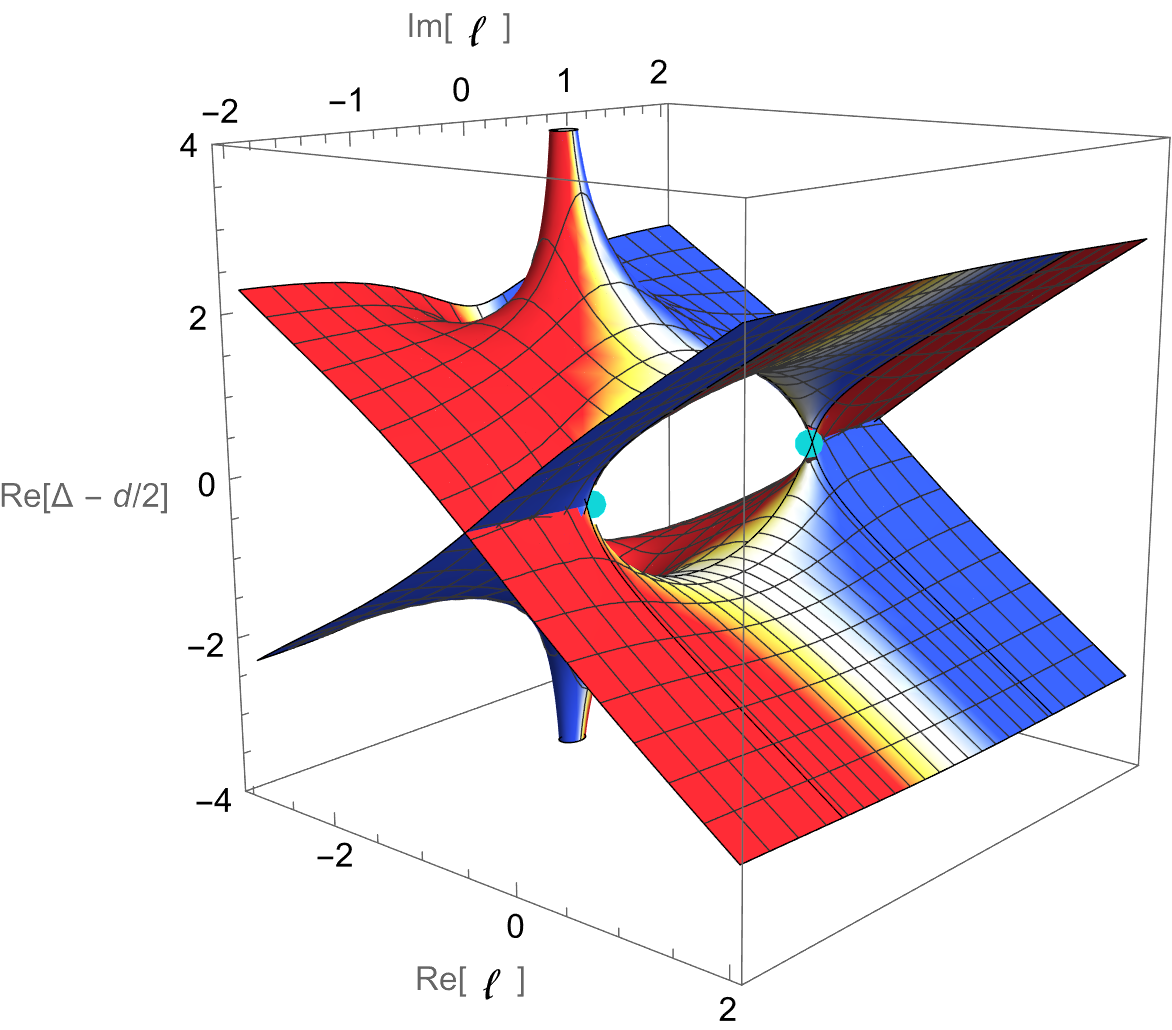}
	\caption{An $\mathbb{R}^3$ projection of the $\mathbb{C}^2$ Regge trajectories of $\mathcal{J}^{(m=0)}_{\ell, \,\text{trivial} }$ at $N=3$. The peaks at $\text{Re}[\ell]=-2$ and $\text{Im}[\ell]=0$ correspond to the $1/(\ell+2)$ pole in the anomalous dimension. We indicate the imaginary part of $\D-d/2$ by color. The regions with $\text{Im}[\D-d/2]>0$ are in red, and the regions with $\text{Im}[\D-d/2]<0$ are in blue. Two of the solutions to $\D=d/2$ are $\ell_{\pm}$ in \eqref{CF-lpm}. They form a complex conjugate pair, and are indicated by the two turquoise points. We use $k=1$ and $\e=0.3$.}
	\label{Delta-l}
\end{figure}

\begin{figure}[h]
	\centering
	\begin{subfigure}{.5\textwidth}
		\raggedright
		\includegraphics[width=0.975\linewidth]{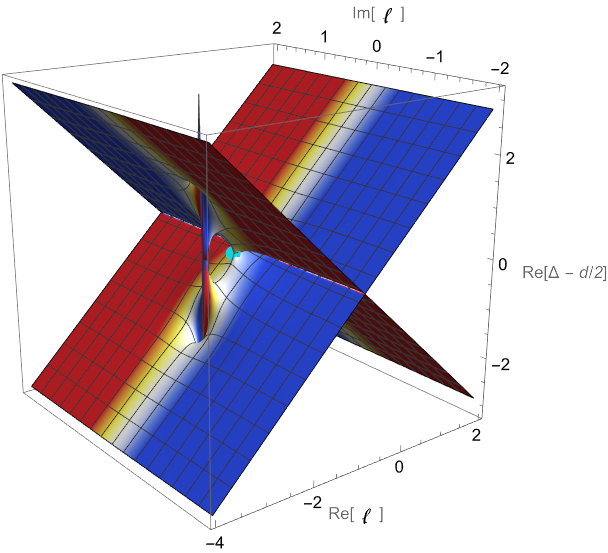}
		\caption{A view showing $\ell_-^*$.}
	\end{subfigure}%
	\begin{subfigure}{.5\textwidth}
		\raggedright
		\includegraphics[width=1.02\linewidth]{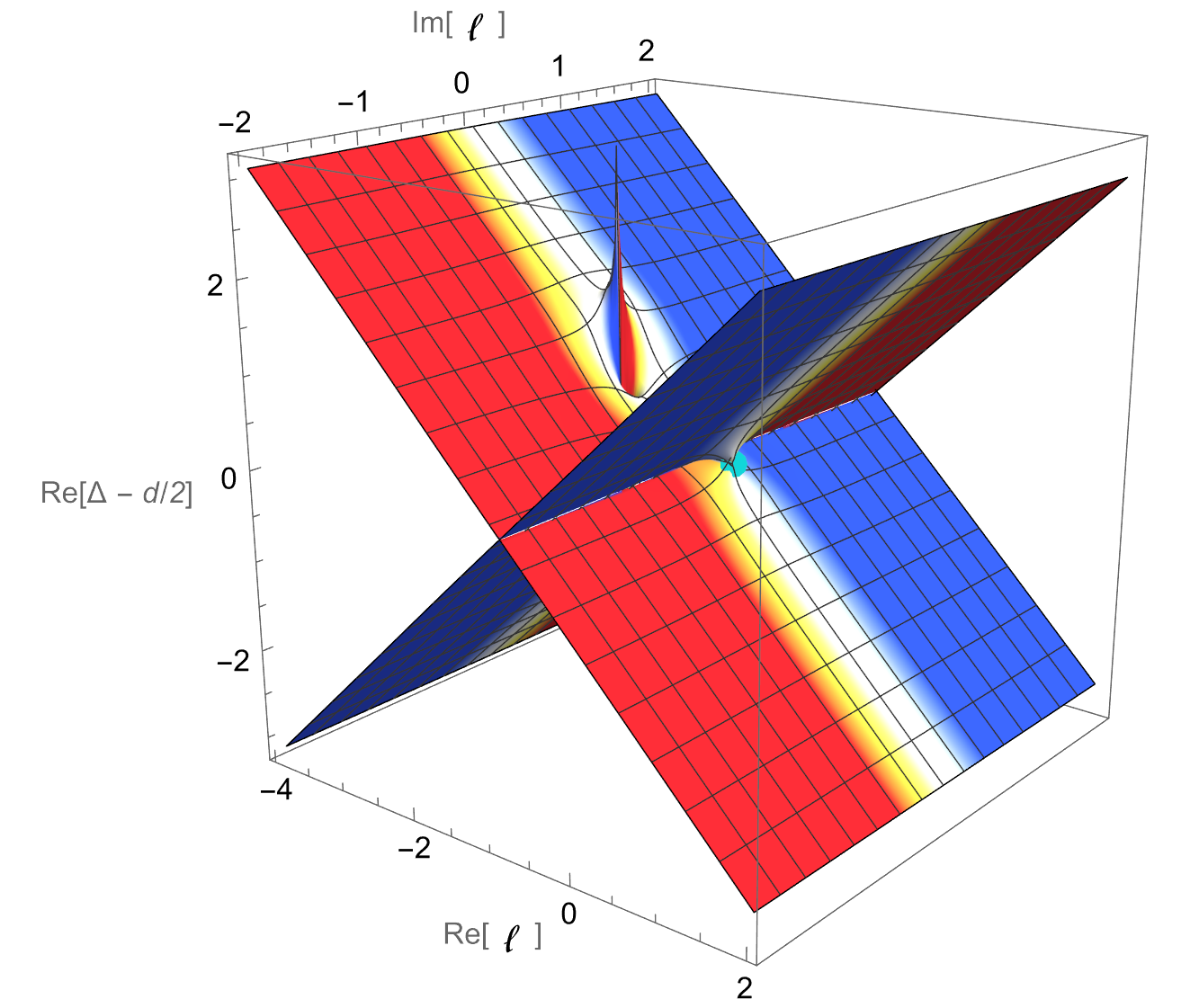}
		\caption{A view showing $\ell_+^*$.}
	\end{subfigure}
	\caption{An $\mathbb{R}^3$ projection of the $\mathbb{C}^2$ Chew-Frautschi plot for $\mathcal{J}^{(m=0)}_{\ell, \,\text{trivial} }$ at  $N=6/5$. The peaks at $\text{Re}[\ell]=-2$, $\text{Im}[\ell]=0$ is due to the $1/(\ell+2)$ pole in the anomalous dimension. The imaginary part of $\ell$ is shown by color: red region implies $\text{Im}[\ell] > 0$, and blue regions indicate $\text{Im}[\ell] < 0$. The turquoise points mark the solutions $\ell_{\pm}^*$ in \eqref{CF-lpm}. The plot is made at $k=1$ and $\e=0.3$.}
	\label{Delta-l-smallN}
\end{figure}

There are also intersections with the vertical axis.
We substitute $\D=d/2$ into \eqref{trajectory-Potts} and obtain three solutions
\begin{align}\label{CF-lpm}
	\ell^*_\pm&=-1
	\pm i \, \sqrt{
	\frac{2R'}{3N-7}
	} \; \e^{1/2}
	+\ldots
	\,, \\
	\ell^*_3&=-2
	-\frac{2R'}{3N-7} \, \e
	+\frac{2 R' (12R'+11N-23)}{3 (3N-7)^2}\e^2
	+\ldots
	\,.
\end{align}
The fact that the leading real intercept is negative should have interesting physical implications. 
For instance, in the Regge limit, the connected part of the four-point correlator decays exponentially fast at large boost if the Regge intercept is below unity \cite{Costa:2012cb,Caron-Huot:2017vep,Liu:2020tpf,Caron-Huot:2020ouj,Caron-Huot:2022eqs}.\footnote{The anomalous dimension \eqref{gammaJ-Potts} is regular at $\ell=0$, which corresponds to a local spin-0 operator.
The analytic continuation above leads to both the local operator and its shadow, 
but the analytic continuation in conformal spin only gives the local operator.
This is similar to the case of the $\phi^{2n}$ theory with $n>2$ \cite{Guo:2023qtt}. 
}

Let us discuss the case of the trivial representation, where $R'=4(N-1)$.
For $N>7/3$, the solutions $\ell^*_\pm$ correspond to complex spins, so they do not show up in the real $(\D-d/2,\ell)$ plane.
For $1<N<7/3$, there are three intersection points, as in figure \ref{CF-smallN}, since $R'$ is negative and $\ell^*_\pm$ are real.
At $N=1$, we have $R'=0$ and $\g_\phi=0\e+\ldots$, so the anomalous dimensions of the bilinear operators are zero to order $\e^1$.
So the $N=1$ trajectory reduces to that in the free theory.\footnote{Strictly speaking, the free trajectory should not include the horizontal line $\ell=-2$.}
The $N<1$ plots are similar to the $N>7/3$ ones, where $\ell^*_\pm$ are complex.
Furthermore, we show the projections of $\mathbb{C}^2$ Chew-Frautschi plots in figures \ref{CF-3d}, \ref{Delta-l}, \ref{CF-3d-smallN} and \ref{Delta-l-smallN}.
For the standard representation, $R'$ changes sign at $N=2$.
There is one intersection point with the vertical axis for $N>7/3$ as well as for $N<2$, and there are three intersection points for $2<N<7/3$.
For the remaining two representations, $R'$ is always negative.
There are only two scenarios: $N>7/3$ and $N<7/3$.
The former gives the trajectories with three intersection points on the vertical axis, and the latter yields the trajectories with one intersection point.

\begin{figure}[H]
	\centering
	\includegraphics[width=.5\textwidth,origin=c,angle=0]{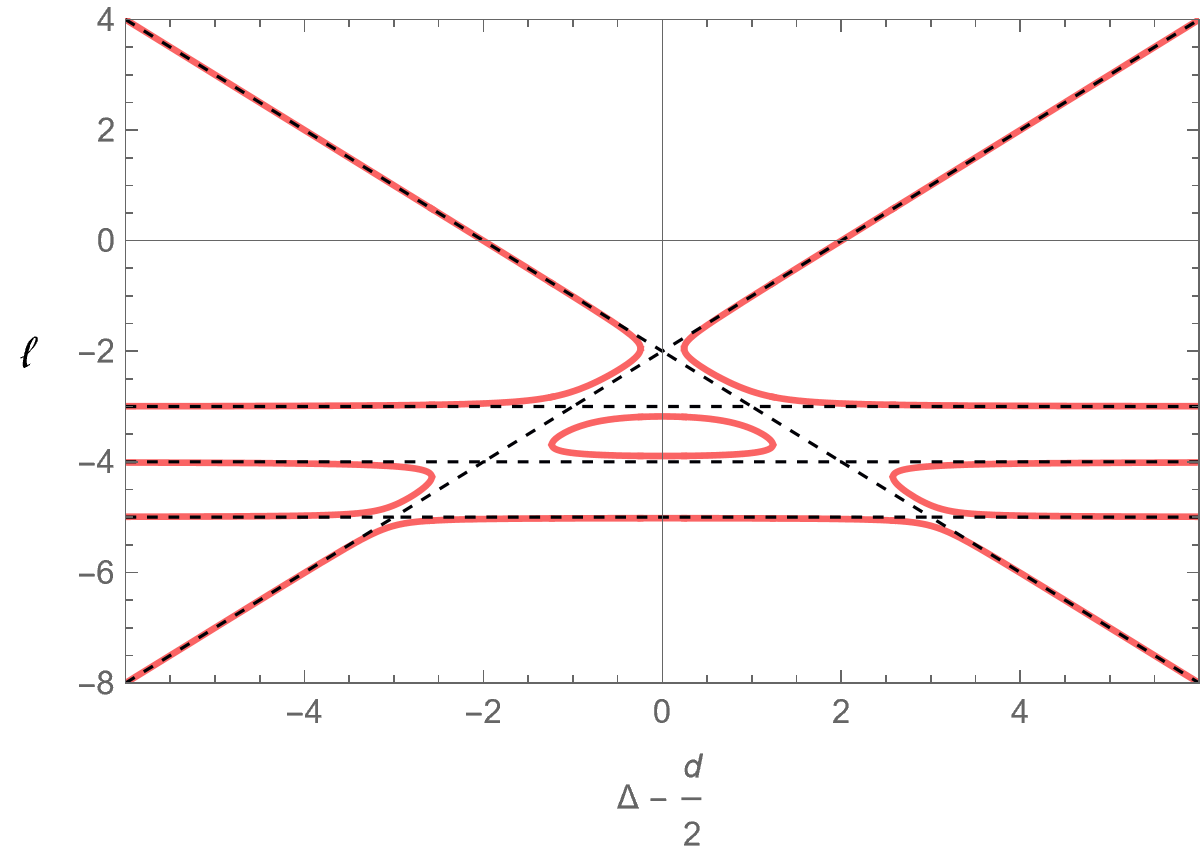}
	\caption{Chew-Frautschi plot for $\mathcal{J}^{(m=0)}_{\ell,\,\text{trivial} }$ in the $k=2$ theory. The interacting case with $N=2$ is shown in red. In the Gaussian limit, there are dashed $45^{\circ}$ and $135^{\circ}$ lines given by $(\ell+2)^2=(\D-6)^2$, and there exist three dashed horizontal lines at $\ell=-3,-4,-5$.}
	\label{CF-k2-m0}
\end{figure}

\begin{figure}[H]
	\centering
	\includegraphics[width=.8\textwidth,origin=c,angle=0]{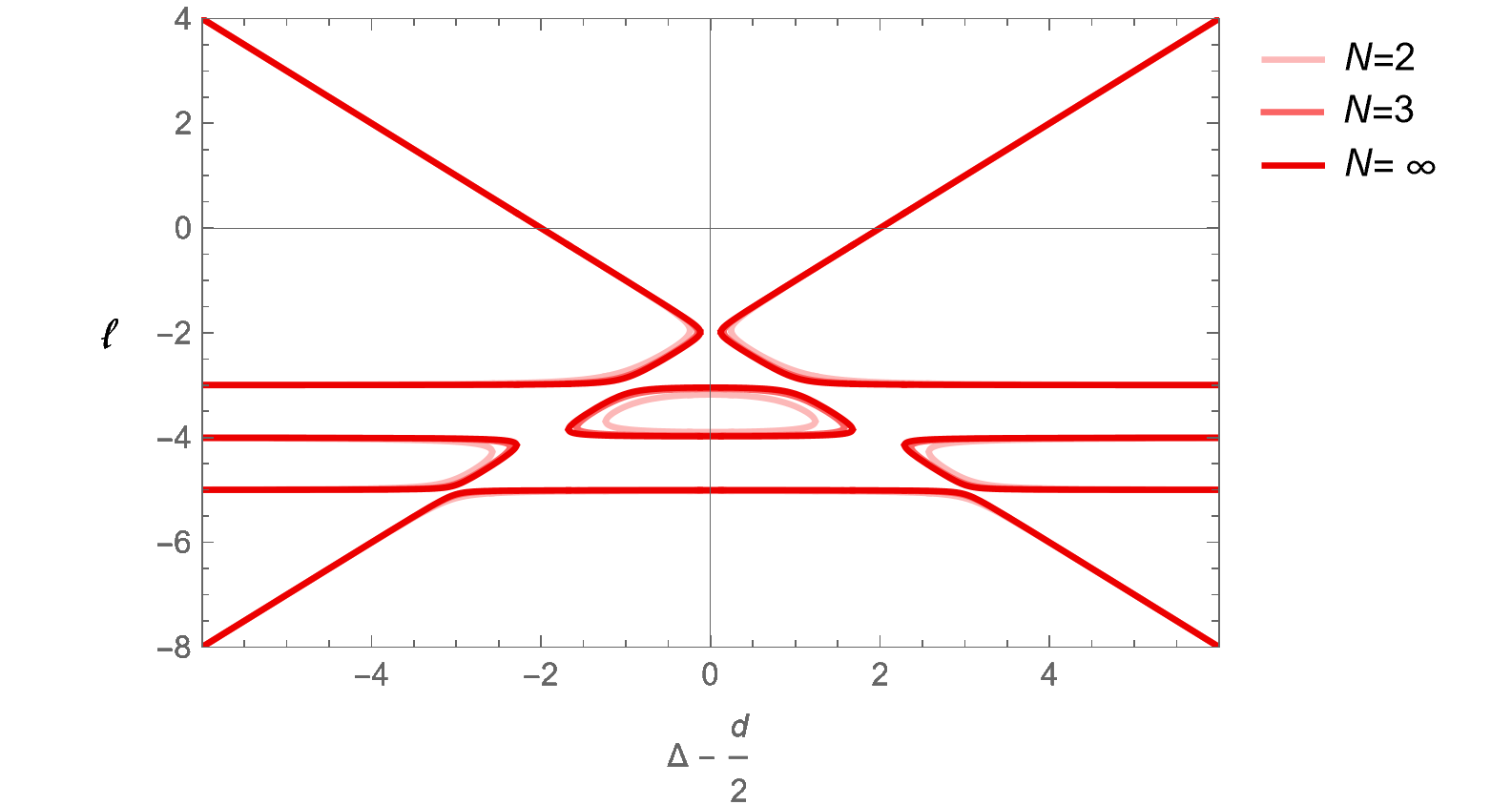}
	\caption{In the $k=2$ theory, the Regge trajectories of $\mathcal{J}^{(m=0)}_{\ell,\,\text{trivial}}$ are shown. These trajectories correspond to the case of $N>65/37$. We set $\e=10^{-3}$.}
	\label{CF-k2-m0-largeN}
\end{figure}

\begin{figure}[H]
	\centering
	\includegraphics[width=.9\textwidth,origin=c,angle=0]{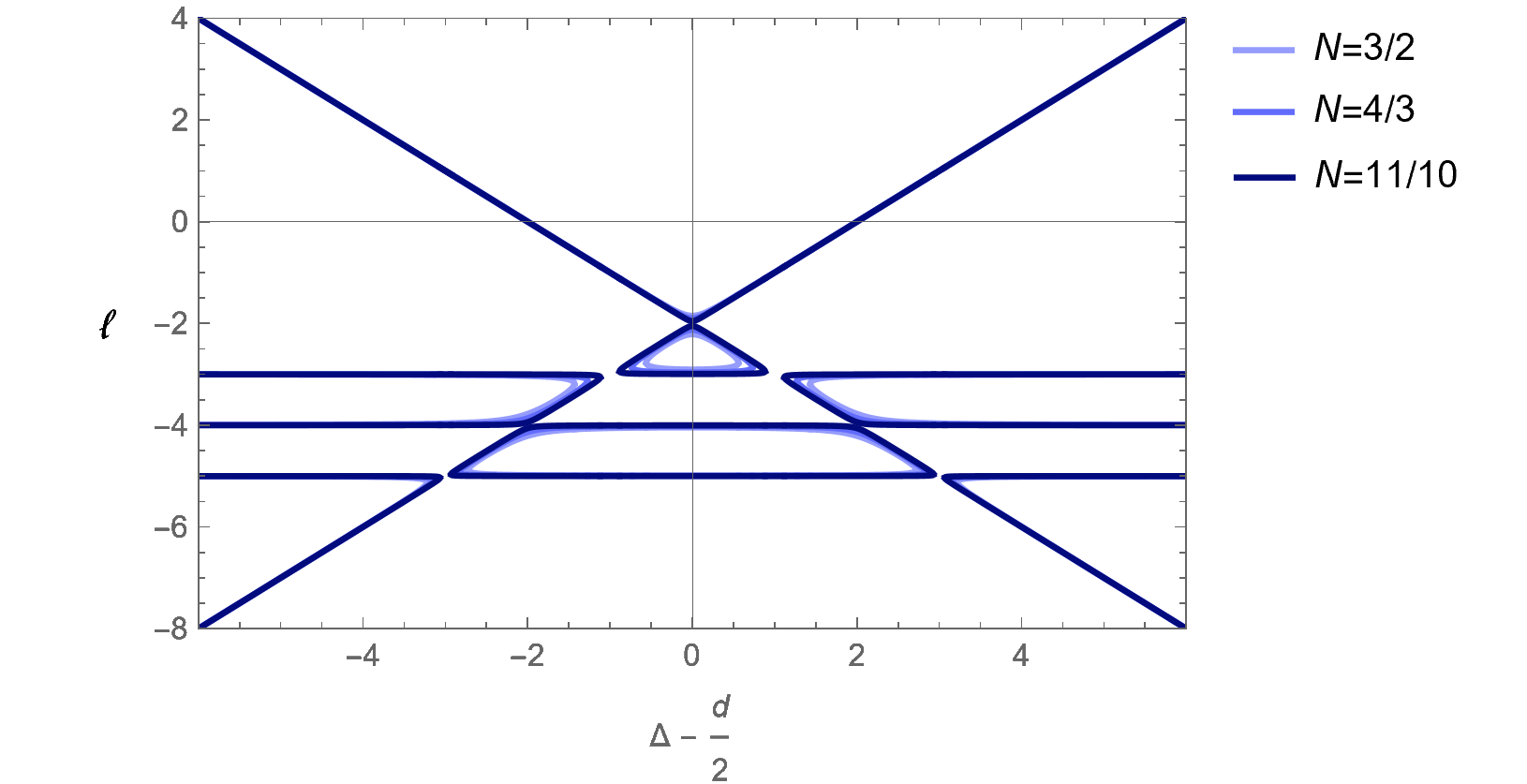}
	\caption{Chew-Frautschi plot of the $k=2$ interacting theory. The Regge trajectories of $\mathcal{J}^{(m=0)}_{\ell,\,\text{trivial} }$ with $1<N<65/37$ are shown. We use $\e=10^{-3}$.}
	\label{CF-k2-m0-smallN}
\end{figure}

\begin{figure}[H]
	\centering
	\includegraphics[width=.8\textwidth,origin=c,angle=0]{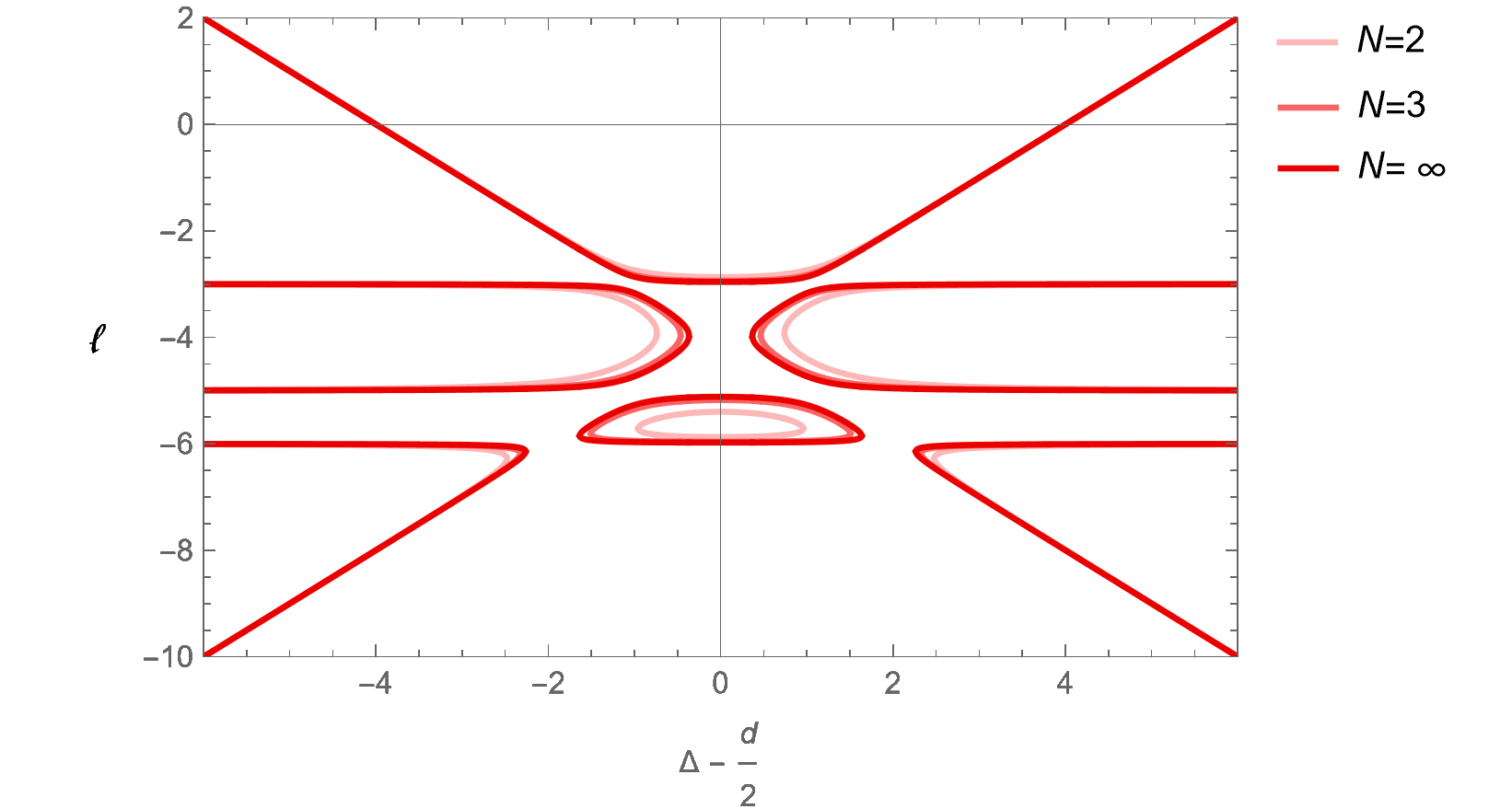}
	\caption{Regge trajectories of $\mathcal{J}^{(m=1)}_{\ell,\,\text{trivial} }$ in the $k=2$ interacting theory. We present the scenario where $N>65/37$. Again, $\e=10^{-3}$ is used. }
	\label{CF-k2-m1-largeN}
\end{figure}

\begin{figure}[H]
	\centering
	\includegraphics[width=.8\textwidth,origin=c,angle=0]{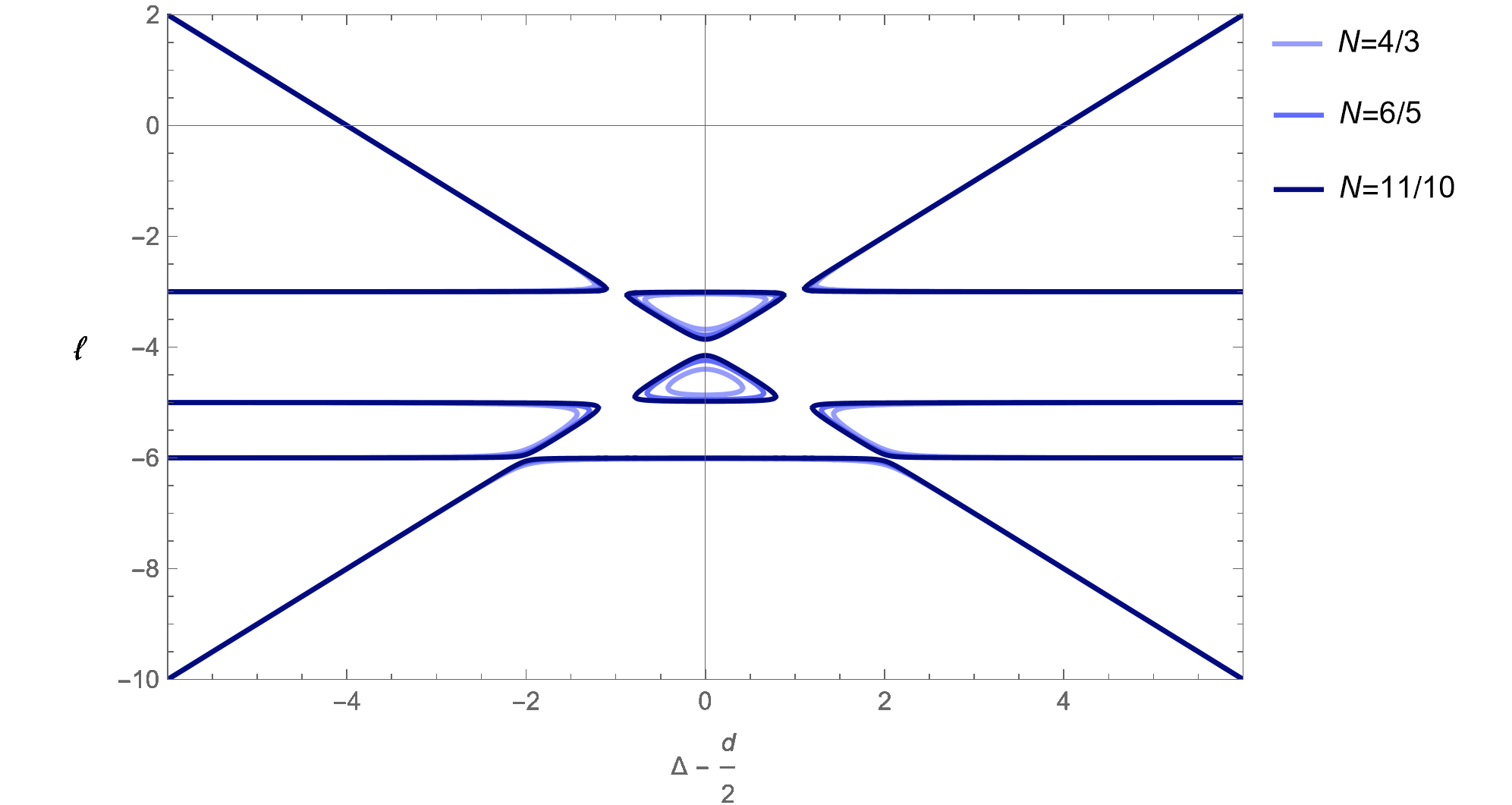}
	\caption{In the $k=2$ interacting theory, we show the Regge trajectories of $\mathcal{J}^{(m=1)}_{\ell,\,\text{trivial} }$ with $1<N<65/37$, where $\e$ is set to $10^{-3}$.}
	\label{CF-k2-m1-smallN}
\end{figure}

Now, we briefly comment on the $k>1$ Chew-Frautschi plots.
One difference is the presence of more curves, which arise from the deformation of more horizontal lines in the Gaussian limit.
Another difference is that we can consider trajectories with higher $m$.
For each $m$, there are more solutions to $\D=d/2$, so the Regge trajectory can have more intersections with the vertical axis. 
Let us first discuss the Gaussian limit, where the scaling dimensions do not depend on the $S_{N+1}$ representations.
For a positive integer $k$, the horizontal lines are given by $\frac{ (k+\ell+m)_{2k} }{k+\ell+2m}=0$, implying that there are $2k-1$ such lines.
The $45^{\circ}$ and $135^{\circ}$ lines are parallel to those in the $k=1$ case.
In the interacting theory, the trajectories are deformed into disconnected pieces as the mixing problems are resolved.
For example, the $\mathcal{J}^{(m=0)}_{\ell, \,\text{trivial} }$ trajectory in the $k=2$ theory and its Gaussian limit are shown in figure \ref{CF-k2-m0}.
As in the $k=1$ case, the Chew-Frautschi plots in the interacting theory depend on $N$.
For both $m=0$ and $m=1$, there are three scenarios: $N>65/37$, $1<N<65/37$ and $N<1$.
We present the Chew-Frautschi plots with different values of $N$ in figures \ref{CF-k2-m0-largeN}, \ref{CF-k2-m0-smallN}, \ref{CF-k2-m1-largeN}, and \ref{CF-k2-m1-smallN}.
The Chew-Frautschi plots with $N<1$ are similar to the case of $N>65/37$.

\subsection{Analytic bootstrap}
\label{Analytic bootstrap Potts}

In this subsection, we use the analytic bootstrap to check the results from the multiplet recombination.
As in section \ref{Analytic bootstrap}, we consider the crossing constraint on the four-point function
\begin{align}
	\< \phi_a(x_{1})\phi_b(x_{2})\phi_c(x_{3})\phi_d(x_{4}) \>
	=\frac{ g_{abcd}(z,\zb) }{ x_{12}^{ 2\D_{ \smash{\phi} } }x_{34}^{ 2\D_{ \smash{\phi} } } }
	\,.
\end{align}
The tensor structures of the four-point function are given by \eqref{4pt-tensors-1}, \eqref{4pt-tensors-2}, \eqref{4pt-tensors-3} and \eqref{4pt-tensors-4} in appendix \ref{Tensor structures}.
We consider
\begin{align}
	g_{abcd}(z,\zb)
	=\sum_{i=1}^4 g^{(i)}(z,\zb) \, t_{abcd}^{(i)}
	\,.
\end{align}
Then, the crossing equations are written as \eqref{crossing-1}, \eqref{crossing-2}, \eqref{crossing-3}, and \eqref{crossing-4}.,
The functions $g^{(i)}(z,\zb)$ decompose into conformal blocks associated with the operators in the trivial, standard, symmetric traceless, and antisymmetric representations.
As in section \ref{Analytic bootstrap}, we focus on the leading order in the $z \rightarrow 0$ expansion, and verify the multiplet-recombination results for the $m=0$ trajectory.

At order $\e^0$, the identity operator contributes to the $\zb \rightarrow 1$ enhanced singularity on the right-hand sides of the crossing equations.
Since the identity operator lives in the trivial representation, the four types of free squared OPE coefficients are
\begin{align}\label{free-OPE-Potts}
	P_{ \mathcal{J}^{(0)}_{\ell},\text{f} }^{ \text{ generic } N } = P_{ \mathcal{J}^{(0)}_{\ell},\text{f} }^{ \;N=1 }  \times
	\begin{cases}
		\; \frac{1}{N} \qquad & \text{trivial} \\
		\; \frac{1}{ (N-1)(N+1)^2 } & \text{standard} \\
		\; 1 & \text{symmetric traceless} \\
		\; -1 & \text{antisymmetric}
		\,,
	\end{cases}
\end{align}
where $P_{ \mathcal{J}^{(0)}_{\ell},\text{f} }^{ \;N=1 }$ is the free squared OPE coefficient at $N=1$.\footnote{Here $N=1$ corresponds to the free theory with a single fundamental scalar, not to be confused with the Ising model, which is an interacting theory.}
The above relation means the following.
The matching of the lightcone singularity is the same as that in \eqref{free-OPE}, except that here we need to account for some $N$-dependent factors in the crossing equations.

At order $\e^{1}$, the enhanced singularities on the right-hand sides of the crossing equations are given by the operator $\phi_a$ in the standard representation.
Then, the anomalous dimensions are related to those in the Yang-Lee case by
\begin{align}\label{gamma-relation}
	\tilde{\g}_{ \mathcal{J}_\ell^{(0)} }
	= \frac{ P_{\phi}^\text{Potts} }{ P_{\phi}^{ \text{YL} } }
	\tilde{\g}_{ \mathcal{J}_\ell^{(0)} }^{ \text{YL} }
	\times
	\begin{cases}
		\; (N-1)(N+1)^2 \qquad & \text{trivial} \\
		\; (N-2)(N+1)^2 & \text{standard} \\
		\; -(N+1)^2 & \text{symmetric traceless} \\
		\; -(N+1)^2 & \text{antisymmetric}
		\,,
	\end{cases}
\end{align}
where $P_{\phi}^{ \text{YL} }$ and $\tilde{\g}_{ \mathcal{J}_\ell^{(0)}}^{ \text{YL} }$ are the Yang-Lee data.
This means that we consider the matching of the lightcone singularity in the Yang-Lee case, and then we add the $N$-dependent factors and replace the Yang-Lee OPE coefficients with those in the Potts model.
Using the OPE coefficients \eqref{lambda} and \eqref{lambda-Potts},\footnote{According to \eqref{OPE-normalization}, the OPE coefficients with different normalizations are related by $P_{\phi}^{ \text{YL} }=\( \l_{\phi\phi\phi}^{ \text{YL} } \)^2$ and $P_{\phi}= \l_{\phi\phi\phi}^2$.} we obtain
\begin{align}\label{crossing gamma Potts}
	\tilde{\g}_{ \mathcal{J}_\ell^{(0)} }
	=-\frac{ R\,\a^2}{2^{4 k-1} (2 k-1)! (k+\ell)_{2 k}}+\ldots
	\,,
\end{align} 
which is in agreement with the multiplet-recombination result \eqref{gammaJ-Potts} at $m=0$.
We also examine the $m=1$ trajectory in appendix \ref{Analytic bootstrap at subleading twist}, and confirm that the multiplet-recombination results are consistent with crossing symmetry.
Similarly to \eqref{crossing gamma YL}, the agreement is valid for arbitrary $\a$.

\section{Generalized \boldmath{$\phi^{2n+1}$} theory (\boldmath{$n>1$})}
\label{Generalized phi-2n+1 theory}

In this section, we study the $\phi^{2n+1}$ deformation of the free higher-derivative theory, where $n$ is larger than $1$.
If $k$ and $2n-1$ have a common divisor, 
there may exist $\mathbb Z_2$-odd marginal derivative interactions, 
which can mix with $\phi^{2n+1}$.
To avoid this, we assume that ($k,2n-1)$ is a pair of coprime integers. 
\footnote{This is similar to the case of the $\phi^{2n}$ theory, where $k$ and $n-1$ have no common divisor \cite{Safari:2017irw,Safari:2017tgs}.
The conditions for the $\phi^{2n+1}$ and $\phi^{2n}$ theories can be unified as follows.
For a $\phi^p$ theory with integer $p$, we require that $k$ and $p-2$ have no common divisor other than 2.}
In this way, we can consistently focus on the pure $\phi^{2n+1}$ interaction.
The action at $d=d_{ \text{u} }-\e$ takes the form
\begin{align}\label{action-phi-2n+1}
	S\propto\int \mathrm{d}^dx
	\( \phi\,\Box^k\phi
	+g\m^{(2n-1)\e/2} \phi^{2n+1} \)
	\,,
\end{align}
where the upper critical dimension is given by
\begin{align}\label{du-phi-2n+1}
	d_\text{u}=
	2k \, \frac{2n+1}{2n-1}
	\,.
\end{align}
This theory describes the generalized multicritical Yang-Lee edge singularity \cite{vonGehlen:1994rp,Lencses:2022ira,Lencses:2023evr,Lencses:2024wib}.
For $k=1$, it was argued that the Lagrangian in \eqref{action-phi-2n+1} can be obtained from the Landau-Ginzburg Lagrangian of the multicritical Ising model.
For the latter Lagrangian, a shift of the order parameter field by a purely imaginary constant generates $\mathbb{Z}_2$-odd terms.
Then, one can tune the coupling constants to reach the multicritical point given by \eqref{action-phi-2n+1}.\footnote{
This procedure can also lead to a multicritical point with an interaction term consisting of an even number of $\phi$'s. 
See \cite{Lencses:2024wib} for more details.
}
This generalizes Fisher's argument in \cite{Fisher:1978pf}.

For $n>1$, the CFT data is not completely fixed by the multiplet recombination.
We determine the set of anomalous dimensions up to one unknown parameter $\a$, which can be fixed using the diagrammatic method in the canonical case.
For arbitrary $\a$, we also check the consistency of the results with crossing symmetry using the analytic bootstrap.

\subsection{Multiplet recombination}

The generalization of \eqref{boxk-phi} reads
\begin{align}\label{boxk-phi-2n+1}
	\lim_{\e \rightarrow 0} \(
	\a^{-1} \Box^{k}\phi \)
	=\phi^{2n}_{ \text{f} }
	\,.
\end{align}
For $n>1$, the normalization $\a$ is not determined by the matching conditions generated from this equation, and $\a$ appears in the results for the anomalous dimensions, as well as the OPE coefficients.
This distinction from the $n=1$ case is due to the vanishing of the free-theory correlator $\< \phi^{2n}_\text{f}(x_1)\phi_\text{f}(x_2)\phi_\text{f}(x_3) \>$ for $n>1$.
In the previous sections, we have seen that the free-theory data is an important input in the multiplet recombination approach.
As there is less free-theory input available here, the set of matching conditions is less constraining.
Nevertheless, we can still obtain a consistent set of data using \eqref{boxk-phi-2n+1}.

For two-point functions, the matching condition takes the form
\begin{align}
	\lim_{\e \rightarrow 0}
	\( \a^{-2} \< \Box^k\phi(x_1) \Box^k\phi(x_2) \> \)
	= \< \phi^{2n}_{ \text{f} }(x_1) \phi^{2n}_{ \text{f} }(x_2) \>
	\,.
\end{align}
The left-hand side has the same form as that of \eqref{boxk-2pt}.
The right-hand side is a free correlator, which can be computed using Wick contractions.
The result is
\begin{align}
	\lim_{\e \rightarrow 0} \[
	\a^{-2} \, 4^{2k} \(\D_{ \smash{\phi} }\)_{2k}
	\(\D_{ \smash{\phi} }-\frac{d-2}{2}\)_{2k}
	\frac{1}{ x_{12}^{ 2\D_{ \smash{\phi} }+4k } } \, \]
	=\frac{(2n)!}{ x_{12}^{ 2 \D_{ \smash{\phi^{2n}_\text{f} } } } }
	\,.
\end{align}
The functional forms match, since $\lim_{\e \rightarrow 0}(2\D_{ \smash{\phi} }+4k)=2\D_{ \smash { \phi^{2n}_{ \text{f} } } }$.
The definition of the anomalous dimension is 
\begin{align}
	\g_{\phi}=\D_{ \smash{\phi} }-\frac{d-2k}{2}
	\,, \qquad
	d=d_{ \text{u} }-\e
	\,.
\end{align}
The constraint on the anomalous dimension reads
\begin{align}
	 (-1)^{k-1}4^{2k} k!(k-1)! \( \frac{ d_{ \text{u} }-2k }{2} \)_{2k}\, \lim_{\e \rightarrow 0} \( \a^{-2} \g_{\phi} \) = (2n)!
	\,,
\end{align}
where the upper critical dimension $d_{ \text{u} }$ is given by \eqref{du-phi-2n+1}.
This generalizes \eqref{alpha-gamma1} in the $n=1$ case.
We obtain a relation between the anomalous dimension $\g_{\phi}$ and the unknown parameter $\a$:
\begin{align}\label{gamma1-phi-2n+1}
	\g_{\phi}=
	\frac{ (-1)^{k-1}(2n)! \, \a^2 }
	{ 4^{2k} k!(k-1)! \( \frac{2k}{2n-1} \)_{2k} }
	+\ldots
	\,,
\end{align}
The next step is to consider three-point functions.
Unfortunately, the matching condition for $\< \Box^k\phi(x_1)\phi(x_2)\phi(x_3) \>$ is not useful as the corresponding free correlator $\< \phi^{2n}_{ \text{f} }(x_1)\phi_{ \text{f} }(x_2)$ $\phi_{ \text{f} }(x_3) \>$ vanishes for $n>1$.
The matching condition for $\< \Box^k\phi(x_1)\Box^k\phi(x_2)\Box^k\phi(x_3) \>$ involves $\a$ and an unknown OPE coefficient $\l_{\phi\phi\phi}$, 
so we cannot determine $\a$ and $\g_{\phi}$ by following the procedure in section \ref{Multiplet recombination}.
Nevertheless, we can still deduce the relation between $\l_{\phi\phi\phi}$ and $\a$ using the matching condition
\begin{align}
	\lim_{\e \rightarrow 0} \( \a^{-3}
	\< \Box^{k}\phi(x_{1})\Box^{k}\phi(x_{2})\Box^{k}\phi(x_{3}) \> \)
	=\< \phi^{2n}_{ \text{f} }(x_{1})
	\phi^{2n}_{ \text{f} }(x_{2})
	\phi^{2n}_{ \text{f} }(x_{3}) \>
	\,.
\end{align}
Using \eqref{3-box}, we obtain a constraint on the OPE coefficient:
\begin{align}\label{3-box-2n+1}
	(-1)^k 4^{3k}\(\frac{d_\text{u}-6k}{4}\)_{3k}\(\frac{d_\text{u}-2k}{4}\)_k^3\;\lim_{\e\rightarrow 0}\(\a^{-3} \l_{\phi\phi\phi}\)
	=\(\frac{(2n)!}{n!}\)^3
	\,.
\end{align}
The equation above does not apply to $n=1$ 
because the factor $(\frac{d_\text{u}-6k}{4})_{3k}$ would vanish. 
As in \eqref{3box n=1}, the correction to $\D_\phi$ should be taken into account for $n=1$, 
but the $n>1$ constraint \eqref{3-box-2n+1} does not involve the anomalous dimension of $\phi$.
The solution for \eqref{3-box-2n+1} reads
\begin{align}
	\l_{\phi\phi\phi}=
	\frac{ (-1)^k \((2n)!\)^3 \a^3}
	{4^{3k}\(n!\)^3\(\frac{k}{2 n-1}\)_k^3 \(-2 k\frac{ n-1}{2 n-1}\)_{3k}}
	+\ldots
	\,,
\end{align}
which agrees with the $k=1$ result in \cite{Codello:2017qek}.

To obtain more useful constraints on the anomalous dimensions, we should avoid the matching conditions where the left-hand sides involve vanishing OPE coefficients in the Gaussian limit.
In this way, we avoid introducing more unknown OPE coefficients as we examine more correlation functions.
Additionally, the free correlators on the right-hand sides of the matching conditions should not be zero.
Since \eqref{boxk-phi-2n+1} changes the $\mathbb{Z}_2$ parity under $\phi \rightarrow -\phi$, we have to consider the matching conditions with two $\Box^k\phi$'s. 
An example is
\begin{align}\label{boxk-phi2}
	\lim_{\e \rightarrow 0}
	\( \a^{-2} \< \Box^k\phi(x_1) \Box^k\phi(x_2) \phi^2(x_3) \> \)
	= \< \phi^{2n}_{ \text{f} }(x_1) \phi^{2n}_{ \text{f} }(x_2) \phi^2(x_3) \>
	\,,
\end{align}
which yields a relation between the anomalous dimensions of $\phi$ and $\phi^2$.
This is a special case of the matching conditions for bilinear operators.

Let us consider the more general case.
The matching condition involving a primary bilinear operator $\mathcal{J}^{(m)}_{\ell}$ reads
\begin{align}
	\lim_{\e \rightarrow 0}
	\( \a^{-2} \< \Box^k\phi(x_1) \Box^k\phi(x_2) \mathcal{J}^{(m)}_{\ell}(x_3,z) \> \)
	= \< \phi^{2n}_{ \text{f} }(x_1) \phi^{2n}_{ \text{f} }(x_2) \mathcal{J}^{(m)}_{\ell}(x_3,z) \>
	\,.
\end{align}
As in \eqref{boxk-current-result}, the constraint is extracted from the leading term in the $x_3 \rightarrow \infty$ expansion.
We have
\begin{align}\label{boxk-current-result-phi-2n+1}
	&\lim_{\e\rightarrow 0}\Bigg[ \a^{-2}4^{2k}
	\(\D_{ \smash{\phi} }-\frac{ \D_{ \mathcal{J} } }{2}+\frac\ell 2\)_{2k}
	\(\D_{ \smash{\phi} }-\frac{ \D_{ \mathcal{J} } }{2} -\frac \ell 2-\frac{d-2}{2} \)_{2k} \nn
	& \times \frac{ \l_{ \phi\phi\mathcal{J} } \(z \cdot x_{12}\)^{\ell}  }
	{ x_{12}^{ 2\D_{ \smash{\phi} }+4k-\D_{ \mathcal{J} }+\ell }|x_{3}|^{ 2\D_{\mathcal{J} } } }
	+O ( \, |x_{3}|^{-2\D_{\mathcal{J} }-1 } ) \Bigg]
	=\frac{ \l_{ \phi^{2n}\phi^{2n}\mathcal{J},\text{f} } \(z \cdot x_{12}\)^{\ell}  }
	{ x_{12}^{ 2\D_{ \smash{ \phi^{2n}_\text{f} } }-\D_{ \mathcal{J}\smash{, \text{f} } }+\ell }|x_{3}|^{ 2\D_{\mathcal{J}, \text{f} } } }
	+O ( \, |x_{3}|^{-2\D_{\mathcal{J}\smash{, \text{f} } }-1 } )
	\,.
\end{align}
Again, one can check that the functional forms match using $\lim_{\e \rightarrow 0}(2\D_{ \smash{\phi} }+4k)=2\D_{ \smash{\phi^{2n}_\text{f} } }$.
According to the definition of the anomalous dimension $\g_{ \mathcal{J} }=\D_{ \mathcal{J} }-(d-2k+2m+\ell)$, the equation above leads to
\begin{align}
	4^{2k}(2k-m-1)! \, (-m)_{m} \(1-\frac{ d_{\text{u} } }{2}-\ell-m \)_{2k}
	\lim_{\e \rightarrow 0} 
	\[ \a^{-2}\( \g_{\phi}-\frac{ \g_{ \mathcal{J} } }{2} \) \]
	=\frac{ \l_{ \phi^{2n}\phi^{2n}\mathcal{J},\text{f} } }{ \l_{ \phi\phi\mathcal{J},\text{f} } }
	\,,
\end{align}
where the ratio of OPE coefficients $\frac{ \l_{ \phi^{2n}\phi^{2n}\mathcal{J},\text{f} } }{ \l_{ \phi\phi\mathcal{J},\text{f} } }=2n(2n)!$ is computed in appendix \ref{Ratios of OPE coefficients}.
The solution for the anomalous dimension of the bilinear operator is
\begin{align}\label{gammaJ-phi-2n+1}
	\g_{ \mathcal{J}^{(m)}_\ell }=
	2\g_{\phi}-
	\frac{(-1)^m n^2 (2n-1)! \, \a^2 }
	{ 2^{4k-3}m! (2k-m-1)! \(\ell+m-k\frac{2n-3}{2n-1}\)_{2k}}
	+\ldots
	\,,
\end{align}
where $\g_{\phi}$ is given in \eqref{gamma1-phi-2n+1}.
This expression generalizes \eqref{gammaJ} in the case of $n=1$.
For $k=1$, this result can also be derived from eq.(4.19) in \cite{Gliozzi:2017gzh}. 
The special case \eqref{boxk-phi2} is associated with $\ell=m=0$.
One can check that, to order $\a^2$, the anomalous dimension of the stress tensor $(m=k-1,\; \ell=2)$ vanishes, even though we do not know the explicit expression for $\a$.
In other words, we cannot determine $\a$ by requiring that the stress tensor does not acquire an anomalous dimension.\footnote{We believe that the spinning recombination cannot fix $\a$ either.
For example, at $k=1$, we find that $\lim_{\e\rightarrow0}(\a^{-2}\<\Box\phi(x_1)\phi(x_2)\pa\mathcal{J}(x_3))\>$ is proportional to $\lim_{\e\rightarrow0}(\a^{-2}\g_\mathcal{J})$, so the matching condition should only yield a relation between $\g_\mathcal{J}$ and $\a$, and this relation ought to be the same as \eqref{gammaJ-phi-2n+1}.}

Since the anomalous dimensions are not fixed by our CFT analysis, we resort to inputs from the diagrammatic method.
At $k=1$, Codello \textit{et al}. computed the anomalous dimensions of some scalar operators in the $n>1$ theory \cite{Codello:2017epp}.
The result for $\g_\phi$ is expressed in terms of the fixed-point coupling constant in their equation (5.7), which is consistent with our formula \eqref{gamma1-phi-2n+1}.
Furthermore, they also obtained the fixed-point coupling constant, so they fully determined the anomalous dimension $\g_\phi$.
We use their results to fix $\a$
\footnote{In \cite{Codello:2017epp}, the general expression for the fixed-point coupling constant $g$ in equation (5.6) may contain a typo, as it does not seem to reproduce their $n=2$ result in equation (3.2). Here we use their beta functionals to solve for $g$, and the result is consistent with their equation (3.2).}
\begin{align}\label{alpha-2n+1}
	\a^2=\frac{96\e}{
	(2n-1)(2n)!\(
	6-(2n+1)!(2n)!\,\G(\frac{2n+1}{2n-1})\mathcal{S}(n)
	\)
	}+\ldots
	\,,
\end{align}
which involves a sum
\begin{align}
\label{Sn}
	\mathcal{S}(n)=\hspace{-0.75em}\sum_{
	\substack{r+s+t=2n+1 \\ r,s,t\in\mathbb{Z}_+}
	}\!\!
	\;\frac{
	\G(\frac{s+t-r}{2n-1})\G(\frac{t+r-s}{2n-1})\G(\frac{r+s-t}{2n-1})
	}{r!s!t!(2n+1-r-s)!(2n+1-s-t)!(2n+1-t-r)!\,\G(\frac{2r}{2n-1})\G(\frac{2s}{2n-1})\G(\frac{2t}{2n-1})}
	\,.
\end{align}
Here $\mathbb{Z}_+$ denotes the set of positive integers.
If we use this input from the diagrammatic method, then the anomalous dimensions of $\phi$ and $\mathcal{J}$ are completely fixed for $k=1$.
For $n=1$, there is only one term in \eqref{Sn}, which is associated with $r=s=t=1$. 
We verify that the $n=1$ case of 
\eqref{alpha-2n+1} matches the Yang-Lee result \eqref{alpha2}.
For $n=2,3,4$, our results for $\g_\phi$ are in agreement with those of \cite{Gracey:2017okb}.\footnote{However, our results for $\g_{\phi^2}$ do not agree with \cite{Gracey:2017okb}. Some disagreement was pointed out in \cite{Codello:2017epp} as well, and it was shown that the results of \cite{Codello:2017epp} are supported by the CFT analysis \cite{Codello:2017qek}.}

\subsection{Analytic bootstrap}

In this subsection, we verify the multiplet-recombination results using the analytic bootstrap.
As noted in the generalized Yang-Lee and Potts theories  below \eqref{crossing gamma YL} and \eqref{crossing gamma Potts}, the multiplet-recombination results are consistent with the leading crossing constraints for arbitrary $\a$. 
Therefore, as in the multiplet-recombination approach, 
we are not able determine $\a$ by the leading crossing analysis.
As in section \ref{Analytic bootstrap}, we consider the four-point function
\begin{align}
	\< \phi(x_{1})\phi(x_{2})\phi(x_{3})\phi(x_{4}) \>
	=\frac{ g(z,\zb) }{ x_{12}^{ 2\D_{ \smash{\phi} } }x_{34}^{ 2\D_{ \smash{\phi} } } }
	\,,
\end{align}
and the crossing equation
\begin{align}
	g(z,\zb)=\frac{ z^{ \D_{ \smash{\phi} } }\zb^{ \D_{ \smash{\phi} } } }
	{ (1-z)^{ \D_{ \smash{\phi} } }(1-\zb)^{ \D_{ \smash{\phi} } } } g(1-\zb,1-z)
	\,.
\end{align}
At order $\a^0$, this analysis is the similar to that in section \ref{Analytic bootstrap}.
The difference is that the scaling dimension of $\phi_{ \text{f} }$ takes a different  value $\D_{ \smash{ \phi_{ \text{f} } } }=\frac{2k}{2n-1}$.
We derive the free OPE coefficient 
\begin{align}\label{free-OPE-phi-2n+1}
	P_{ \smash{\mathcal{J}_{\ell}^{(0)}},\text{f} } = 2S_{ \frac{2k}{1-2n} }(h)
	=2\frac{ \( \frac{2k}{2n-1} \)_\ell^2 }{ \ell! \( \frac{4k}{2n-1}+\ell-1 \)_{\ell} }
	\,,
\end{align}
where we have used the relation \eqref{match-singularity} and the formula \eqref{S-formula}.

\begin{figure}[h]
	\centering
	\begin{tikzpicture}[x=0.75pt,y=0.75pt,yscale=-1,xscale=1]
		\draw [line width=1.5]    (144.3,22.55) -- (214.63,94.03) ;
		\draw [line width=1.5]    (143.8,163.55) -- (214.63,94.03) ;
		\draw  [line width=1.5]  (214.63,94.03) .. controls (214.63,56.54) and (263,26.15) .. (322.67,26.15) .. controls (382.33,26.15) and (430.7,56.54) .. (430.7,94.03) .. controls (430.7,131.52) and (382.33,161.91) .. (322.67,161.91) .. controls (263,161.91) and (214.63,131.52) .. (214.63,94.03) -- cycle ;
		\draw  [line width=1.5]  (214.63,94.03) .. controls (214.63,65.78) and (263,42.87) .. (322.67,42.87) .. controls (382.33,42.87) and (430.7,65.78) .. (430.7,94.03) .. controls (430.7,122.29) and (382.33,145.2) .. (322.67,145.2) .. controls (263,145.2) and (214.63,122.29) .. (214.63,94.03) -- cycle ;
		\draw [line width=1.5]    (430.7,94.03) -- (501.23,163.55) ;
		\draw [line width=1.5]    (430.7,94.03) -- (500.73,22.55) ;
		\draw  [line width=1.5]  (214.63,94.03) .. controls (214.63,74.35) and (263,58.39) .. (322.67,58.39) .. controls (382.33,58.39) and (430.7,74.35) .. (430.7,94.03) .. controls (430.7,113.72) and (382.33,129.68) .. (322.67,129.68) .. controls (263,129.68) and (214.63,113.72) .. (214.63,94.03) -- cycle ;
		\draw [line width=1.5]  [dash pattern={on 5.63pt off 4.5pt}]  (322.5,2.05) -- (322.5,4.55) -- (322.5,189.85) ;
		\draw [line width=2.25]  [dash pattern={on 2.25pt off 7.5pt}]  (293.5,77.35) -- (293.5,110.35) ;
		\draw [line width=2.25]  [dash pattern={on 2.25pt off 7.5pt}]  (349,75.85) -- (349,89.85) -- (349,108.85) ;
		
		\draw (128,157) node [anchor=north west][inner sep=0.75pt]  [font=\normalsize]  {$\phi $};
		\draw (128,12) node [anchor=north west][inner sep=0.75pt]  [font=\normalsize]  {$\phi $};
		\draw (504,12) node [anchor=north west][inner sep=0.75pt]  [font=\normalsize]  {$\phi $};
		\draw (504,157) node [anchor=north west][inner sep=0.75pt]  [font=\normalsize]  {$\phi $};
		\draw (317,188.9) node [anchor=north west][inner sep=0.75pt]    {$\phi^{2n-1}$};
		\draw (196,87) node [anchor=north west][inner sep=0.75pt]    {$g$};
		\draw (440,87) node [anchor=north west][inner sep=0.75pt]    {$g$};
	\end{tikzpicture}
	\caption{Feynman diagram in the Lagrangian description \eqref{action-phi-2n+1}. Due to the $g\phi^{2n+1}$ vertices, the operator $\phi^{2n-1}$ contributes at order $g^2 \sim \a^2$. This is similar to the case of the $\phi^{2n}$ theory discussed in \cite{Henriksson:2020jwk}.}
	\label{feynman-diagram}
\end{figure}
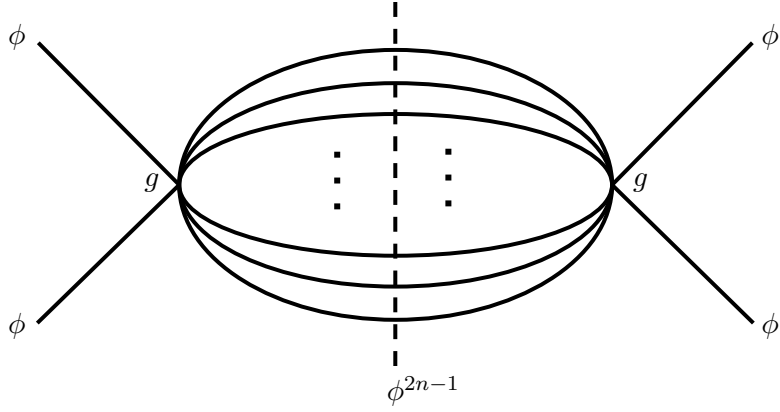

At order $\a^2$, the scalar $\phi^{2n-1}$ contributes to the enhanced singularity on the right-hand side of the crossing equation, due to the squared OPE coefficient $P_{ \phi^{2n-1} } \sim \a^2$ (see figure \ref{feynman-diagram}).
The $\log z$ part of the $\phi^{2n-1}$ block is given by \eqref{scalar-block}, with $\mathcal{O}=\phi^{2n-1}$.
The corresponding term takes the form
\begin{align}\label{phi-2n-1-contribution}
	&\frac{ \zb^{ \D_{ \smash{\phi} } } }{ (1-\zb)^{ \D_{ \smash{\phi} } } }
	P_{ \phi^{2n-1} }
	G_{ \phi^{2n-1} }(1-\zb,1-z)\Big|_{ \log z \text{ part} } \nn
	=\;&-\frac{ n(2n)! \, \G(2k) \a^2 }
	{ 2^{4k-1}\G(k)^2(k)_k^2\( -k\frac{2n-3}{2n-1} \)_k^2 } \,
	\zb^{ \frac{2k}{2n-1} }
	(1-\zb)^{ k\frac{2n-3}{2n-1} } 
	{}_2F_1 \! \( k,k;k\frac{2n-3}{2n-1}+1;1-\zb \)
	+\ldots
	\,,
\end{align}
where we have used the squared OPE coefficient \eqref{OPE-2n-1-crossing} derived in appendix \ref{An OPE coefficient in the generalized phi-2n+1 theory}.
We should be more careful, as there are infinitely many terms contributing to the enhanced singularity.
Expanding \eqref{phi-2n-1-contribution} in terms of $\frac{1-\zb}{\zb}$, we can derive the large-spin expansion of the anomalous dimensions $\tilde{\g}_{\smash {\mathcal J^{(0)}_\ell}}$ order by order using the asymptotic relation \eqref{match-singularity}.
A more elegant approach is to use the Lorentzian inversion formula, which gives the complete result as an analytic function in spin.
In the lightcone limit, this leads to an SL($2,\mathbb{R}$) inversion integral \cite{Caron-Huot:2017vep,Alday:2017zzv}
\begin{align}\label{SL2R-inversion}
	C(z,h)=
	\frac{(2h-1)\G(h)^{4}}{\pi^{2}\G(2h)^2}
	\int_{0}^{1}
	\frac{\mathrm{d}\zb}{\zb^{2}}
	k_{h}(\zb) \,
	\text{dDisc}[g(1-\zb,1-z)]\Big|_{ \text{leading term in $z$} }\,,
\end{align}
where we consider the leading term in the $z\rightarrow 0$ expansion.
The double discontinuity (dDisc) is defined as
\begin{align}
	\text{dDisc}[f(\zb)]=f(\zb)-\frac{1}{2}f^{\circlearrowleft}(\zb)-\frac{1}{2}f^{\circlearrowright}(\zb)
	\,,
\end{align}
where $f^{\circlearrowleft}$ and $f^{\circlearrowright}$ are two analytic continuations around $\zb=1$.
The result of the integral is expanded as
\begin{align}
	C(z,h)
	=z^{\D_\phi} P_{\mathcal{J} } \,
	\( 1+\frac {1}{2}\tilde\g_{ \mathcal{J} }\log z
	+O( \tilde{\g}^2_{ \mathcal{J} } ) \)
	+O( z^{\D_\phi+1} )
	\,.
\end{align}
The calculation of the dDisc associated with \eqref{phi-2n-1-contribution} amounts to multiplying by the factor $2\sin^2 ( \p k \, \frac{2n-3}{2n-1} )$.
Substituting this dDisc into the inversion integral \eqref{SL2R-inversion} and dividing by the free OPE coefficient \eqref{free-OPE-phi-2n+1}, we obtain the anomalous dimensions of the bilinear operators on the $m=0$ trajectory:
\begin{align}
	\tilde{\g}_{\mathcal {J}_\ell^{(0)}}=-\frac{n^2 (2 n-1)!\,\a^2}{2^{4 k-3}(2 k-1)! \(\ell-k\frac{2 n-3}{2
   n-1}\)_{2 k}}+\ldots
	\,.
\end{align}
The result agrees with \eqref{gammaJ-phi-2n+1} at $m = 0$ and confirms the compatibility of the multiplet recombination with OPE associativity.
In appendix \ref{Analytic bootstrap at subleading twist}, the analytic bootstrap is also carried out at subleading twist, where the multiplet-recombination results are again consistent with crossing symmetry.

\section{Discussion}

In this paper we computed the anomalous dimensions of $\phi$ and bilinear operators $\mathcal{J}_\ell^{(m)}\sim\phi\pa^\ell\Box^m\phi$ in the generalized $\phi^{2n+1}$ theories, using the conformal multiplet recombination.
For $n=1$, we also considered the $\phi^3$ theory with global symmetry $S_{N+1}$. 
Our main results are summarized in table \ref{summary}.
We verified that they are consistent with crossing symmetry using the analytic bootstrap. 
To the best of our knowledge, new results in the canonical case $k=1$ include the anomalous dimensions of spinning bilinear operators in the Potts model.
In the generalized case with a higher-derivative kinetic term, i.e., $k>1$, all of the results for the anomalous dimensions and OPE coefficients should be new. 

We considered the $S_{N+1}$ global symmetry, but restricted to the $\phi^3$ case.
It might be interesting to study the $S_{N+1}$-symmetric $\phi^{p}$ theories with higher $p$, as they should provide a variety of universality classes.
In particular, the canonical $\phi^4$ and $\phi^5$ theories with $S_{N+1}$ symmetry were studied in \cite{Rong:2017cow,Codello:2018nbe,Codello:2020mnt}.
These theories correspond to the quartic and quintic Potts models.
It would also be interesting to investigate their higher-derivative generalizations. 

Since the set of anomalous dimensions is fixed up to an unknown parameter $\a$ in the $n>1$ theory, another important question is 
if we can determine $\a$ in the $n>1$ theory by conformal symmetry and consistency requirements.
In the canonical case $k=1$, 
we fix  the value of $\a$ by the input from \cite{Codello:2017epp}, which is given in \eqref{alpha-2n+1}. 
Given the complicated summation in \eqref{Sn}, 
it seems unlikely that we can determine $\alpha$ by the leading analysis of the multiplet recombination method.  
As \eqref{Sn} shares some similarities with the subleading corrections in the $\phi^{2n}$ theory \cite{Codello:2017epp}, 
we expect that $\alpha$ can be determined by a higher-order study, 
which requires a merging of the multiplet recombination and crossing analyses. 
As a necessary step, we verified that the multiplet-recombination results are consistent with crossing symmetry for the leading corrections. 

As the generalized free theory is nonunitary, 
we expect that the $\phi^{2n+1}$ deformations are nonunitary as well. The nonperturbative bootstrap study of their integer $d$ versions requires  bootstrap methods that do not rely on positivity constraints 
\cite{Gliozzi:2013ysa,Gliozzi:2014jsa,El-Showk:2016mxr,Esterlis:2016psv,Li:2017agi,Li:2017ukc,Li:2021uki,Kantor:2021kbx,Kantor:2021jpz,Afkhami-Jeddi:2021iuw,Laio:2022ayq,Li:2023tic}. 
In \cite{Guo:2023qtt} and the present work, 
we investigated the $\phi^{2n}$ and $\phi^{2n+1}$ deformations of the generalized free theory, 
i.e., $\phi^m$ interactions with integer power $m\geq 3$.  
\footnote{Can the generalized $\phi^m$ theories be realized by lattice models?}
A natural extension is to consider noninteger $m$. 
Some simple examples are the fractional power $m=\frac {m_1} 2, \frac {m_1} 3, \frac {m_1} 4,\dots$, 
where $m_1$ is a large enough integer. 
The $\phi^m$ deformations with irrational power or even complex power are also interesting from the perspective of analytic continuation,
as they can connect the standard integer-power cases in the theory space, 
but we expect that they are nonunitary even for the canonical case of $k=1$. 
The $\mathcal {PT}$ symmetric theories are particularly relevant because they can satisfy the physical requirement of a bounded-from-below spectrum in a continuous range of power \cite{Bender:1998ke,Bender:2007nj,Bender:2023cem}. 
\footnote{It is not clear to us how to capture the $\mathcal {PT}$-symmetric CFTs  with even power $m$ using the $\e$ expansion,  
as one needs to be more careful about the quantization scheme and  
it is not an easy task to handle multidimensional Stokes sectors, i.e., Lefschetz thimbles. 
See \cite{Felski:2021evi} and references therein. 
Furthermore, it seems more natural to use the combination $i\phi$ as the building block for $\mathcal{PT}$-symmetric theories. 
}
Recently, the low-dimensional $\mathcal {PT}$ symmetric models were revisited by bootstrap methods in 
\cite{Li:2022prn,Khan:2022uyz,Li:2023nip,John:2023him,Li:2023ewe,Li:2024rod}, 
using the equations of motion in the Lagrangian or Hamiltonian formalism. 
It would be interesting to extend the low-dimensional bootstrap methods to higher dimensions.
Along the lines of analytic continuation, one may also 
bootstrap the non-perturbative conformal field theory by the homotopy continuation method \cite{Li:2024ggr}, 
where the $\epsilon$ expansion can provide regular boundary conditions near but away from 
the potentially dangerous Gaussian limit. 
\footnote{Conformal symmetry implies the existence of zero-norm states associated 
with specific Gaussian scaling dimensions. 
Usually, the zero-norm states should decouple to avoid inconsistency, such as divergences in the Gaussian limit. } 

In addition, the $\phi^4$ deformation of a canonical free theory with boundaries or defects was studied using the multiplet recombination and equation of motion \cite{Yamaguchi:2016pbj,Soderberg:2017oaa,Dey:2020jlc,Nishioka:2022odm,Nishioka:2022qmj,Bissi:2022bgu,SoderbergRousu:2023pbe}.
An interesting future direction is to consider the $\phi^m$ deformations and higher-derivative generalizations, as well as other types of defects.
Another direction would be to study CFTs on nontrivial manifolds.
The case of real projective space was carried out for the $\phi^3$ and $\phi^4$ theory \cite{Hasegawa:2016piv,Hasegawa:2018yqg}.
It would also be interesting to apply the multiplet-recombination method to CFTs at finite temperature.

\section*{Acknowledgments}

This work was supported by the Natural Science Foundation of China (Grant No. 12205386) and the Guangzhou Municipal Science and Technology Project (Grant No. 2023A04J0006).

\appendix

\section{The action of Laplacians}
\label{The action of Laplacians}

In this appendix, we compute the action of Laplacians on a generic scalar three-point function
\begin{align}
	\<\mathcal{O}_1(x_1)\mathcal{O}_2(x_2)\mathcal{O}_3(x_3)\>
	\propto
	\frac{1}{ 
	x_{12}^{ h_{123} }
	x_{13}^{ h_{132} }
	x_{23}^{ h_{231} }
	}
	\,,
\end{align}
where $h_{ijk}$ are defined by
\begin{align}\label{hijk}
	h_{ijk}=\D_{i}+\D_{j}-\D_{k}
	\,.
\end{align}
Here $\D_{i}$ denotes the scaling dimension of $\mathcal{O}_i$.
The result of the $\Box_{x_1}$ action takes the form
\begin{align}
	&\Box_{x_1} \frac{1}{ 
	x_{12}^{ h_{123} }
	x_{13}^{ h_{132} }
	x_{23}^{ h_{231} }
	} 
	\nn=\;&\Bigg[
	\frac{
	h_{123} \( h_{123}+h_{132}+2-d \)
	}{ x_{12}^2 }
	+\frac{
	h_{132} \( h_{123}+h_{132}+2-d \)
	}{ x_{13}^2 }
	-\frac{ h_{123}h_{132} \, x_{23}^2 }
	{ x_{12}^2 x_{13}^2 }
	\Bigg] \frac{1}{ 
	x_{12}^{ h_{123} }
	x_{13}^{ h_{132} }
	x_{23}^{ h_{231} }
	}
	\,.
\end{align}
From this we derive the formula for the action of $\Box_{x_1}^k$:
\begin{align}\label{one-boxk-3pt}
	\Box_{x_1}^k \frac{1}{ 
	x_{12}^{ h_{123} }
	x_{13}^{ h_{132} }
	x_{23}^{ h_{231} }
	}=
	\sum^{k}_{
	\substack{i,j=0 \\
	i+j \geqslant k }
	} \[ B_{i,j} C_{i,j}^{ h_{123},h_{132} }
	\frac{ x_{23}^{2(i+j-k)} }
	{ x_{12}^{2i} x_{13}^{2j} }
	\] \frac{1}{ 
	x_{12}^{ h_{123} }
	x_{13}^{ h_{132} }
	x_{23}^{ h_{231} }
	}
	\,,
\end{align}
where the coefficients are
\begin{align}
	\label{Bij}
	B_{i,j}&=\frac{(-1)^{i+j-k}2^{2k}k!}
	{(k-i)! \, (k-j)! \, (i+j-k)!}
	\,, \\
	\label{Cij}
	C_{i,j}^{a,b}&=\bigg( \frac{ a }{2} \bigg)_{i} \bigg( \frac{ b }{2} \bigg)_{j}
	\( \frac{ a+b }{2}-\frac{d-2}{2}+i+j-k \)_{2k-i-j}
	\,.
\end{align}
Since each summand on the right-hand side of \eqref{one-boxk-3pt} can be seen as a new three-point function, the formula associated with $\Box_{x_1}^k\Box_{x_2}^k\Box_{x_3}^k$ follows:
\begin{align}\label{3-box}
	&\Box_{x_1}^k\Box_{x_2}^k\Box_{x_3}^k
	\frac{1}{ 
	x_{12}^{ h_{123} }
	x_{13}^{ h_{132} }
	x_{23}^{ h_{231} }
	} \nn
	=\;&\sum^{k}_{ \substack{i_{1},j_{1}=0 \\
	i_{1}+j_{1} \geqslant k } } \;\; \sum^{k}_{ \substack{i_{2},j_{2}=0 \\
	i_{2}+j_{2} \geqslant k } } \;\; \sum^{k}_{ \substack{i_{3},j_{3}=0 \\
	i_{3}+j_{3} \geqslant k } } \Bigg[
	B_{ i_{1},j_{1} } C_{ i_{1},j_{1} }^{ h_{123},h_{132} }
	B_{ i_{2},j_{2} } C_{ i_{2},j_{2} }^{ h_{123}+2i_{1} , h_{231}+2(k-i_{1}-j_{1}) } \nn
	&\hspace{4.5cm}\times B_{ i_{3},j_{3} } C_{ i_{3},j_{3} }^{ h_{231}+2(j_{2}+k-i_{1}-j_{1}) , h_{132}+2(j_{1}+k-i_{2}-j_{2}) } x_{12}^{ 2(i_{3}+j_{3}-i_{1}-i_{2}-k) } \nn
	&\hspace{4.5cm}\times x_{13}^{ 2(i_{2}+j_{2}-j_{1}-j_{3}-k) }
	x_{23}^{ 2(i_{1}+j_{1}-j_{2}-i_{3}-k) } \Bigg]
	\frac{1}{ 
	x_{12}^{ h_{123} }
	x_{13}^{ h_{132} }
	x_{23}^{ h_{231} }
	}
	\,.
\end{align}

\section{Analytic bootstrap at subleading twist}
\label{Analytic bootstrap at subleading twist}

In this appendix, we check the consistency between the multiplet-recombination results and crossing symmetry at subleading twist $m=1$ using the analytic bootstrap.

Let us start with the generalized Yang-Lee edge singularity.
We examine the subleading term of the crossing equation \eqref{crossing} in the $z\rightarrow 0$ expansion.
On the left-hand side of the crossing equation, both $\mathcal{J}^{(0)}_{\ell}$ and  $\mathcal{J}^{(1)}_{\ell}$ contribute to the infinite sum.
The lightcone expansion of the conformal blocks is written as
\begin{align}\label{conformal-block-z1}
	G_{ \mathcal{O} }(z,\zb)=z^{ (\D_{ \mathcal{O} }-\ell)/2 } k_{h}(\zb) 
	+\sum_{i=-1}^{1}c_i \, z^{ (\D_{ \mathcal{O} }-\ell)/2+1 } k_{h+i}(\zb)
	+ O (z^{ (\D_{ \mathcal{O} }-\ell)/2+2 } )
	\,.
\end{align}
For $\mathcal{O}=\mathcal{J}^{(0)}_{\ell}$, the relevant part is the subleading term in this expansion.
Here, $c_i$ are obtained by requiring that \eqref{conformal-block-z1} solves the Casimir equation \cite{Dolan:2003hv}.
The results are
\begin{align}
	c_{-1}=\frac{(d-2)\ell}{d+2\ell-4}\,,\quad
	c_0=\frac{ \D_{\mathcal{O} }-\ell }{4}\,,\quad
	c_1=-\frac{(d-2) (\D_{\mathcal{O} } -1) (\D_{\mathcal{O} } +\ell)^2}{16 (d-2 \D_{\mathcal{O} } -2) (\D_{\mathcal{O} } +\ell-1) (\D_{\mathcal{O} } +\ell+1)}
	\,.
\end{align}
We expand \eqref{conformal-block-z1} in the anomalous dimension \eqref{gamma-crossing}:
\begin{align}
	&G_{ \mathcal{J}^{(0)}_{\ell} }(z,\zb)=z^{ \D_{ \smash{\phi} } } \( k_{h^{(0)}}(\zb)+\frac{1}{2}\tilde{\g}_{ \mathcal{J}^{(0)}_{\ell} } k_{h^{(0)}}(\zb)\log z+ O( \tilde{\g}_{ \smash{ \mathcal{J}^{(0)}_{\ell} } }^{2} ) \) \nn
	&+z^{ \D_{ \smash{\phi} }+1 }\(
	\sum_{i=-1}^{1}c_i k_{h^{(0)}+i}(\zb)
	+\frac{1}{2}\tilde{\g}_{ \mathcal{J}^{(0)}_{\ell} }\sum_{i=-1}^{1}c_i k_{h^{(0)}+i}(\zb)\log z + O( \tilde{\g}_{ \smash{ \mathcal{J}^{(0)}_{\ell} } }^{2})
	\)
	+ O(z^{ (\D_{ \smash{ \mathcal{J}^{(0)}_{\ell} } }-\ell)/2+2 } )
	\,,
\end{align}
where the conformal spin is $h^{(m)}\equiv h_{ \smash{ \mathcal{J}^{(m)}_{\ell} } }=(\D_{ \smash{ \mathcal{J}^{(m)}_{\ell} } }+\ell)/2$.
For $\mathcal{O}=\mathcal{J}^{(1)}_{\ell}$, it is sufficient to consider the leading term in the $z\rightarrow 0$ expansion:
\begin{align}
	G_{ \mathcal{J}^{(1)}_{\ell} }(z,\zb)=z^{ \D_{ \smash{\phi} }+1 } \( k_{h^{(1)}}(\zb)+\frac{1}{2}\tilde{\g}_{ \mathcal{J}^{(1)}_{\ell} } k_{h^{(1)}}(\zb)\log z+ O\big( \tilde{\g}_{ \smash{ \mathcal{J}^{(1)}_{\ell} } }^{2} \big) \) 
	+ O(z^{ (\D_{ \smash{ \mathcal{J}^{(1)}_{\ell} } }-\ell)/2+1 } )
	\,.
\end{align}

At order $\e^0$, the left-hand side of the crossing equation contains the infinite sums associated with $\mathcal{J}^{(1)}_{\ell}$ and $\mathcal{J}^{(0)}_{\ell}$:
\begin{align}
	\sum_{\ell=0,2,4,\ldots}P_{ \mathcal{J}^{(1)}_{\ell} }k_{h^{(1)}}(\zb)
	+\sum_{\ell=0,2,4,\ldots}P_{ \mathcal{J}^{(0)}_{\ell} }\sum_{i=-1}^{1}c_i k_{h^{(0)}+i}(\zb)
	\,,
\end{align}
which is present at subleading order in the $z\rightarrow 0$ expansion, i.e., at order $z^{\D_\phi+1}$.
To make use of the asymptotic relation \eqref{match-singularity}, we rewrite it as a single infinite sum of $\text{SL}(2,\mathbb{R})$ blcoks
\begin{align}\label{infinite-sum}
	\sum_{\ell=0,2,4,\ldots}\[
	P_{ \mathcal{J}^{(1)}_{\ell} }+ \sum_{i=-1}^{1}\(P_{ \mathcal{J}^{(0)}_{\ell} }c_i\)_{\ell\rightarrow\ell-i+3}
	\]k_{h^{(1)}}(\zb)+O(\e)
	\,.
\end{align}
The enhanced singularity on the right-hand side of the crossing equation is $\D_{ \smash{ \phi_\text{f} } }(\frac{1-\zb}{\zb})^{-\D_{ \smash{ \phi_\text{f} } } }$, which corresponds to the identity-operator contribution.
Since $\D_{ \smash{ \phi_\text{f} } }=2k$, the asymptotic relation implies that the coefficient in \eqref{infinite-sum} is given by $2S_{-2k}(h^{(1)})$ at order $\e^0$.
Subtracting the contribution from the leading-twist operators $\mathcal{J}^{(0)}_{\ell}$, we find the free squared OPE coefficients
\begin{align}\label{OPE-subleading-free}
	P_{ \mathcal{J}^{(1)}_{\ell},\text{f} }
	=-\frac{(k-1) (2 k)_{\ell+1}^2}{\ell! (k+\ell+1) (3 k+\ell) (4 k+\ell+1)_\ell}
	\,,
\end{align}
where we have used the free squared OPE coefficients at leading twist \eqref{free-OPE} .

For the leading correction term in the expansion in the anomalous dimensions, the infinite sum is given by
\begin{align}\label{gamma-infinite-sum}
	\sum_{\ell=0,2,4,\ldots}\[
	P_{ \mathcal{J}^{(1)}_{\ell} }	\tilde{\g}_{ \smash{ \mathcal{J}^{(1)}_{\ell} } }+ \sum_{i=-1}^{1}\(P_{ \mathcal{J}^{(0)}_{\ell} }\tilde{\g}_{ \smash{ \mathcal{J}^{(0)}_{\ell} } }c_i\)_{\ell\rightarrow\ell-i+3}
	\]k_{h^{(1)}}(\zb)\log z+O(\e)
	\,.
\end{align}
On the right-hand side of the crossing equation, the leading correction term is of order $\e^1$ and is associated with the contribution of $\phi$.
This term corresponds to the infinite sum involving anomalous dimensions on the left-hand side.
For a scalar operator $\mathcal{O}$, the $\log z$ part of the conformal block takes the form \cite{Li:2020ijq}
\begin{align}\label{scalar-block-z2}
	&G_{ \mathcal{O} }(1-\zb,1-z)\big|_{\log z}\nn
	=&\;(1-\zb)^{ \frac{ \D_{ \mathcal{O} } }{2} }\frac{\Gamma(\D_{ \mathcal{O} })}{\Gamma(\D_{ \mathcal{O} }/2)^{2}}\Big\{-\,_{2}F_{1}\(\frac{\D_{ \mathcal{O} }}{2},\frac{\D_{ \mathcal{O} }}{2};\D_{ \mathcal{O} }-\frac{d-2}{2};1-\zb\)\nn
	&+\frac{\D_{ \mathcal{O} }}{4}\[(2-\D_{ \mathcal{O} })\,_{2}F_{1}\(\frac{\D_{ \mathcal{O} }}{2},\frac{\D_{ \mathcal{O} }}{2};\D_{ \mathcal{O} }-\frac{d-2}{2};1-\zb\)\right.\nn
	&+\left.\frac{(d-2)\D_{ \mathcal{O} }}{d-2\D_{ \mathcal{O} }-2}(1-\zb)\,_{2}F_{1}\(\frac{\D_{ \mathcal{O} }}{2}+1,\frac{\D_{ \mathcal{O} }}{2}+1;\D_{ \mathcal{O} }-\frac{d-4}{2};1-\zb\)\]z +O(z^2)\Big\}
	\,.
\end{align}
The contribution from $\phi$ reads $z^{\D_\phi}(1+\D_\phi\, z+O(z^2))\frac{ \zb^{ \D_{ \smash{\phi} } } }{ (1-\zb)^{ \D_{ \smash{\phi} } } }P_{\phi}G_{\phi}(1-\zb,1-z)|_{ \log z \text{ part} }$.
At subleading order in the lightcone expansion, this leads to the enhanced singularity given by
\begin{align}\label{enhanced-singularity-subleading}
	&P_{\phi} \frac{\G(2k)}{ \G(k)^{2} }\Bigg[\(
	\D_\phi+\frac{\D_\phi\(\D_\phi-2\)}{4}
	\)  \sum_{i=0}^{k-1}(-1)^{i+1}
	\frac{ (1-2k)_{i}(k)_{i} }{ i! \, (1-k)_{i} }
	\( \frac{1-\zb}{\zb} \)^{i-k} 
	-\frac{\(d-2\)\D_\phi^2}{4\(d-2\D_\phi-2\)}\nn
	&\times\sum_{i=0}^{k-2}(-1)^{i+1}
	\frac{ (1-2k)_{i}(k+1)_{i} }{ i! \, (2-k)_{i} }
	\( \frac{1-\zb}{\zb} \)^{i-k+1}
	+\text{Casimir-regular terms}\Bigg]+\ldots
	\,,
\end{align}
where we have only kept the terms of order $\e^1$.
Using \eqref{lambda}, \eqref{OPE-normalization}, and \eqref{match-singularity}, we find the coefficients in the infinite sum \eqref{gamma-infinite-sum}.
The anomalous dimensions $\tilde{\g}_{ \smash{ \mathcal{J}^{(1)}_{\ell} } }$ are obtained by subtracting the contribution from $\mathcal{J}^{(0)}_{\ell}$.
The results agree with those from the multiplet recombination.

For the generalized Potts model, the free squared OPE coefficients are related to those at $N=1$ by
\begin{align}
	P_{ \mathcal{J}^{(1)}_{\ell},\text{f} }^{ \text{ generic } N } = P_{ \mathcal{J}^{(1)}_{\ell},\text{f} }^{ \;N=1 }  \times
	\begin{cases}
		\; \frac{1}{N} \qquad & \text{trivial} \\
		\; \frac{1}{ (N-1)(N+1)^2 } & \text{standard} \\
		\; 1 & \text{symmetric traceless} \\
		\; -1 & \text{antisymmetric}
		\,,
	\end{cases}
\end{align}
which is the $m=1$ version of \eqref{free-OPE-Potts} at leading twist.
For the anomalous dimensions, we consider the enhanced singularity given by \eqref{enhanced-singularity-subleading}.
From this, we extract the coefficients in the infinite sum of the form \eqref{gamma-infinite-sum}.
Here we need to account for the $N$-dependent factors in the crossing equations \eqref{crossing-1}--\eqref{crossing-4}.
As a result, the coefficients in the infinite sums have the following relation with those in the Yang-Lee case:
\begin{align}
	\text{Potts coefficients}=\text{Yang-Lee coefficients}\times\begin{cases}
		\; \frac{ (N-1)(N+1)^2 }{N} \qquad & \text{trivial} \\
		\; \frac{N-2}{N-1} & \text{standard} \\
		\; -(N+1)^2 & \text{symmetric traceless} \\
		\; (N+1)^2 & \text{antisymmetric}
		\,.
	\end{cases}
\end{align}
We subtract the contribution from $\mathcal{J}^{(0)}_{\ell}$, and obtain the anomalous dimensions in the Potts model.
The results are again in agreement with those from the multiplet recombination.

Finally, let us consider the generalized $\phi^{2n+1}$ theory with $n>1$.
At order $\a^0$, the enhanced singularity at subleading order in the lightcone expansion is given by $\D_{ \smash{ \phi_\text{f} } }(\frac{1-\zb}{\zb})^{-\D_{ \smash{ \phi_\text{f} } } }$, with the scaling dimension
$\D_{\phi_\text{f}}=\frac{2k}{2n-1}$.
We obtain the free squared OPE coefficient
\begin{align}\label{free-OPE-subleading}
	P_{ \smash{\mathcal{J}_{\ell}^{(1)}},\text{f} } = \frac{(k-1) (2 n-1)^2 \(\frac{2 k}{2 n-1}\)_{\ell+1}^2}{\ell! (k-\ell+2n(k +\ell ))(\ell-3 k+2 n(k-\ell-1)+1)  \(\frac{4k}{2 n-1}+\ell+1\)_\ell}
	\,.
\end{align}
At order $\a^2$, we consider the inversion formula, which is associated with the double discontinuity given by
\begin{align}
	z^{\D_{\phi^{2n-1}}}(1+\D_{\phi^{2n-1}}\, z+O(z^2))\frac{ \zb^{ \D_{\phi^{2n-1}} } }{ (1-\zb)^{ \D_{\phi^{2n-1}} } }P_{\phi^{2n-1}}G_{\phi^{2n-1}}(1-\zb,1-z)\Big|_{ \log z \text{ part} }
	\,.
\end{align}
Multiplying by the factor $2\sin^2 ( \p k \, \frac{2n-3}{2n-1} )$, we obtain the corresponding dDisc, as in \eqref{phi-2n-1-contribution}.
At subleading order in the $z\rightarrow 0$ expansion, we find the anomalous dimensions of $\mathcal{J}^{(1)}_{\ell}$ using the leading-twist anomalous dimension, as well as the free squared OPE coefficients \eqref{free-OPE-phi-2n+1} and \eqref{free-OPE-subleading}.
The results are consistent with those obtained from the multiplet recombination.

In conclusion, we confirm the consistency between the multiplet-recombination results and crossing symmetry at subleading twist.

\section{Representation theory of \boldmath{$S_{N+1}$}}
\label{Representation theory}

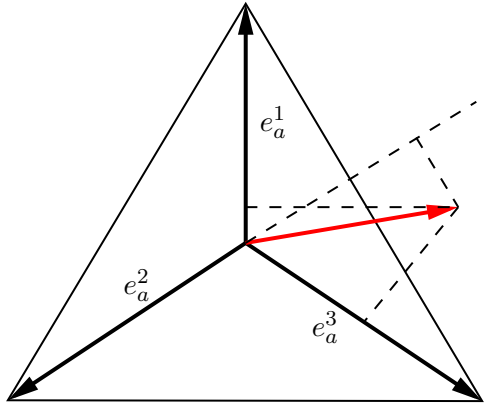
\begin{figure}[h]
	\centering
	\tikzset{every picture/.style={line width=0.75pt}}
	\begin{tikzpicture}[x=0.75pt,y=0.75pt,yscale=-1,xscale=1]
		\draw   (350.5,13.08) -- (468.52,212.96) -- (231.77,212.55) -- cycle ;
		\draw [fill={rgb, 255:red, 0; green, 120; blue, 255 }  ,fill opacity=1 ][line width=1.5]    (350.5,133.4) -- (350.5,17.08) ;
		\draw [shift={(350.5,13.08)}, rotate = 90] [fill={rgb, 255:red, 0; green, 0; blue, 0 }  ][line width=0.08]  [draw opacity=0] (15.6,-3.9) -- (0,0) -- (15.6,3.9) -- cycle    ;
		\draw [fill={rgb, 255:red, 0; green, 117; blue, 255 }  ,fill opacity=1 ][line width=1.5]    (350.5,133.4) -- (235.1,210.33) ;
		\draw [shift={(231.77,212.55)}, rotate = 326.31] [fill={rgb, 255:red, 0; green, 0; blue, 0 }  ][line width=0.08]  [draw opacity=0] (15.6,-3.9) -- (0,0) -- (15.6,3.9) -- cycle    ;
		\draw [fill={rgb, 255:red, 0; green, 117; blue, 255 }  ,fill opacity=1 ][line width=1.5]    (350.5,133.4) -- (465.21,210.73) ;
		\draw [shift={(468.52,212.96)}, rotate = 213.98] [fill={rgb, 255:red, 0; green, 0; blue, 0 }  ][line width=0.08]  [draw opacity=0] (15.6,-3.9) -- (0,0) -- (15.6,3.9) -- cycle    ;
		\draw [color={rgb, 255:red, 255; green, 0; blue, 0 }  ,draw opacity=1 ][line width=1.5]    (350.5,133.4) -- (452.66,116.07) ;
		\draw [shift={(456.6,115.4)}, rotate = 170.37] [fill={rgb, 255:red, 255; green, 0; blue, 0 }  ,fill opacity=1 ][line width=0.08]  [draw opacity=0] (15.6,-3.9) -- (0,0) -- (15.6,3.9) -- cycle    ;
		\draw  [dash pattern={on 4.5pt off 4.5pt}]  (350.32,115.4) -- (456.6,115.4) ;
		\draw  [dash pattern={on 4.5pt off 4.5pt}]  (456.6,115.4) -- (409.51,173.18) ;
		\draw [color={rgb, 255:red, 0; green, 0; blue, 0 }  ,draw opacity=1 ] [dash pattern={on 4.5pt off 4.5pt}]  (350.5,133.4) -- (465.6,62.4) ;
		\draw  [dash pattern={on 4.5pt off 4.5pt}]  (435.6,80.4) -- (456.6,115.4) ;
		
		\draw (356,63.4) node [anchor=north west][inner sep=0.75pt]    {$e^1_a$};
		\draw (288,145.4) node [anchor=north west][inner sep=0.75pt]    {$e^2_a$};
		\draw (382,166.4) node [anchor=north west][inner sep=0.75pt]    {$e^3_a$};
	\end{tikzpicture}
	\caption{Projection of a two-component vector. Using the vectors $e^\a_a$ in the 2-simplex (equilateral triangle), the 2D vector in red is associated with three components.}
	\label{2-simplex}
\end{figure}

The ($N+1$)-component vectors transform naturally under $\tilde{P}^{\a\b}\in S_{N+1}$, where $\tilde{P}^{\a\b}$ is a permutation matrix.
However, this does not lead to an irreducible representation, since the ($N+1$)-component vector $(1,1,\ldots,1)$ forms the basis for a 1D invariant subspace.
To construct a nontrivial irreducible representation, it is convenient to think of $S_{N+1}$ as the symmetry group of an $N$-simplex, which has $N+1$ vertices.
The vertices are in one-to-one correspondence with the vectors $e_a^\a$, which connect the center and vertices.
The latin index $a$ ranges from 1 to $N$ as the simplex is embedded in an $N$-dimensional space.
The greek index $\a$ ranges from 1 to $N+1$ because there are $N+1$ vertices.
The vectors satisfy the relations \cite{Zia:1975ha}\footnote{The normalization in \cite{Zia:1975ha} is different from ours: $e_{a,\text{ours} }^\a=\sqrt{N} \; e_{a,\text{theirs} }^\a$.}
\begin{align}\label{e-relation}
	\sum_{\a=1}^{N+1}e_{a}^{\a}=0
	\,, \qquad
	\sum_{\a=1}^{N+1}e_{a}^{\a}e_{b}^{\a}=(N+1)\d_{ab}
	\,, \qquad
	\sum_{a=1}^{N}e_{a}^{\a}e_{a}^{\b}=(N+1)\d^{\a\b}-1
	\,.
\end{align}
An $N$-component vector $v_{a}$ can be projected onto the vectors $e_a^1,e_a^2,\ldots,e_a^{N+1}$ (see figure \ref{2-simplex}), and the results can be viewed as an $(N+1)$-component vector that transforms under $\tilde{P}^{\a\b}$.
Then, we write the vector after the transformation as an $N$-component vector again.
In short, the transformation rule for $N$-component vectors is given by
\begin{align}\label{v-transform}
	v_{a}=\sum_{b=1}^{N} P_{a b}\, v_{b}
	\,, \qquad
	P_{a b}=\frac{1}{N+1}\sum_{\a,\b=1}^{N+1} e^{\a}_{a} \tilde{P}^{\a\b} e^{\b}_{b}
	\,.
\end{align}
In this way, the space of the $N$-component vectors furnishes an irreducible representation of $S_{N+1}$, which is isomorphic to the standard representation of $S_{N+1}$.\footnote{The standard representation of $S_{N+1}$ consists of ($N+1$)-component vectors $u^{\a}$.
They are subject to the condition that the sum of the components vanishes $\sum_{\a=1}^{N+1}u^{\a}=0$.}
We do not distinguish between the two isomorphic vector spaces and refer to the space of $N$-component vectors as the standard representation.

The dimensions of the corresponding representations $V_{(\ldots)}$ are given by the hook length formula $\frac{(N+1)!}{\prod(\text{hook lengths})}$.
We have
\begin{align}
	\label{trivial-dim}
	\dim V_{(N+1)}
	&=\frac{(N+1)!}
	{(N+1)\times N \times \ldots \times 1}
	=1
	\,, \\
	\label{standard-dim}
	\dim V_{(N,1)}
	&=\frac{(N+1)!}
	{(N+1) \times (N-1) \times (N-2) \times \ldots \times 1 \times 1}
	=N
	\,, \\
	\label{ST-dim}
	\dim V_{(N-1,2)}
	&=\frac{(N+1)!}
	{N \times (N-1) \times (N-3) \times \ldots \times 1 \times 2 \times 1}
	=\frac{(N+1)(N-2)}{2}
	\,, \\
	\label{antisymmetric-dim}
	\dim V_{(N-1,1^2)}
	&=\frac{(N+1)!}
	{(N+1) \times (N-2) \times (N-3) \times \ldots \times 1 \times 2 \times 1}
	=\frac{N(N-1)}{2}
	\,.
\end{align}

More concretely, using the projectors in appendix \ref{Tensor structures}, we obtain some examples of the bilinear operators:
\begin{align}
	\label{trivial-example}
	\sum_{c,d=1}^{N} t^{(1)}_{abcd} \, \phi_c\phi_d &\rightarrow \sum_{a=1}^{N}\phi_a\phi_a = \phi^2
	\,, \\
	\label{standard-example}
	\sum_{c,d=1}^{N} t^{(2)}_{abcd} \, \phi_c\phi_d &\rightarrow \sum_{b,c=1}^{N} T_{abc} \, \phi_{b}\phi_{c}
	\,,	 \\
	\label{ST-example}
	\sum_{c,d=1}^{N} t^{(3)}_{abcd} \, \phi_c\phi_d &\rightarrow  (N+1)(N-1)\phi_a\phi_b+\frac{N+1}{N}\d_{ab}\phi^2-\sum_{\a=1}^{N+1} \sum_{c,d=1}^{N} e_a^\a e_b^\a e_c^\a e_d^\a \phi_c \phi_d
	\,, \\
	\label{antisymmetric-example}
	\sum_{c,d=1}^{N} t^{(4)}_{abcd} \, \phi_c \pa_\m \phi_d &\rightarrow  \phi_{a}\pa_{\m}\phi_{b}
	-\phi_{b}\pa_{\m}\phi_{a}
	\,.
\end{align}
The last two lines follow directly from the definition of $t_{abcd}^{(i)}$.
The first line is obtained by removing the redundancy $\d_{ab}$.
The second line follows from
\begin{align}
	\sum_{c,d=1}^{N} t^{(2)}_{abcd}\phi_{c}\phi_{d}
	=(N+1)\(
	\sum_{\a=1}^{N+1} \sum_{c,d=1}^{N} e_a^\a e_b^\a e_c^\a e_d^\a \phi_c \phi_d
	-(N+1)\d_{ab}\phi^2
	\)=\sum_{c,i,j=1}^{N}T_{abc}T_{cij}\phi_{i}\phi_{j}
	\,,
\end{align}
which implies that $\sum_{c,d=1}^{N} t^{(2)}_{abcd}\phi_{c}\phi_{d}$ is a redundant description of $\sum_{b,c=1}^{N}T_{abc}\phi_{b}\phi_{c}$.
We thus focus on the latter one.

Comparing the dimensions with those of \eqref{trivial-dim}, \eqref{standard-dim}, \eqref{ST-dim}, and \eqref{antisymmetric-dim}, we conclude that \eqref{trivial-example}, \eqref{standard-example}, \eqref{ST-example}, and \eqref{antisymmetric-example} correspond to the Young diagrams $(N+1)$, $(N,1)$, $(N-1,2)$, and $(N-1,1^2)$.\footnote{The dimension of \eqref{ST-example} follows from the symmetric tracelessness and the constraints \begin{align}
	\sum_{a,b=1}^{N}T_{iab}\sum_{c,d=1}^{N} t^{(3)}_{abcd}\phi_c\phi_d=0
	\,,
\end{align}
	where $i=1,2,\ldots,N$.}
From these examples, we see that the tensor product decomposition resembles that in the O($N$) model, except that here the symmetric traceless part further decomposes into two irreducible representations $V_{(N,1)}$ and $V_{(N-1,2)}$.
Since $V_{(N,1)}$ is the standard representation, we use the term ``symmetric traceless representation'' to refer to $V_{(N-1,2)}$.
Moreover, we refer to $V_{(N+1)}$ and $V_{(N-1,1^2)}$ as the ``trivial representation'' and the ``antisymmetric representation.''\footnote{The antisymmetric representation $V_{(N-1,1^2)}$ is the exterior power $\L^2 V_{(N,1)}$ of the standard representation $V_{(N,1)}$.}

\section{Tensor structures}
\label{Tensor structures}

The tensor product decomposition of $(N,1)^{\otimes 4}$ contains the trivial representation four times, so there are four linearly independent rank-4 invariant tensors \cite{wallace1978spin}.
Using $e_a^\a$ as defined in \eqref{e-relation}, we identify four rank-4 $S_{N+1}$-invariant tensors: $\d_{ab} \d_{cd}$, $\d_{ac} \d_{bd}$, $\d_{ad} \d_{bc}$, and $\sum_{\a=1}^{N+1} e_{a}^{\a}e_{b}^{\a}e_{c}^{\a}e_{d}^{\a}$.
We then use these tensors to define a basis for the four-point function $\< \phi_a(x_{1})\phi_b(x_{2})\phi_c(x_{3})\phi_d(x_{4}) \>$:
\begin{align}
	\label{4pt-tensors-1}
	t_{abcd}^{(1)}&=\d_{ab} \d_{cd} 
	\,,\\
	\label{4pt-tensors-2}
	t_{abcd}^{(2)}&=(N+1) \( \sum_{\a=1}^{N+1} e_{a}^{\a}e_{b}^{\a}e_{c}^{\a}e_{d}^{\a}
	-(N+1) \d_{ab} \d_{cd} \)
	\,,\\
	\label{4pt-tensors-3}
	t_{abcd}^{(3)}&=\frac{1}{2}\( \d_{ac} \d_{bd} + \d_{ad} \d_{bc} \)
	+\frac{1}{ N(N-1) } \d_{ab} \d_{cd}
	-\frac{1}{ (N+1)(N-1) } \sum_{\a=1}^{N+1} e_{a}^{\a}e_{b}^{\a}e_{c}^{\a}e_{d}^{\a}
	\,,\\
	\label{4pt-tensors-4}
	t_{abcd}^{(4)}&=\frac{1}{2}\( \d_{ac} \d_{bd} - \d_{ad} \d_{bc} \)
	\,.
\end{align}
For $i=1,2,3,4$, the tensor structures $t_{abcd}^{(i)}$ are associated with the intermediate operators in the trivial, standard, symmetric tracless, and antisymmetric representations.
The four tensors can also be used as projectors, which satisfy
\begin{align}
	\sum_{c,d=1}^N t_{a_1 b_1 c \, d}^{(i)} \, t_{c \, d \, a_2 b_2}^{(j)}
	\propto
	\d^{ij} t_{a_1 b_1 a_2 b_2}^{(i)}
	\,.
\end{align} 
The normalization of $t_{abcd}^{(i)}$ is chosen as follows.
The free correlator $\< \phi_{a, \text{f} }(x_{1})\phi_{b, \text{f} }(x_{2})\phi_{c, \text{f} }(x_{3})$ $\phi_{d, \text{f} }(x_{4}) \>$ is computed using Wick contractions, yielding four parts that correspond to the four tensor structures. We require that the conformal block decomposition of each part is precisely \eqref{block-decomposition}, with the same normalization of OPE coefficients and conformal blocks.
We use these projectors to construct bilinear operators in the irreducible representations, as in \eqref{trivial-example}, \eqref{standard-example}, \eqref{ST-example}, and \eqref{antisymmetric-example}.

Using the tensor structures defined above, the four-point function is written as
\begin{align}
	\< \phi_a(x_{1})\phi_b(x_{2})\phi_c(x_{3})\phi_d(x_{4}) \>
	=\frac{1}{ x_{12}^{ 2\D_{ \smash{\phi} } }x_{34}^{ 2\D_{ \smash{\phi} } } }
	\sum_{i=1}^4 g^{(i)}(z,\zb) \, t_{abcd}^{(i)}
	\,.
\end{align}
The crossing equations read
\begin{align}
	\label{crossing-1}
	g^{(1)}(z,\zb)=\;
	&\frac{ z^{ \D_{ \smash{\phi} } }\zb^{ \D_{ \smash{\phi} } } }
	{ (1-z)^{ \D_{ \smash{\phi} } }(1-\zb)^{ \D_{ \smash{\phi} } } }
	\Bigg( \frac{ g^{(1)}(1-\zb,1-z) }{N}
	+\frac{ (N-1)(N+1)^2 }{N} g^{(2)}(1-\zb,1-z)
	\nn
	&\hspace{6em}+\frac{ (N-2)(N+1) }{ 2N^2 } g^{(3)}(1-\zb,1-z)
	+\frac{1-N}{2N} g^{(4)}(1-\zb,1-z) \Bigg)
	\,, \\
	\label{crossing-2}
	g^{(2)}(z,\zb)=\;
	&\frac{ z^{ \D_{ \smash{\phi} } }\zb^{ \D_{ \smash{\phi} } } }
	{ (1-z)^{ \D_{ \smash{\phi} } }(1-\zb)^{ \D_{ \smash{\phi} } } }
	\Bigg( \frac{ g^{(1)}(1-\zb,1-z) }{ (N-1)(N+1)^2 }
	+\frac{N-2}{N-1} g^{(2)}(1-\zb,1-z)
	\nn
	&\hspace{6em}-\frac{N-2}{ 2N(N+1)(N-1)^2 } g^{(3)}(1-\zb,1-z)
	+\frac{ g^{(4)}(1-\zb,1-z) }{ 2(N-1)(N+1)^2 } \Bigg)
	\,, \\
	\label{crossing-3}
	g^{(3)}(z,\zb)=\;
	&\frac{ z^{ \D_{ \smash{\phi} } }\zb^{ \D_{ \smash{\phi} } } }
	{ (1-z)^{ \D_{ \smash{\phi} } }(1-\zb)^{ \D_{ \smash{\phi} } } }
	\Bigg( g^{(1)}(1-\zb,1-z)
	-(N+1)^2 g^{(2)}(1-\zb,1-z)
	\nn
	&\hspace{10em}+\frac{N^2-N+2}{ 2N(N-1) } g^{(3)}(1-\zb,1-z)
	+\frac{ g^{(4)}(1-\zb,1-z) }{2} \Bigg)
	\,, \\
	\label{crossing-4}
	g^{(4)}(z,\zb)=\;
	&\frac{ z^{ \D_{ \smash{\phi} } }\zb^{ \D_{ \smash{\phi} } } }
	{ (1-z)^{ \D_{ \smash{\phi} } }(1-\zb)^{ \D_{ \smash{\phi} } } }
	\Bigg( -g^{(1)}(1-\zb,1-z)
	+(N+1)^2 g^{(2)}(1-\zb,1-z)
	\nn
	&\hspace{9em}+\frac{(N-2)(N+1)}{ 2N(N-1) } g^{(3)}(1-\zb,1-z)
	+\frac{ g^{(4)}(1-\zb,1-z) }{2} \Bigg)
	\,.
\end{align}
The analytic bootstrap in section \ref{Analytic bootstrap Potts} is based on these crossing equations.

\section{Ratios of OPE coefficients}
\label{Ratios of OPE coefficients}

In the matching conditions \eqref{boxk-current-result} and \eqref{boxk-current-result-phi-2n+1}, we encounter the ratios of OPE coefficients $\frac{ \l_{ \phi^{2n}\phi^{2n}\mathcal{J},\text{f} } }{ \l_{ \phi\phi\mathcal{J},\text{f} } }$, where $n \geqslant 1$.
Moreover, the ratio $\frac{ \l_{ KK\mathcal{J},\text{f} } }{ \l_{ 1,1,\mathcal{J},\text{f} } }$ in \eqref{boxk-current-result-Potts} depends on $N$, the number of components of $\phi_a$.
In this appendix, we determine these ratios by inspecting the conformal block decompositions of four-point functions.

Let us first focus on the $N=1$ case.
We only need the free-theory data here, and we can consider the ratios with $n=1$ and $n>1$ together.
The four-point function of identical scalars reads
\begin{align}\label{4pt-1-1-1-1}
	\< \phi_{ \text{f} }(x_1)
	\phi_{ \text{f} }(x_2)
	\phi_{ \text{f} }(x_3)
	\phi_{ \text{f} }(x_4) \>
	=\[
	1+ \( z\zb \)^{
	\D_{ \phi_{ \text{f} } }
	}+\( \frac{ z\zb }{
	(1-z)(1-\zb) }
	\)^{
	\D_{ \phi_{ \text{f} } }
	} \] \mathcal{K}(\D_i,x_i)
	\,.
\end{align}
The mixed correlator of interest is
\begin{align}\label{4pt-2n-2n-1-1}
	&\< \phi_{ \text{f} }^{2n}(x_1)
	\phi_{ \text{f} }^{2n}(x_2)
	\phi_{ \text{f} }(x_3)
	\phi_{ \text{f} }(x_4) \> 
	\nn=\;&\[ (2n)!
	+(2n)(2n)!\(
	\( z\zb \)^{
	\D_{ \phi_{ \text{f} } }	
	}+\( 
	\frac{z\zb}{(1-z)(1-\zb)}
	\)^{
	\D_{ \phi_{ \text{f} } }	
	} \) \] \mathcal{K}(\D_i,x_i)
	\,,
\end{align}
where the factors $(2n)!$ and $(2n)(2n)!$ come from Wick contractions.
The kinematic factors are given by
\begin{align}
	\mathcal{K}(\D_i,x_i)=
	\frac{1}{
	x_{12}^{\D_1+\D_2}
	x_{34}^{\D_3+\D_4}
	}\(
	\frac{ x_{24} }{ x_{14} }
	\)^{\D_1-\D_2}
	\(
	\frac{ x_{14} }{ x_{13} }
	\)^{\D_3-\D_4}
	\,,
\end{align}
where $\D_i$ is the scaling dimension of the operator at position $x_i$.
In \eqref{4pt-1-1-1-1} and \eqref{4pt-2n-2n-1-1}, the terms in the square brackets have the same functional form in $z$ and $\zb$, so their conformal block decompositions differ by a factor $(2n)(2n)!$.
In other words, there exists a relation between the OPE coefficients
\begin{align}
	(2n)(2n)! \l_{ \phi\phi\mathcal{J} }^2=
	\l_{ \phi^{2n}\phi^{2n}\mathcal{J} }\l_{ \phi\phi\mathcal{J} }
	\,,
\end{align}
which yields the ratio
\begin{align}
	\frac{ \l_{ \phi^{2n}\phi^{2n}\mathcal{J},\text{f} } }
	{ \l_{ \phi\phi\mathcal{J},\text{f} } }
	=(2n)(2n)!
	\,.
\end{align}
In particular, we have $\frac{ \l_{ \phi^{2n}\phi^{2n}\mathcal{J},\text{f} } }{ \l_{ \phi\phi\mathcal{J},\text{f} } }=4$ at $n=1$.

Now we consider the case of $N$-component fields.
The relevant four-point functions are
\begin{align}\label{4pt-1-1-1-1-N}
	&\< \phi_{ a, \text{f} }(x_1)
	\phi_{ b, \text{f} }(x_2)
	\phi_{ c, \text{f} }(x_3)
	\phi_{ d, \text{f} }(x_4) \> \nn
	=\;&\Bigg[
	\(
	1+\frac{
	\(z\zb\)^{
	\D_{ \phi_{ \text{f} } }
	}\(
	1+ \( (1-z)(1-\zb) \)^{
	-\D_{ \phi_{ \text{f} } }
	} \) }{N}
	\)t_{abcd}^{(1)}
	+\frac{
	\(z\zb\)^{
	\D_{ \phi_{ \text{f} } }
	}\(
	1+\( (1-z)(1-\zb) \)^{
	-\D_{ \phi_{ \text{f} } }
	} \) }
	{(N-1)(N+1)^2}t_{abcd}^{(2)} \nn
	&+
	\(z\zb\)^{
	\D_{ \phi_{ \text{f} } }
	}\(
	1+\( (1-z)(1-\zb) \)^{
	-\D_{ \phi_{ \text{f} } }
	} \) t_{abcd}^{(3)} \nn
	&+\(z\zb\)^{
	\D_{ \phi_{ \text{f} } }
	}\(
	1-\( (1-z)(1-\zb) \)^{
	-\D_{ \phi_{ \text{f} } }
	} \) t_{abcd}^{(4)}
	\Bigg] \mathcal{K}(\D_i,x_i)
\end{align}
and
\begin{align}\label{4pt-2-2-1-1-N}
	&
	\< K_{a,\text{f}}(x_1)
	K_{b,\text{f}}(x_2)
	\phi_{ c, \text{f} }(x_3)
	\phi_{ d, \text{f} }(x_4) \> \nn
	=\;&\Bigg[
	\(
	2(N-1)(N+1)^2+\frac{
	4(N-1)(N+1)^2 \(z\zb\)^{
	\D_{ \phi_{ \text{f} } }
	}\(
	1+ \( (1-z)(1-\zb) \)^{
	-\D_{ \phi_{ \text{f} } }
	} \) }{N}
	\)t_{abcd}^{(1)} \nn
	&\quad+\frac{
	4(N-2)\(z\zb\)^{
	\D_{ \phi_{ \text{f} } }
	}\(
	1+\( (1-z)(1-\zb) \)^{
	-\D_{ \phi_{ \text{f} } }
	} \) }
	{N-1}t_{abcd}^{(2)} \nn
	&\quad
	-4(N+1)^2 \(z\zb\)^{
	\D_{ \phi_{ \text{f} } }
	}\(
	1+\( (1-z)(1-\zb) \)^{
	-\D_{ \phi_{ \text{f} } }
	} \) t_{abcd}^{(3)} \nn
	&\quad
	-4(N+1)^2 \(z\zb\)^{
	\D_{ \phi_{ \text{f} } }
	}\(
	1-\( (1-z)(1-\zb) \)^{
	-\D_{ \phi_{ \text{f} } }
	} \) t_{abcd}^{(4)} \;
	\Bigg] \mathcal{K}(\D_i,x_i)
	\,.
\end{align}
For each tensor structure $t_{abcd}^{(i)}$, the two expressions \eqref{4pt-1-1-1-1-N} and $\eqref{4pt-2-2-1-1-N}$ have the same functional form in $z$ and $\zb$.
Then, the ratios of OPE coefficients can be read off from the $N$-dependent factors:
\begin{align}
	R=\frac{ \l_{ KK\mathcal{J},\text{f} } }{ \l_{ \phi\phi\mathcal{J},\text{f} } }=
	\begin{cases}
		\; 4(N-1)(N+1)^{2} \qquad & \text{trivial} \\
		\; 4(N-2)(N+1)^{2}  & \text{standard} \\
		\; -4(N+1)^{2} & \text{symmetric traceless} \\
		\; -4(N+1)^{2} & \text{antisymmetric}
		\,.
	\end{cases}
\end{align}

\section{An OPE coefficient in generalized \boldmath{$\phi^{2n+1}$} theory}
\label{An OPE coefficient in the generalized phi-2n+1 theory}

We are interested in the OPE coefficient $\l_{ \phi\phi\phi^{2n-1} }$.
This coefficient is useful in the analytic bootstrap, as in \eqref{phi-2n-1-contribution}.
We consider the matching condition
\begin{align}
	\lim_{\e \rightarrow 0}
	\( \a^{-2} \< \Box^k\phi(x_1) \phi(x_2) \phi^{2n-1}(x_3) \> \)
	= \< \phi^{2n}_{ \text{f} }(x_1) \phi_{ \text{f} }(x_2) \phi^{2n-1}(x_3) \>
	\,.
\end{align}
At the leading order in the $x_3\rightarrow \infty$ expansion, the condition above yields
\begin{align}
	\lim_{\e\rightarrow 0}&\Bigg[ \a^{-2}2^{2k}
	\(\D_{ \smash{\phi} }-\frac{ \D_{\phi^{2n-1} } }{2}\)_{k}
	\(\D_{ \smash{\phi} }-\frac{ \D_{\phi^{2n-1} } }{2}-\frac{d-2}{2} \)_{k}
	\frac{ \l_{ \phi\phi\phi^{2n-1} } }
	{ x_{12}^{ 2\D_{ \smash{\phi} }+2k-\D_{ \smash{\phi^{2n-1} } } }|x_{3}|^{ 2\D_{ \smash{\phi^{2n-1} } } } } \nn
	&\;\; +O ( \, |x_{3}|^{-2\D_{ \smash{\phi^{2n-1} } }-1 } ) \Bigg]
	=\frac{ \l_{ \phi^{2n}\phi\phi^{2n-1},\text{f} } }
	{ x_{12}^{ 2\D_{ \smash{ \phi^{2n}_\text{f} } }-\D_{ \smash{\phi^{2n-1}_\text{f} } } }|x_{3}|^{ 2\D_{ \smash{\phi^{2n-1}_\text{f} } } } }
	+O ( \, |x_{3}|^{-2\D_{ \smash{\phi^{2n-1}_\text{f} } }-1 } )
	\,.
\end{align}
The constraint on $\l_{ \phi\phi\phi^{2n-1} }$ reads
\begin{align}
	2^{2k} \( 
	\frac{ (3-2n) \(d_{\text{u} }-2k\) }{4} 
	\)_k \(
	\frac{ (3-2n) \(d_{\text{u} }-2k\) }{4}
	-\frac{ d_{ \text{u} }-2 }{2}
	\)_k 
	\lim_{\e \rightarrow 0} \( \a^{-1} \l_{1,1,2n-1}
	\) = (2n)!
	\,,
\end{align}
where the upper critical dimension is given in \eqref{du-phi-2n+1}.
The solution is
\begin{align}\label{OPE-2n-1}
	\l_{ \phi\phi\phi^{2n-1} }=
	\frac{(-1)^k(2n)!\,\a}
	{ 2^{2k}(k)_k \(
	-k\frac{2n-3}{2n-1} \)_k }
	+\ldots
	\,,
\end{align}
which corresponds to \eqref{lambda} at $n=1$.
According to the convention \eqref{OPE-normalization} in the analytic bootstrap, we obtain
\begin{align}\label{OPE-2n-1-crossing}
	P_{ \phi^{2n-1} }=
	\frac{ \l_{ \phi\phi\phi^{2n-1} }^2 }
	{ (2n-1)! }
	\,.
\end{align}
We have used the expressions \eqref{OPE-2n-1} and \eqref{OPE-2n-1-crossing} in \eqref{phi-2n-1-contribution}.

\section{The \boldmath{$N\rightarrow 0$} limit, logarithmic CFTs, and logarithmic recombination}
\label{logCFT}

As mentioned in footnote \ref{logf}, the $N\rightarrow 0$ limit of the Potts model corresponds to a logarithmic CFT.
A key feature is that the tensor structure \eqref{4pt-tensors-3} diverges in the $N\rightarrow 0$ limit.
This singularity can be resolved by mixing bilinear operators.
We will see below that this mixing leads to logarithmic structures and nonvanishing off-diagonal two-point functions.
This means that the primaries are not independent and form logarithmic multiplets, which can be viewed as a logarithmic version of the multiplet recombination.
The discussion in this appendix applies to the canonical case $k=1$ and the higher-derivative cases with $k> 1$.

In the above, the tensor structure \eqref{4pt-tensors-3} is associated with the two-point functions of bilinear operators $\mathcal{J_{\text{ST}}}$ in the symmetric traceless (ST) representation.
To cancel the $1/N$ pole, $\mathcal{J_{\text{ST}}}$ should mix with $\mathcal{J_{\text{trivial}}}$ and 
they form new operators.\footnote{
Although the tensor structure \eqref{4pt-tensors-3} is also singular in the $N \rightarrow -1$ limit, the two-point function of $\phi^2_\text{ST}$ vanishes due to the normalization [see \eqref{ST-ST}].
At order $\e^0$, the OPE coefficient vanishes as well [see \eqref{phi-phi-ST}], but it can be made finite by a singular change in the normalization $\bar{\phi}^2=\frac{1}{N+1}\phi^2_\text{ST}$.
}
For instance, the two-point functions of the spin-0 operators with $m=0$ take the form
\begin{align}
	\<\phi^2_\text{trivial}(x_1)\phi^2_\text{trivial}(x_2)\>
	&=\frac{2N}{
		x_{12}^{ \D_{ \smash{ \phi^2_\text{trivial} } } }
	}\,,\\
	\label{ST-ST}
	\<\phi^2_{ab,\text{ST}}(x_1)\phi^2_{cd,\text{ST}}(x_2)\>
	&=\frac{2(N+1)^2(N-1)^2}{
		x_{12}^{ \D_{ \smash{ \phi^2_\text{ST} } } }
	}t^{(3)}_{abcd}
	\,.
\end{align}
Here the normalization of the two-point functions is the same as that in the free theory, i.e., the two-point functions are given by Wick contractions and $\<\phi_a(x_1)\phi_b(x_2)\>=|x_1-x_2|^{-2\smash{\D_\phi}}\d_{ab}$.
The explicit expressions for $\phi^2_{ab,\text{ST}}$ and $\phi^2_\text{trivial}$ can be found in \eqref{trivial-example} and \eqref{ST-example}.
A linear combination that is regular in the $N\rightarrow 0$ limit reads
\begin{align}\label{mixing}
	\bar{\phi}^2_{ab}=\phi^2_{ab,\text{ST}}-\frac{\d_{ab}}{N}\phi^2_\text{trivial}
	\,.
\end{align}
The coefficient $-1/N$ ensures that the correlation functions $\<\bar{\phi}^2_{ab}(x_1)\bar{\phi}^2_{cd}(x_2)\>$ and $\<\phi_a(x_1)$ $\phi_b(x_2)\bar{\phi}^2_{cd}(x_3)\>$ are regular in the $N \rightarrow 0$ limit. 
This coefficient is not unique 
because one can add more $N$-dependent terms as long as they are not singular in the $N\rightarrow 0$ limit.
In our convention \eqref{mixing}, the two-point function of $\bar{\phi}^2_{ab}$ reads
\begin{align}
	&\lim_{N\rightarrow 0}\<\bar{\phi}^2_{ab}(x_1)\bar{\phi}^2_{cd}(x_2)\> \nn
	=\;&\lim_{N\rightarrow 0}\Bigg[\frac{1}{x_{12}^{2\D_{\smash{\phi^2_{\text{ST}}}}}}\Bigg(
	2\(N^2+N-1\)\d_{ab}\d_{cd}+(N+1)^2(N-1)^2\(\d_{ac}\d_{bd}+\d_{ad}\d_{bc}\)\nn
	&\hspace{3em}-2(N+1)(N-1)\sum_{\a=1}^{N+1} e_{a}^{\a}e_{b}^{\a}e_{c}^{\a}e_{d}^{\a}+\frac{8\e}{7-3N}\log x_{12}\,\d_{ab}\d_{cd}+\ldots
	\Bigg)\Bigg]\nn
	=\;&\frac{1}{x_{12}^{2\D_0}}\(-2\d_{ab}\d_{cd}+2\(\d_{ac}\d_{bd}+\d_{ad}\d_{bc}\)+2\sum_{\a=1}^{N+1} e_{a}^{\a}e_{b}^{\a}e_{c}^{\a}e_{d}^{\a}+\frac{8\e}{7}\log x_{12}\,\d_{ab}\d_{cd}+\ldots\)
	\,.
\end{align}
In the last line, $\D_0$ denotes the degenerate scaling dimension
\begin{align}\label{dengeneracy-scalar-percolation}
	\D_0=\D_{\smash{\phi^2_{\text{ST}}}}\big|_{N\rightarrow 0} =\D_{\smash{\phi^2_{\text{trivial}}}}\big|_{N\rightarrow 0}
	\,,
\end{align}
which can be checked to order $\e^1$ using \eqref{gammaJ-Potts}.
The logarithm arises due to the $\e$ expansion of $x_{12}^{2\D_{\smash{\phi^2_{\text{ST}}}}-2\D_{\smash{\phi^2_{\text{trivial}}}}}$, which comes from the contribution of $\<\phi^2_\text{trivial}(x_1)\phi^2_\text{trivial}(x_2)\>$. 
Here we have used \eqref{gammaJ-Potts}.

Apart from the regularity of the two-point function $\<\bar{\phi}^2_{ab}(x_1)\bar{\phi}^2_{cd}(x_2)\>$, the combination \eqref{mixing} also lead to nonvanishing off-diagonal two-point functions:
\begin{align}\label{2pt-matrix}
	&\lim_{N\rightarrow 0}\begin{pmatrix}
		\<\bar{\phi}^2_{ab}(x_1)\bar{\phi}^2_{cd}(x_2)\> & \d_{cd}\<\bar{\phi}^2_{ab}(x_1)\phi^2_\text{trivial}(x_2)\> \\
		\d_{ab}\<\phi^2_\text{trivial}(x_1)\bar{\phi}^2_{cd}(x_2)\> & \;\d_{ab}\d_{cd}\<\phi^2_\text{trivial}(x_1)\phi^2_\text{trivial}(x_2)\>
	\end{pmatrix} \nn
	=\;&\frac{1}{x_{12}^{2\D_0}}\begin{pmatrix}
		\bar t_{abcd}+\frac{8\e}{7}\log x_{12}\,t^{(1)}_{abcd} & \;-2t^{(1)}_{abcd} \\
		-2t^{(1)}_{abcd} & 0
	\end{pmatrix}
	+\ldots
	\,,
\end{align}
where $t^{(1)}_{abcd}$ is given by \eqref{4pt-tensors-1} and $\bar t_{abcd}$ is defined as
\begin{align}\label{t-bar}
	\bar t_{abcd}=-2\d_{ab}\d_{cd}+2\(\d_{ac}\d_{bd}+\d_{ad}\d_{bc}\)+2\sum_{\a=1}^{N+1} e_{a}^{\a}e_{b}^{\a}e_{c}^{\a}e_{d}^{\a}
	\,.
\end{align}
We interpret the matrix elements as the two-point functions of the primaries in a rank-2 logarithmic multiplet. 
Although the norm of $\phi^2_\text{trivial}$ vanishes in the $N\rightarrow 0$ limit, 
$\phi^2_\text{trivial}$ is not a null state and has nonzero correlation functions with other operators. 

Let us verify that the three-point function $\<\phi_a(x_1)\phi_b(x_2)\bar{\phi}^2_{cd}(x_3)\>$ is regular as well.
We consider the three-point functions of two $\phi$'s and a bilinear operator,
\begin{align}
	\<\phi_a(x_1)\phi_b(x_2)\phi^2_\text{trivial}(x_3)\>&=\(2+O(\e^{1/2})\)\frac{1}{x_{12}^{h_{123}}x_{13}^{h_{132}}x_{23}^{h_{231}}}\d_{ab}\,, \\
	\label{phi-phi-ST}
	\<\phi_a(x_1)\phi_b(x_2)\phi^2_{cd,\text{ST}}(x_3)\>&=\(2(N+1)(N-1)+O(\e^{1/2})\)\frac{1}{x_{12}^{h_{123}}x_{13}^{h_{132}}x_{23}^{h_{231}}}t^{(3)}_{abcd}
	\,.
\end{align}
Here $h_{ijk}=\D_{i}+\D_{j}-\D_{k}$, and $\D_{i}$ is the scaling dimension of the operator at position $x_i$. 
The leading term in the OPE coefficient for $\phi^2_\text{trivial}$ does not depend on $N$, so it remains finite in the $N\rightarrow 0$ limit, as mentioned above. 
However, the tensor structure $t^{(3)}_{abcd}$ in the second line diverges as $N$ approaches zero.
This singular behavior is also resolved by the linear combination \eqref{mixing}, 
yielding the three-point function
\begin{align}
	\lim_{N\rightarrow 0}\<\phi_a(x_1)\phi_b(x_2)\bar \phi^2_{cd}(x_3)\>
	=\;&\(-\bar t_{abcd}+O(\e^{1/2})\)\frac{1}{x_{12}^{h_{123}}x_{13}^{h_{132}}x_{23}^{h_{231}}}\Bigg|_{N\rightarrow 0}
	\,.
\end{align}
Here $\D_{3}$ is understood as $\D_{\smash{\phi^2_{\text{ST}}}}$. 

According to the results for the correlation functions, 
the two independent primaries $\phi^2_\text{trivial}$ and $\phi^2_\text{ST}$ mix and form the logarithmic primary $\bar{\phi}^2_\text{ab}$ in the limit $N\rightarrow 0$.
The two primary states associated with $\bar{\phi}^2_\text{ab}$ and $\phi^2_\text{trivial}$ are not orthogonal to each other, 
giving rise to nonvanishing off-diagonal two-point functions.
In other words, $\{\phi^2_\text{trivial}\}$ and $\{\bar{\phi}^2_\text{ab}\}$ recombine into a logarithmic multiplet.
As the two multiplets are related by the derivative with respect to the scaling dimension $\Delta$, the logarithmic recombination is different from the multiplet recombination based on the derivative with respect to $x^\mu$ in the main text. 
The logarithmic version of the multiplet recombination may deserve further investigations.

Furthermore, the degeneracy \eqref{dengeneracy-scalar-percolation} extends to all bilinear operators in the trivial and symmetric traceless representations.
Using \eqref{gammaJ-Potts}, we obtain
\begin{align}\label{degeneracy}
	&\D_{ \mathcal{J}_{ \text{trivial} } }\big|_{ N \rightarrow 0 }=\D_{ \mathcal{J}_{ \text{ST} } }\big|_{ N \rightarrow 0 }=d-2k+2m+\ell \nn
	&\hspace{5em}+\(
	\frac{1}{3+\frac{(-1)^k k! (3 k)_k}{2^{4 k-5} (1/2)_k^2}}+\frac{(-1)^m(2 k)! (2 k-m)_{k+m}}{\(8 k!+\frac{3\times 2^{4 k-2} (-1)^k (1/2)_k^2}{(3 k)_k}\) m! (k+\ell+m)_{2 k}}
	\)\e+\ldots
\end{align}
These equations hold for arbitrary spin $\ell$, 
so the entire Regge trajectories associated with $\mathcal{J}_{ \text{trivial} }$ and $\mathcal{J}_{ \text{ST} }$ with the same $m$ are degenerate. 
The spinning operators should also exhibit logarithmic structures. 
To order $\e^1$, this agrees with the fact that the $k=1$ crossover exponent is one \cite{wallace1978spin}:
\begin{align}
	\frac{d-\D_{ \phi^2_\text{ST} }|_{ N \rightarrow 0 }}{d-\D_{ \phi^2_\text{trivial} }|_{ N \rightarrow 0 }}=1+0\e+\ldots
	\,.
\end{align}
Note that \eqref{degeneracy} generalizes this relation to $k>1$ and arbitrary spin, to leading order in $\e$.

There is a similar story in the $N\rightarrow 0$ limit of the generalized O($N$)-symmetric $\phi^4$ theory.
For general $N$, the scalar two-point functions of singlet (S) and ST bilinear operators are
\begin{align}\label{ON-tensor}
	\<\phi^2_\text{S}(x_1)\phi^2_\text{S}(x_2)\>&=\frac{2N}{x_{12}^{\D_{\phi_\text{S}}}}\,,\\ 
	\<\phi^2_{ab,\text{ST}}(x_1)\phi^2_{cd,\text{ST}}(x_2)\>&=\frac{2}{x_{12}^{\D_{\phi_\text{ST}}}}\(\frac{\delta_{ac}\delta_{bd}+\delta_{ad}\delta_{bc}}{2}-\frac{1}{N}\delta_{ab}\delta_{cd}\)
	\,,
\end{align}
where $\phi^2_\text{S}=\sum_{i=1}^N\phi_i\phi_i$ and $\phi^2_{ab,\text{ST}}=\phi_a\phi_b-\frac{\d_{ab}}{N}\sum_{i=1}^N\phi_i\phi_i$.
To avoid the singularity in the $N \rightarrow 0$ limit, 
the operator $\phi^2_\text{ST}$ should mix with $\phi^2_\text{S}$  (see e.g. \cite{Hogervorst:2016itc}). 
A regular combination is 
\begin{align}
	\bar{\phi}^2_{ab}=\phi^2_{ab,\text{ST}}+\frac{\d_{ab}}{N}\phi^2_\text{S}
	\,.
\end{align}
As in \eqref{mixing}, the coefficient in front of $\phi^2_\text{S}$ is not unique.
Furthermore, we expect that the degeneracies extend to the spinning operators.
The scaling dimensions of the S and ST bilinear operators indeed become identical in the $N \rightarrow 0$ limit:
\begin{align}
	\D_{\mathcal{J}_\text{S}}\big|_{N \rightarrow 0}=\D_{\mathcal{J}_\text{ST}}\big|_{N \rightarrow 0}
	=\;&d-2k+2m+\ell\nn
	&-\frac{(k-1)! (k+1)_k^2}{128 (k)_{k-m}}\(\frac{(-1)^k}{k! (2 k-m)_{k+m}}+\frac{3(-1)^m}{m! (m+\ell)_{2 k}}\)\e^2+O(\e^3)
	\,.
\end{align}
As in \eqref{degeneracy}, the spin $\ell$ is arbitrary, implying that the entire Regge trajectories are degenerate in the limit $N \rightarrow 0$ for each $m$, 
giving rise to logarithmic structures as in the Potts model.

\section{Generalized self-avoiding walks, loop-erased random walks, and their \boldmath{$\phi^{2n}$} generalizations}
\label{Generalized self-avoiding walks and loop-erased random walks}

As prominent models of random walks, we consider the self-avoiding walks and loop-erased random walks that play important roles in physics.
Interestingly, they can be associated with the O($N$)-symmetric $\phi^4$ theory.
The self-avoiding walks correspond to the $N \rightarrow 0$ limit \cite{deGennes:1972zz} 
\footnote{See \cite{Liu:2012ca} for some Monte Carlo results for the loop model 
at various $N$ in three dimensions, including the case of $N=0$. } and the loop-erased random walks are related to the $N \rightarrow -2$ limit \cite{Nienhuis:1982fx,duplantier1992loop,Wiese:2018dow,Wiese:2019xmu,helmuth2020loop}.
In this appendix, we discuss these limits and their $\phi^{2n}$ generalizations using the results from our previous paper \cite{Guo:2023qtt}.

In \cite{Guo:2023qtt}, we studied the generalized $\phi^{2n}$ theory with O($N$) global symmetry.
The anomalous dimension of $\phi$ at $n=2$ reads
\begin{align}
	\g_{\phi}=
	\frac{(-1)^{k+1}(N+2)(k+1)_k}
	{ 8(N+8)^2 k^2 (2k+1)_{k-1} }\e^2
	+O(\e^3)
	\,,
\end{align}
In the limits $N \rightarrow 0$ and $N \rightarrow -2$, the formula reduces to
\begin{align}
	\g_{\phi}\big|_{N \rightarrow 0}&=
	\frac{(-1)^{k+1}(k+1)_k}
	{ 256 k^2 (2k+1)_{k-1} }\e^2
	+O(\e^3)
	\,, \\
	\g_{\phi}\big|_{N \rightarrow -2}&=
	O(\e^3)
	\,.
\end{align}
For $k=1$, the $N \rightarrow 0$ result agrees with \cite{deGennes:1972zz}. 
Moreover, the anomalous dimension of the symmetric traceless bilinear operator $\phi_{ \text{ST} }^2$ is
\begin{align}
	\g_{\phi_{ \text{ST} }^2}=\frac{2}{N+8}\e+O(\e^2)
	\,,
\end{align}
which is independent of $k$ at leading order in the $\e$ expansion.
We derive the fractal dimension (or Hausdorff dimension) \cite{Shimada:2015gda} 
\begin{align}
	D_{ \text{F} }=d-\D_{\phi_{ \text{ST} }^2}
	=2-\frac{2}{N+8}\e+O(\e^2)
	\,,
\end{align}
where the spacetime dimension is $d=4k-\e$ and the fractal dimension $D_{ \text{F} }$ is $k$-independent to order $\e$.
In particular, the limits $N \rightarrow 0$ and $N \rightarrow -2$ yield
\begin{align}
	D_{ \text{F,self-avoiding} }
	=2-\frac{\e}{4}+O(\e^2)
	\,, \qquad
	D_{ \text{F,loop-erased} }
	=2-\frac{\e}{3}+O(\e^2)
	\,.
\end{align}

Let us further discuss the $n>2$ counterpart of the special $N$ limits.
The $n>2$ theories are expected to be the multicritical generalizations of the critical O($N$) model at $d=4-\e$.
The tensor structures \eqref{ON-tensor} are independent of $n$, so the $N \rightarrow 0$ limit is also singular in the case of $n>2$.
The anomalous dimensions of $\phi$ read
\begin{align}
	\g_\phi\big|_{N \rightarrow 0}&=\frac{(-1)^{k+1} (n-1)^3 ((n-1)!)^2 \left(1+\frac{k}{n-1}\right)_k}{4 k^2 \,
	_3F_2\left(1-n,\frac{1}{2}-\frac{n}{2},-\frac{n}{2};1,\frac{1}{2}-n;1\right)^2 \(1+\frac{nk}{n-1}\)_{k-1}
	(n)_n^2}\e^2+O(\e^3)
	\,.
\end{align}
The scaling dimensions of bilinear operators in the singlet and symmetric traceless representations are degenerate:
\begin{align}
	\D_{_{\mathcal{J}_\text{S}}}\big|_{N \rightarrow 0}
	=\;&\D_{_{\mathcal{J}_\text{ST}}}\big|_{N \rightarrow 0}
	=d-2k+2m+\ell+\frac{((n-1)!)^2 (n-1)^2 \left(1+\frac{k}{n-1}\right)_k}
	{2(n)_n^2 \, _3F_2\left(1-n,\frac{1}{2}-\frac{n}{2},-\frac{n}{2};1,\frac{1}{2}-n;1\right)^2} \nn
	&\hspace{2em}\times\left(\frac{(-1)^{k+1} (n-1)}{k^2 \left(1+\frac{n
	k}{n-1}\right)_{k-1}}-\frac{(-1)^m(k-1)! (2 n-1) \left(1+\frac{k}{n-1}\right)_k}{(k)_{k-m} \,m! \left(\ell+m-\frac{(n-2) k}{n-1}\right)_{2
	k}}\right)+O(\e^3)
	\,.
\end{align}
Besides the singular tensor structure, we can identify more special $N$ limits by requiring that the bilinear operators are degenerate.
There are three possibilities:
\begin{itemize}
	\item
	$\D_{_{\mathcal{J}_\text{S}}}|_{N \rightarrow 0}=\D_{_{\mathcal{J}_\text{ST}}}|_{N \rightarrow 0}$: \\
	These correspond to the case of the singular tensor structure above.
	\item	
	$\D_{_{\mathcal{J}_\text{S}}}|_{N \rightarrow -2}=\D_{_{\mathcal{J}_\text{A}}}|_{N \rightarrow -2}=d-2k+2m+\ell+O(\e^3)$: \\
	Here A denotes the antisymmetric representation.
	These degeneracies occur due to the vanishing of anomalous dimensions.
	\item	
	$\D_{_{\mathcal{J}_\text{all three reps}}}|_{N \rightarrow -4,-6,\ldots,-2n+2}=d-2k+2m+\ell+O(\e^3), \quad(n>2)$: \\
	These degeneracies are similar to those in the $N \rightarrow -2$ case, except that here all three types of bilinear operators are degenerate.
\end{itemize}
We emphasize that the spin $\ell$ is arbitrary, so all of the degeneracies above involve the entire Regge trajectories.
The latter two scenarios also lead to
\begin{align}
	\g_\phi|_{N\rightarrow -2,-4,\ldots,-2n+2}=O(\e^3)
	\,.
\end{align}

We also comment on the Chew-Frautschi plots.
In the $k=1$ case, the positive vertical-axis intercepts associated with $\D=d/2$ are
\begin{align}
	\ell^*_\pm=\frac{n-2}{n-1}+\(\frac{1}{2}\pm\frac{ n \sqrt{r}}{ 2^{n-3} (n)_n \,
	_3F_2\left(\frac{1}{2}-\frac{n}{2},-\frac{n}{2},-n-\frac{N}{2}+1;1,\frac{1}{2}-n;1\right)}\)\e+O(\e^2)
	\,,
\end{align}
where the ratio of OPE coefficients in the O($N$)-symmetric $\phi^{2n}$ theory is
\begin{align}
	r=\begin{cases}
		4^{n-1}(2n-1)(n-1)!\(\frac{N}{2}+1\)_{n-1}&\quad\!\!\!\text{singlet,}\\
		2^{2n-3}(4n+N-2)(n-1)!\(\frac{N}{2}+2\)_{n-2}&\quad\!\!\!\text{symmetric traceless,}\\
		4^{n-1}(n-1)!\(\frac{N}{2}+1\)_{n-1}&\quad\!\!\!\text{antisymmetric.}
	\end{cases}
\end{align}
From this perspective, the limit $N \rightarrow -2$ is special in the sense that the two intercepts $\ell^*_\pm$ become identical for the singlet and antisymmetric representations, respectively, and the limits $N \rightarrow -4,\ldots,-2n+2$ are special because $\ell^*_+=\ell^*_-$ for all three representations.

For $k>1$, there are three qualitative differences.
First, the Chew-Frautschi plots contain more curves.
Second, there exist trajectories associated with higher $m$.
Third, for each $m$, the equation $\D=d/2$ has more solutions, indicating that more intersections of the trajectory with the vertical axis can be found.
These features are similar to those in the generalized Potts model, which are discussed at the end of section \ref{Multiplet recombination Potts}.

\bibliographystyle{JHEP}
\bibliography{references}

\end{document}